\documentclass[11pt, a4paper]{article}
\usepackage{fancyhdr}
\usepackage{graphicx}
\usepackage{color}
\usepackage{ccaption}
\usepackage{booktabs}
\usepackage{multirow}
\usepackage{moresize}
\usepackage{subfig}
\usepackage{amsmath}
\usepackage{amssymb}
\usepackage{amsfonts}

\advance\textwidth +40pt 
\advance\oddsidemargin -20pt

\title{}
\author{Tomas Kasemets\\
Supervisor: Torbjorn Sj\"{o}strand}

\fancypagestyle{plain}{
\fancyhf{}
\fancyfoot[LE,RO]{\thepage}
}

\newcommand{\stext}[1]{\mbox{\ssmall{#1}}}

\captionnamefont{\bfseries}
\captiontitlefont{\small}
\captiondelim{: }
\definecolor{Farg}{rgb}{0.7,0.7,0.7}

\begin{document}
\thispagestyle{empty}
\begin{flushright}
LU TP 10-03\\
MCnet/10/03\\
February 15, 2010
\end{flushright}
\vfill
\begin{center}
{\Large\bf Inclusion of \\[0.2cm] Parton Distribution Functions in PYTHIA8}\\[1.5cm]
Tomas Kasemets\\[0.3cm]

Master Thesis in Theoretical High Energy Physics \\Department of Astronomy and
Theoretical
Physics, Lund University \\S\"olvegatan 14A, SE 223 62 Lund, Sweden\\[0.3cm]

Thesis Advisor: Torbj\"{o}rn Sj\"{o}strand
\end{center}
\vfill
\begin{abstract}
A selection of the latest and most frequently used PDFs is incorporated in
\textsc{Pythia8}, including the MC-adapted PDFs
from the MSTW and CTEQ collaborations. This thesis examines the differences
in PDFs as well as the effect they have on results of simulations. The results are also compared to data
collected by the CDF experiment.
\end{abstract}
\vfill
\newpage
\tableofcontents
\newpage
\section{Introduction}
In order to learn more about the inner essence of nature, particle physicists smash together
small particles, such as protons, at large energies, and use huge
detectors to detect whatever comes out. The higher the energy, the smaller
distances can be examined, which also means that nature at smaller scales can
be studied and reveal its secrets. The story is in fact much more
complicated. 

The protons that clash together are composite particles made
up by quarks and gluons that cannot be isolated and studied in their
own, but are always confined inside the proton. The collision
between two protons is therefore better described as a collision of two approaching
bunches of particles, and for the outcome to tell us about nature we need to know what takes part in the collision. Therefore, so called
parton distribution functions (PDFs) describe how the momentum of the proton
is shared between partons, i.e. quarks and gluons. The PDFs make up
one of the ingredients of computer simulations, which combine theories and
models in order to predict
the outcome, and make it possible to test theory against experiment.

In this thesis we have included the very
latest parton distribution functions in the Monte Carlo event generator \textsc{Pythia8}
\cite{Pythia8}. We compare them and examine how the difference in PDFs affect results
of simulations. We also compare the results to real data collected by experiments
at Tevatron \cite{Tevatron} and examine how the differences change when
increasing the energy up to the level of a fully operational LHC (14~TeV)
\cite{LHC}.

The
structure of the thesis is as follows. In section 2 we introduce some
of the concepts of the theory of strong interactions, Quantum Chromo Dynamics (QCD). In section 3 we move on to describe Monte Carlo
generators (such as \textsc{Pythia8}) and section
4 tells the story of parton distribution functions. Section 5
describes how the PDFs were included in \textsc{Pythia8} and we compare the
different PDFs in section 6. Subsequently we study results from simulations of minimum bias
events and hard QCD events in section 7 and 8 respectively. Finally we
conclude with summary and conclusions in section 9.

\section{Quantum Chromo Dynamics}
The Standard Model of particle physics is the joint theory of three of the four
fundamental forces of nature: electromagnetism, weak and strong force. The
part describing the strong interactions, Quantum Chromo Dynamics, is
responsible for holding together protons and neutrons in the nucleus as well
as 
quarks inside the nucleons. The strong force is mediated by gluons, which
interact with particles that carry color charge, and it grows stronger with
distance. Colored particles are therefore confined inside
colorless hadrons such as the proton and can never be observed and studied as
free particles.

Most calculations in QCD rely on a perturbative expansion in the coupling of
the theory. Perturbative QCD uses the Feynman diagrams and rules to calculate matrix elements and
thereby obtain cross sections for different processes. This technique
relies on the coupling of the strong force, $\alpha_S$, to be small enough to make the
expansion converge and allow the first terms to be a good approximation. At
high energies the strong force is weak, the quarks experience asymptotic
freedom and the approximation of truncating the expansion at low orders
is good. As one moves down to lower energy, increasingly higher orders would be
required, but the perturbation expansion
in $\alpha_S$ quickly becomes very complicated and at small enough scales it finally breaks down completely. There are many possible Feynman graphs for each
process and all of them will contribute to the cross section. At
next-to-leading order the diagrams start to involve loops which introduce integrals
over the phase space of the internal lines. These loop integrals diverge, but
at each order the real (extra particle in the final state) and virtual (extra particle internal) divergences 
 combine, in a far from trivial way, to cancel out
all infinities and leave finite results. This results in very complicated
calculations and therefore many processes have only been calculated to leading
order. The matrix element calculation gives rise to divergences in two cases, when two partons
are collinear and when the energy of a parton is small, soft
divergence. How accurate the approximation of the low order expansion is depends on the size
of the coupling $\alpha_S$.


\subsection{Running Coupling}
The running of the coupling, $\alpha_S=\frac{g^2}{4\pi}$, is necessary in
order to absorb infinities in the theory. This is called renormalization and the running is determined by the renormalization group 
equation \cite{Peskin}
\begin{equation}
\frac{d}{d\log(Q/M)}g = \beta (g), 
\end{equation}
where 
\begin{equation}
\beta(g)=-\frac{b_0}{(4\pi)^2}g^3 - \frac{b_1}{(4\pi)^4}g^5 + \ldots
\end{equation}
The two constants depend on the number of flavors, $n_f$, that have their threshold
below the energy scale, $b_0=11-\frac{2}{3}n_f$ and  $b_1=\frac{153-19n_f}{2\pi (33-2n_f)}$. Solving this
equation and introducing the mass scale $\Lambda$ yields
\begin{equation}
\alpha_s(Q^2) = \frac{4\pi}{b_0}\frac{1}{\log(Q^2/\Lambda^2)} - \frac{4\pi
  b_1}{b_0^3}\frac{\log\log(Q^2/\Lambda^2)}{(\log(Q^2/\Lambda^2))^2} + \ldots
\end{equation}
The first term is the first order expression and the dots are terms that
decrease in relative importance for large $Q^2$. From this it can be seen that
$\alpha_S$ decreases at large $Q^2$ as $1/\log(Q^2)$ and therefore become very small,
but also that the coupling increases towards infinity as $Q^2$ approaches $\Lambda^2$ .

\subsection{Cross Section}
The cross section describes how likely it is that an interaction will take
place. In the regions where the matrix elements can be calculated these give the
cross sections for hard sub-processes, such as $q(p_1)g(p_2)\rightarrow
q(p_3)g(p_4)$. To describe the cross section for the sub-process we introduce the Mandelstam
variables,
\begin{eqnarray}
\hat{s}&=&(p_1+p_2)^2 \\
\hat{t}&=&(p_1-p_3)^2 \\
\hat{u}&=&(p_2-p_3)^2,
\end{eqnarray}
where $p_1$-$p_4$ are the four-momenta of the particles. The differential cross section is given by 
\begin{equation}
\frac{E_3E_4d^6\hat{\sigma}}{d^3p_3d^3p_4}=\frac{1}{2\hat{s}}\frac{1}{16\pi}\bar{\sum}\left|{\mathcal{M}}\right|^2\delta^4(p_1+p_2-p_3-p_4),
\end{equation}
where $E_3$ and $E_4$ are energies of the two outgoing partons and the
$\delta$-function ensures conservation of energy and momentum.
$\bar{\sum}$ is the sum (average) over initial- (final-) \linebreak state spins and
colors. $\mathcal{M}$ is the matrix element and for the $qg\rightarrow
qg$ process, assuming massless partons,
\begin{equation}
\frac{1}{g^4}\bar{\sum}\left|{\mathcal{M}}\right|^2=+\frac{\hat{s}^2+\hat{u}^2}{\hat{t}^2}-\frac{4}{9}\frac{\hat{s}^2+\hat{u}^2}{\hat{s}\hat{u}},
\end{equation}
to leading order.
The sub-process cross
section, $\hat{\sigma}(\hat{s}, \hat{t}, \alpha_S(Q^2),
\mu^2)$, is a function of the momenta of the partons, the value of the strong coupling at a relevant
energy scale $Q$ of the process
 and the factorization scale $\mu$. $\mu$ can
be seen as the scale which separates long- and short-distance physics \cite{LHCphysics}, i.e.
partons with transverse momentum less than $\mu$ are considered as part of the
proton and are absorbed into the PDF. The standard choice is to
set $\mu=Q$ \cite{Peskin}, \cite{QCDbok}.

Assuming no $p_{\perp}$ of the interacting partons, the center-of-mass energy squared for the entire
collision, $s$, and for the colliding sub-system, $\hat{s}$, are related by
\begin{equation}
\hat{s}=(x_1P_1+x_2P_2)^2\approx x_1x_2s.
\end{equation}
Hence the product $x_1x_2$ determines the energy fraction available in the
sub-process, while the ratio gives the rapidity
\begin{equation}
y=\frac{1}{2}\ln\left(\frac{x_1}{x_2}\right),
\end{equation}
 and thereby determines the direction of
motion of the colliding sub-system. In large rapidity events the interacting
partons therefore have
very different momentum fractions. The average transverse momentum is often
good as an energy scale of the process and an upper limit on the transverse
energy, $p_{\perp}$, is set by the energy available in the subprocess
\begin{equation}
p^2_{\perp} \leq \frac{x_1x_2s}{4} ,
\end{equation}
with equality only for back to back scattering of two partons perpendicular to
the beam axis. In minimum bias events a typical $p_{\perp}=2$~GeV, setting a
lower limit on the product, $x_1x_2 \geq \frac{4p_{\perp}^2}{s} \sim 10^{-7}$ at
the LHC, for such an event.

The cross section from the matrix element calculations divergence as 
\begin{equation}
\frac{d\sigma}{dp_{\perp}^2} \sim \frac{1}{p_{\perp}^4},
\end{equation}
when $p_{\perp}
\rightarrow 0$. This must be regularized for low transverse momentum which is
accomplished by introducing a parameter, $p_{\perp0}$, (to be discussed more in section \ref{sec:minbias})  such that 
\begin{equation}
\label{eqn:pt0}
\frac{d\sigma}{dp_{\perp}^2} \sim \frac{1}{(p^2_{\perp} + p^2_{\perp0})^2}.
\end{equation}

\begin{figure}[tp]
\begin{center}
\includegraphics[width=0.6\textwidth]{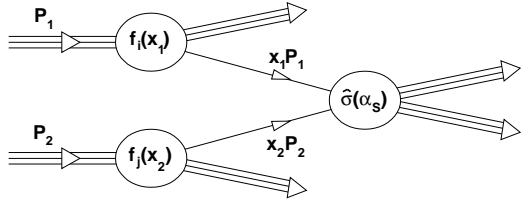}
\caption{Two incoming protons
  with momentum $P_1$ and $P_2$. One parton from each respective proton, with
  momentum fractions $x_1$ and $x_2$, take part
  in a hard scattering sub-process producing final-state particles.}
\label{fig:feynman}
\end{center}
\end{figure}

To obtain the total cross section for the process when two
protons collide the probability to find the partons with momentum fraction $x$ inside the proton have to be taken into account, see Fig.~\ref{fig:feynman}. The total cross
section is then obtained by summing over the different possible partons and
integrating over the allowed momentum fractions $x$,
\begin{equation}
\label{eqn:cross}
\sigma(P_1, P_2) = \sum_{ij}
\int dx_1dx_2f_i(x_1,Q^2)f_j(x_2,Q^2)\hat{\sigma}(x_1P_1, x_2P_2, \alpha_S)
\end{equation}
where $P_1$ and $P_2$ are the momenta of the two incoming protons. $f_i$ is
the parton distribution function of parton i, which will be described in more
detail in section \ref{chap:pdf}.



\section{Monte Carlo Generators}
Experiments such as proton collisions at high energy are too complicated to
make predictions based solely on calculations from first principles of the Standard
Model and therefore one has to rely on Monte Carlo generators. The generators
start with a central hard collision and then combine
matrix elements, parton showers, multiple parton-parton interactions and
hadronization models to predict the
outcome.\cite{LesHouches}

As described in the previous chapter, matrix-element calculations do not
describe QCD accurately over the entire phase space. In regions where higher
order terms are necessary to give good approximations the MC generators use
the technique of parton showers, which approximates the higher order terms. At each step in the shower the emission
probability is calculated and an emission is generated, building up an event
with many outgoing partons. This method is a good approximation when the
shower can be strongly ordered, i.e. the ordering variable such as virtuality decreases with each
emission. That is the case for the largest part of the cross section but not
for the hard interactions, where outgoing partons escape at large angles \cite{Lavesson}. The parton showers can be subdivided into two types, initial-state radiation (ISR)
and final-state radiation (FSR). As the names suggest, ISR describes how the
incoming proton branches into many quarks and gluons before the collision while FSR takes the
products of the collision and describes how they branch into more
particles. FSR is evolved in strictly ordered virtuality where at each
step the virtuality is smaller than in the previous. ISR is more complicated. It
is ordered in increasing virtuality but since the simulation starts with the hard interaction the ISR has to be
evolved backwards \cite{ISR-Sjostrand}. This backward
evolution of ISR is one of the most challenging parts of a
generator \cite{MCGen}.

The showers and matrix elements are good approximations in two entirely
different regions. Matrix elements are good for hard collisions, producing
particles at large transverse momentum while parton showers are good at low
energies and for collinear partons. In order to make use of their respective
advantages they need to be matched at some energy scale to
cover the entire phase space and in a way that assures there is no double
counting \cite{Friberg}, \cite{Lavesson}.

 \textsc{Pythia8} uses showers ordered in $p_{\perp}$ which has an advantage
over mass ordered showers (which were used in the previous version \textsc{Pythia6.4} \cite{Pythia6}) because it automatically takes care of some effects due to coherence
between emissions. These, however, are not the only possible choices and for
example HERWIG \cite{HERWIG}, the other large general Monte Carlo generator,
uses energy weighted angular ordering. This has advantages related to coherence effects but can not
be used over the full phase space and hence leaves some regions to be filled in
by high order matrix elements \cite{MCGen}.

After the FSR the hadronization starts, and the partons produced in the
previous stages split up and combine into colorless hadrons than can survive and
possibly reach a detector. It is also at this level that experiments collect
their data. The hadronization has not yet been possible to determine from the
theory and therefore models are required. \textsc{Pythia8} uses the Lund string model
\cite{LundString} where partons tie color connections to each other which
break by forming quark--antiquark pairs, while HERWIG uses cluster fragmentation \cite{ClusterFrag}.

There is no reason why, in a proton collision, there should only be one
parton from each proton that interacts, and in fact it is not even the case
most of the time. Therefore the so called multiple interaction framework
simulates the additional interactions. As the collision energy increases,
partons at smaller $x$ become involved and more
particles can be produced in the ISR. Multiple interactions are therefore
increasing and the models for multiparticle interactions are put to the
test as the energy at LHC rises.

PDFs are used in Monte Carlo generators at several stages. First in
calculating the cross section for the hard collision to take place, which we
have seen in eqn.~\ref{eqn:cross}. Secondly, ISR use the PDFs since it is the partons inside the proton that
will shower, so the PDFs are needed to describe what we have to start
with and is used in the backwards evolution of the ISR. Finally, PDFs also enter
through cross section calculations in the  multiple interactions framework.

\section{Parton Distribution Functions}
\label{chap:pdf}
The static properties of the
proton are dominated by the valence quarks, i.e. two up quarks and one down
quark. The dynamic picture of the proton is less
simple. Quarks interact by exchanging gluons, which can split up into
quark--antiquark pairs and send out additional gluons. The proton thus consists
of three valence
quarks, gluons and sea quarks which all go under the common name of
partons. Parton distribution functions (PDFs) describe how the momentum of the proton is
distributed among partons and at leading order they can be
interpreted as the probability to find a parton inside the proton with a
certain
momentum fraction. At higher order this simple picture fails since, for
example, the NLO PDFs can be negative. 

There are six quarks with antiquarks
and adding the gluon, the proton has,
potentially, 13 PDFs. However, the heavy top quark/antiquark is not included taking the number down to eleven. When a
gluon splits it is always into a quark--antiquark pair, and hence the
distributions of quarks and antiquarks should be the same (once the valence
quarks are excluded). Therefore strange, charm and bottom quarks used to be
considered to have exactly the
same distribution as their respective antiquark, setting the total number of
distributions to 8. However this is not necessarily true and this symmetry can
be broken, which is also the case (at least) for the strange
quark/antiquark. The proton can be seen as accompanied by a kaon cloud, since
the proton can split into a $\Lambda(uds)$ and a $K^+(u\bar{s})$, where the $s$ quark sits in
the first and the $\bar{s}$ in the second. There is no reason to believe that the
$s$ quark in the $\Lambda$ should have the same momentum as the $\bar{s}$ in the $K^+$ \cite{Chen}. Therefore some of the
newer PDFs have a small difference between the two distributions. 

The PDFs obey a number of relations such as the the momentum sum
rule 
\begin{equation}
\int_0^1dx\sum_i xf_i(x) = 1,
\end{equation}
where the sum is over all partons in the PDF, which states that the total
momentum of all the partons must equal the momentum of the proton. 

The PDFs are functions of the fraction of the proton's momentum $x$ carried by
the parton and of $Q^2$, which can be interpreted as a measure of resolving
power. 
Once the parton distributions are known for one specific $Q^2$, they can be evolved
to higher $Q^2$ by use of the DGLAP equation \cite{Martin}
\begin{equation}
\frac{\partial f_i(x,Q^2)}{\partial Q^2} =
\frac{\alpha_s}{2\pi}\sum_j\int_x^1\frac{dy}{y}f_j(y,Q^2)P_{j\rightarrow i}\left(\frac{x}{y}\right),
\end{equation}
where $f_q$ is the parton distribution function for parton $q$ and
$P\left(\frac{x}{y}\right)$ is the splitting function. 

The outcome of experiments depend on how the momentum is distributed
but there is no way to directly calculate what the distribution should
be. Therefore several groups work on parameterizing and fitting PDFs to
available data \cite{mstw2008}-\cite{NLOPDF}. The large $x$ behavior of the PDFs
are constrained by fixed target experiments. At smaller $x$ the constraints
come from deep inelastic scattering (DIS) at electron-proton colliders such
as HERA. The DIS data have high statistics and therefore dominate the
PDF fits. This gives good knowledge of the quark distributions but the gluon
distribution is harder to obtain, since gluons do not directly interact with
electrons. In hadron colliders the picture is much more messy, since there is
now two composite particles and the gluon distribution is therefore the least
constrained part of the PDFs. For more details on the data sets of PDF fits see \cite{LHCPDF-Huston}
and \cite{mstw2008}.

\subsection{Monte Carlo-Adapted PDFs}
There are PDFs of several different orders, LO, NLO and some NNLO. The general-purpose generators are all leading order and therefore one would like to
combine them with leading-order PDFs. However, we know for example that higher
orders generally give positive contributions to cross sections and in recent years some modified LO PDFs have
been released, specifically tailored for leading-order Monte Carlo
generators. These try to simulate some of the effects of NLO calculations by
compensating for known shortcomings of the leading-order. Among other things these PDFs allow for a non-conservation
of momentum by relaxing the momentum sum rule, i.e. the partons inside the
proton are allowed to have a total momentum larger than the momentum of the
proton. This permits the PDFs to grow large in some regions without decreasing
in others and thereby simulate the effects of some of the next-to-leading-order
corrections, in particular allowing a large value of the gluon distribution at small
$x$ without compromising the quark distributions at large $x$.
The MC-adapted PDFs released so far are LO* and LO** \cite{mrstmod} from the MRST group and
MC1, MC2 and MCS \cite{cteqmc} from the CTEQ group. All of them except MCS have
relaxed the momentum sum rule. MC1 use a leading-order running of $\alpha_S$
while LO*, LO** and MC2 use next to leading-order running. LO** also has a change
in argument, to $p_{\perp}^2$ rather than $Q^2$, for $\alpha_S$ for high-x evolution. MCS has more freedom in the
parameterization and allows for change with scale, to simulate NLO cross section
calculations (a feature we do not make use of in \textsc{Pythia8}). MC1/2/S are all fitted to a combination of real data and NLO
pseudo data in an attempt to obtain the ideal PDF for leading-order MC generators.




\section{PDFs in PYTHIA8}
\textsc{Pythia8} \cite{Pythia8}
has so far been distributed with the option to choose between two PDFs,
GRV94L \cite{grv} and CTEQ5L \cite{cteq5l}, which are both fairly old. Many
new and improved PDFs have been released and made available to \textsc{Pythia8}
simulations only through LHAPDF \cite{lhapdf}. The LHAPDF package has grown quite large and in that process also a
bit slow, also the code is written in Fortran while the community is changing to C++. It is desirable to include some PDFs directly into \textsc{Pythia8} because it can speed up simulations, make \textsc{Pythia8} more complete and make it
easier to switch between different frequently used PDFs. Furthermore some of the latest PDFs
have not yet been included in LHAPDF. Therefore we incorporate
ten new PDFs from the MRST \cite{mrstmod}-\cite{mrstmod2}, MSTW 2008 \cite{mstw2008}, CTEQ6
\cite{cteq6} and CTEQ MC \cite{cteqmc} distributions into \textsc{Pythia8}. Two of them
are NLO which are not intended for MC use, but included for comparison. The main
danger with them is for low-$p_{\perp}$ processes. Inclusion of the PDFs was done in
cooperation with the MSTW and CTEQ collaborations, \cite{Thorne}-\cite{Huston}, and
the PDFs are listed in Tab.~\ref{pdftab}.
\\
\begin{table}[tp]
  \center\small
  \begin{tabular}[h]{lll}    \toprule
    \multirow{2}{*}{Previous} & \multicolumn{2}{c}{New} \\
    \cmidrule(r){2-3}
    & MRST/MSTW & CTEQ \\
    \midrule
    GRV94L &  MRST LO* & CTEQ6L \\
    CTEQ5L &  MRST LO** & CTEQ6L1   \\
    & MSTW LO & CTEQ66  \\
    & MSTW NLO &  CT09MC1 \\
    & & CT09MC2   \\
    & & CT09MCS   \\
    \bottomrule
  \end{tabular}
  \caption{PDFs that are now included in \textsc{Pythia8}. We included all but two, which were already available in \textsc{Pythia8}.}
  \label{pdftab}
\end{table}
Including additional PDFs proved to be less straightforward than might first
be expected. A major reason for this is the need to, in MC simulations, go outside the range of the PDF grids. Specifically we need to go down to
smaller $x$ and $Q^2$ values than many of the distributions. At LHC energies, $x$
values as low as $10^{-8}$ are desirable, while some of the PDFs only range
down to $10^{-6}$, and multiple interactions take place at low
$Q^2$. MSTW provides routines not only for interpolation but also
for extrapolation outside this grid while the CTEQ collaboration has
recommended a freeze of the PDFs at the value just inside the
grid. The range of the grids for the different PDFs are shown in Tab.~\ref{pdfrange}.

The code supplied by the authors had to be modified to fit natively
into \textsc{Pythia8} and we also did extensive tests. When possible the tests included
comparisons to the corresponding PDFs in the LHAPDF package. We then found that
the \textsc{Pythia8} included PDFs run about a factor two faster than they do going the
way via the LHAPDF package.
 
The $s$ and $\bar{s}$ distributions were set equal in previous versions of
\textsc{Pythia8} and since that was not the case in some of the new PDFs, P\
\textsc{Pythia8} was modified to support such a difference.
The different PDFs have different values of $\alpha_S$ and also use
different orders of the running, as listed in Tab.~\ref{pdfrange}. The \textsc{Pythia8} default is to use first order
running for all $\alpha_S$ but this can be changed in the settings and we
examine the effects that such a change can have.

\begin{table}[tp]
  \center\small
  \begin{tabular}[h]{lllll}
    \toprule
    PDF & $x$ range & $Q^2$ range [GeV$^2$] & $\alpha_S$ & $\alpha_S(M_Z)$
    \\
    \midrule
    GRV94L & $10^{-5} - 1$ & $0.40 - 10^6$ & LO & 0.128 \\ 
    CTEQ5L & $10^{-6} - 1$ & $1.00 - 10^8$ & LO & 0.127 \\
    MRST LO* & $10^{-6} - 1$ & $1.00 - 10^9$ & NLO & 0.12032 \\
    MRST LO** & $10^{-6} - 1$ & $1.00 - 10^9$ & NLO & 0.11517 \\
    MSTW LO & $10^{-6} - 1$ & $1.00 - 10^9$ & LO & 0.13939 \\
    MSTW NLO & $10^{-6} - 1$ & $1.00 - 10^9$ & NLO & 0.12018 \\
    CTEQ6L & $10^{-6} - 1$ & $1.69 - 10^8$ & NLO & 0.1180 \\
    CTEQ6L1 & $10^{-6} - 1$ & $1.69 - 10^8$ & LO & 0.1298 \\
    CTEQ66 (NLO)  & $10^{-8} - 1$ & $1.69 - 10^{10}$ & NLO & 0.1180 \\
    CT09MC1 & $10^{-8} - 1$ & $1.69 - 10^{10}$ & LO & 0.1300 \\
    CT09MC2  & $10^{-8} - 1$ & $1.69 - 10^{10}$ & NLO & 0.1180 \\
    CT09MCS  & $10^{-8} - 1$ & $1.69 - 10^{10}$ & NLO & 0.1180 \\
    \bottomrule
  \end{tabular}
  \caption{The $x$ and $Q^2$ ranges of the grids for the different parton
    distribution functions, as well as the order of the running of $\alpha_S$
    and the value at $M_Z$.}
  \label{pdfrange}
\end{table}

\subsection{MRST/MSTW}
\begin{figure}[tp]
  \centering
  \subfloat[]{\label{negGluon1}\includegraphics[width=0.5\textwidth]{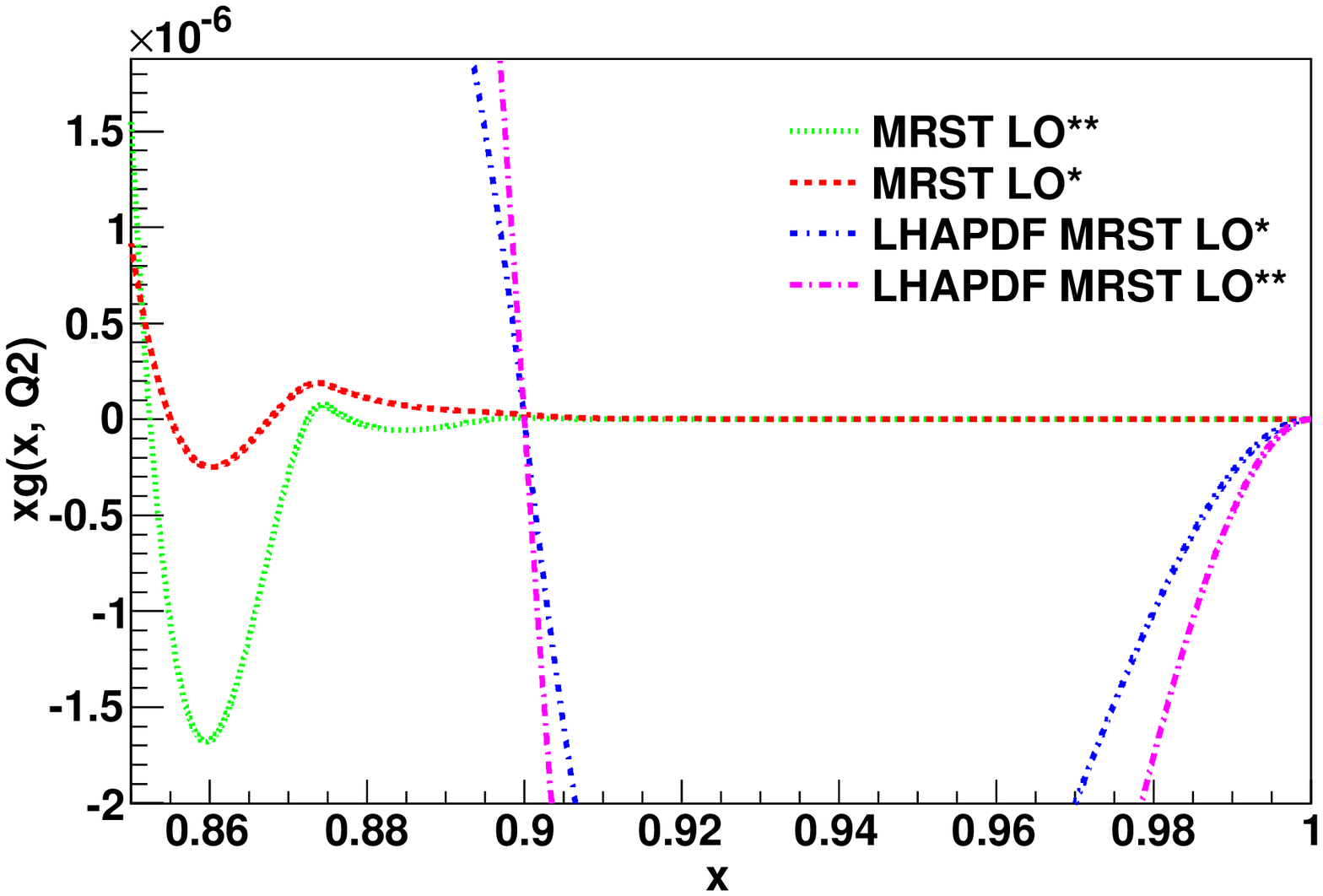}}                
  \subfloat[]{\label{negGluon2}\includegraphics[width=0.5\textwidth]{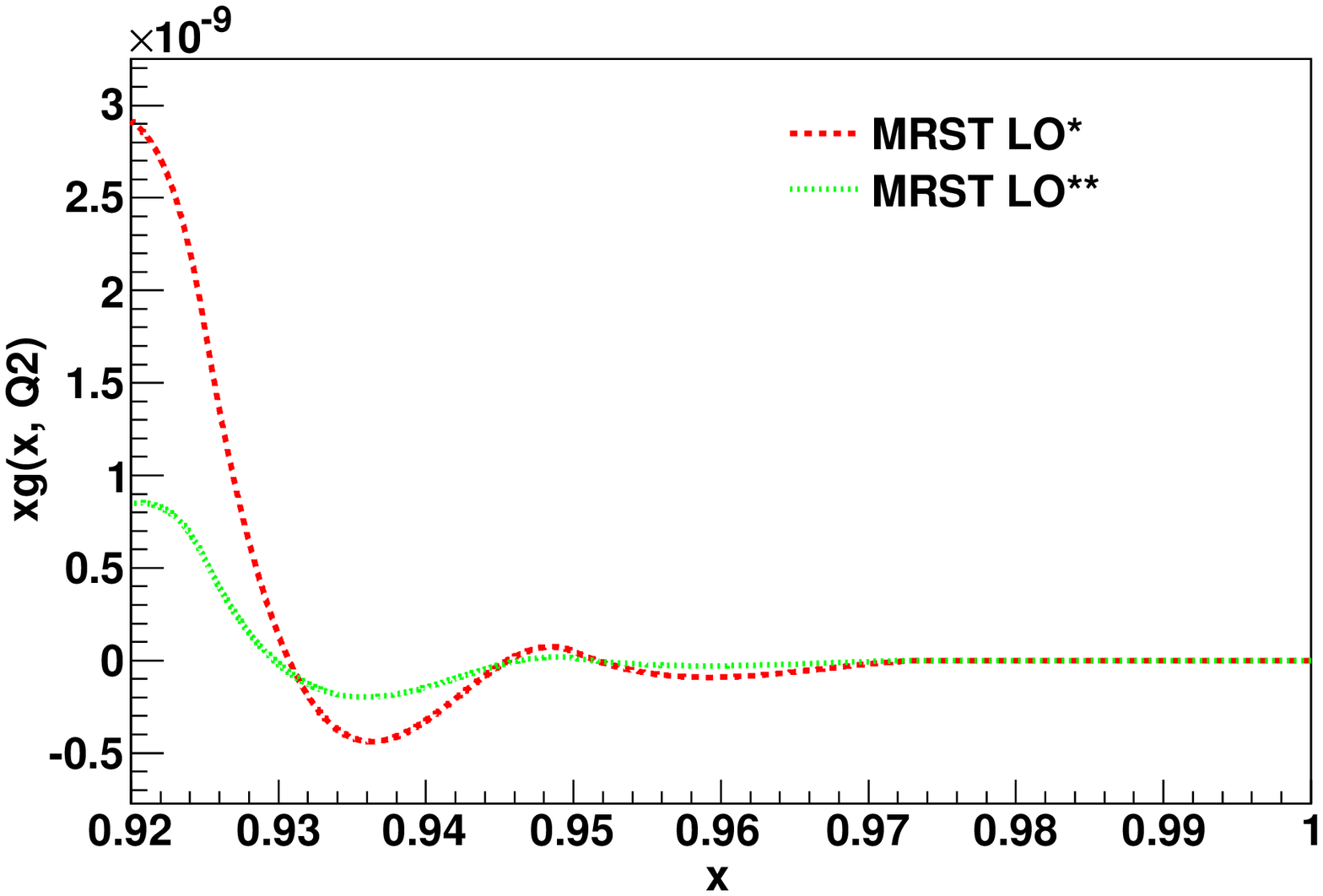}}
  \caption{Negative gluon distributions for large $x$ values.}
  \label{negGluon}
\end{figure}

The PDFs supplied to us
from MSTW have in some respects been improved compared to the versions available
in LHAPDF. Our implementation for the
MRST LO* and LO** PDFs make use of the new MSTW grid ($64\times48$) ranging
down to $x=10^{-6}$ while the LHAPDF
versions use the original grid with fewer ($49\times37$) grid points and
shorter $x$ range ($10^{-5}$). The values of $\alpha_S$ are a bit different in the new grid files of LO* and LO**
than in the corresponding LHAPDF grid files. LHAPDF versions use
$\Lambda_{QCD}$ for four active flavors which introduce
possible round off errors while stepping into the five flavor region of
$\alpha_S(M_Z)$ and the change to $\Lambda_{QCD}$ for five active flavors yields a slightly different
value for $\alpha_S(M_Z)$. Also worth noticing is that LO* and LO** both use
the unorthodox value of the $Z$ boson mass, $M_Z = 91.71$~GeV, unlike the MSTW 2008 distribution which uses $M_Z=91.19$~GeV \cite{lhapdf}.
For MSTW 2008 LO the LHAPDF interpolation gave negative values
between the last two grid points (as previously discovered by the HERWIG group) in $x$, i.e. $0.975 \leq x \leq 1$ and we therefore changed to a linear interpolation between the two
points.  However,
the problem at large $x$ values is not limited to the very last interval on
the grid but extends over more $x$ values and over a wide range of $Q^2$. Both
LO* and LO** give negative values of the gluon distribution in several intervals where
$x>0.85$, as shown in Fig.~\ref{negGluon}. The corresponding LHAPDF PDFs, which use the older grid with fewer
grid points also have this problem but not in the same intervals.  In simulations with \textsc{Pythia8} any negative PDF value will automatically
be put to zero and therefore the negative values of the gluon distribution
(where it is very small) do not affect the results of simulations. LHAPDF also gives
negative values for the up quark, which is worse since the up quark dominates
for these $x$ values. This
indicates that improvements of the numerical stability are needed in the large
x region. We also found a large difference between the two distributions for LO* and
LO**, where the old distributions gave much steeper PDFs at small $x$ and at $x=10^{-8}$ the differences reached a factor of two.

The MSTW NLO distribution gave very large negative values for the anomalous dimension $\frac{d\log(xf)}{d\log(Q^2)}$ at small
$Q^2$ and $x$-values around $10^{-5}$, which resulted in a huge
$\bar{s}$-distribution when extrapolated to low $Q^2$. This could also be a problem for the gluon which could
get large negative anomalous dimensions in
the $x$ region where the distribution is negative. To avoid this the anomalous
dimension is manually forced to be larger than $-2.5$. Although this does fix
this issue at hand, it is also an example of the dangers of using NLO PDFs in LO MC simulations, and an indication that
one has to be very careful with such use.

\subsection{CTEQ 6/MC}
The CTEQ distributions work well inside the grid but outside or near
the edges some problems occurred. The $tv = \log(\log(Q))$ values in the grid file were discovered to not exactly
correspond to the $Q$ values and hence, some points that were inside the $Q$
grid would end up outside the $tv$ grid. This caused some large issues, for example
the $b$ distribution, after being zero below the threshold, suddenly became
huge at $Q^2$ values just inside the grid. Therefore we choose to read in only the
$Q$  grid points and then calculate $tv$.

There can be differences between the CTEQ6 PDFs in \textsc{Pythia8} and the corresponding
ones in LHAPDF outside the grid. This is because LHAPDF provides the option to
use extrapolation
routines where \textsc{Pythia8}, by recommendation from the CTEQ authors, freezes the
values. The CTEQ MC distributions are not included in the current LHAPDF
package, which is therefore, to the best of our knowledge currently available for
simulations only in \textsc{Pythia8}. To freeze the PDFs can be dangerous both at small $x$ and small
$Q^2$. The region below $Q_{min}$  is
populated by multiple interactions so simulations are affected by the behavior
of the PDFs in this region and some of the PDFs do not range down to small enough $x$
if LHC reaches its full energy. There should be no problems above $Q_{max}$ which is
already large enough and in a region where the PDF evolution is slow.



\section{Comparison of PDFs}
The gluon distribution is dominating in the region of small $x$ while the
valence quarks, and then especially the up quarks, dominate for large $x$. We
therefore choose to focus mainly on these two distributions since they will
affect the results the most.

\begin{figure}[tp]
  \centering
  \subfloat[]{\label{fig:momsum1}\includegraphics[width=0.5\textwidth]
    {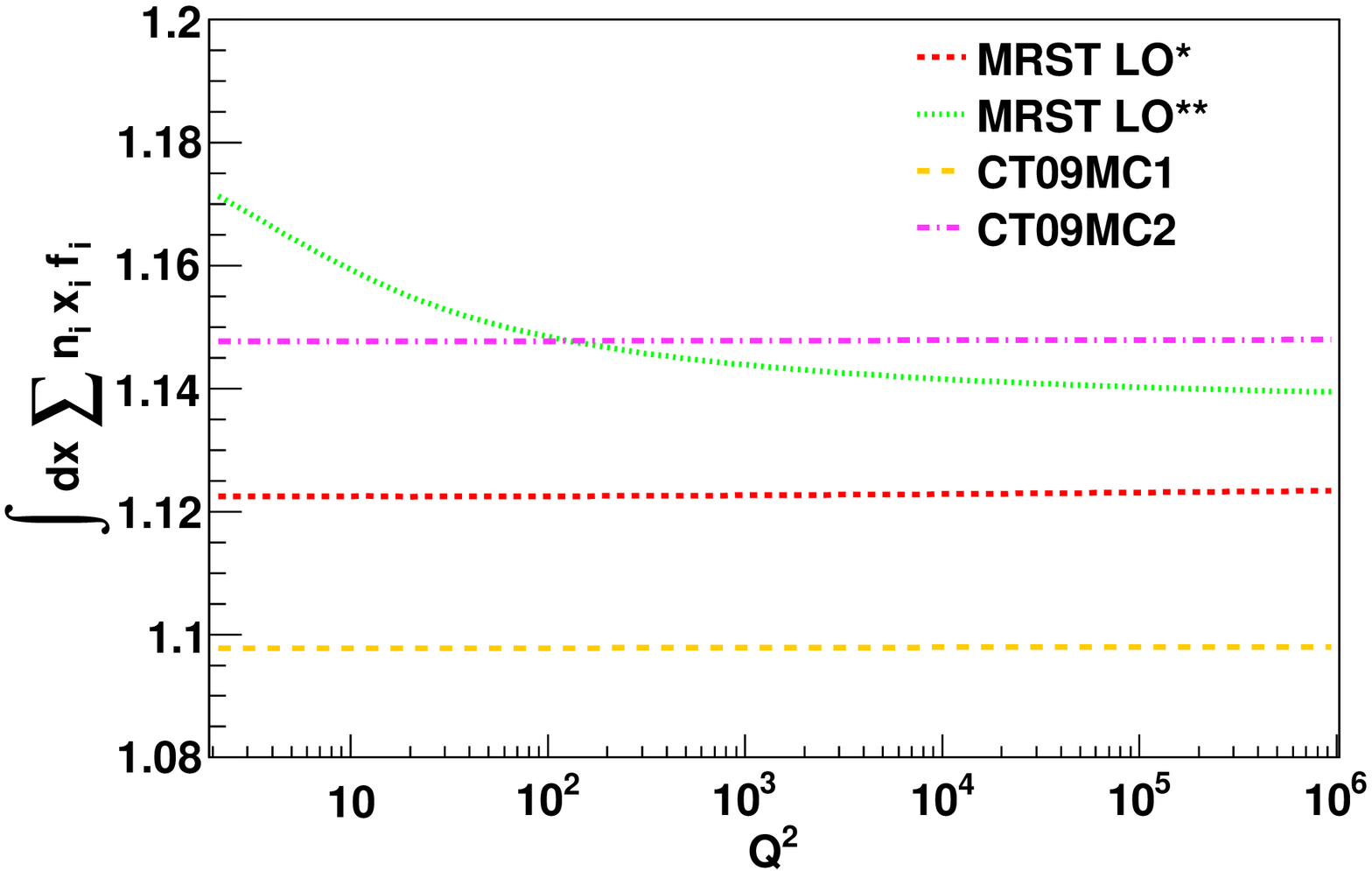}}                
  \subfloat[]{\label{fig:momsum2}\includegraphics[width=0.5\textwidth]
    {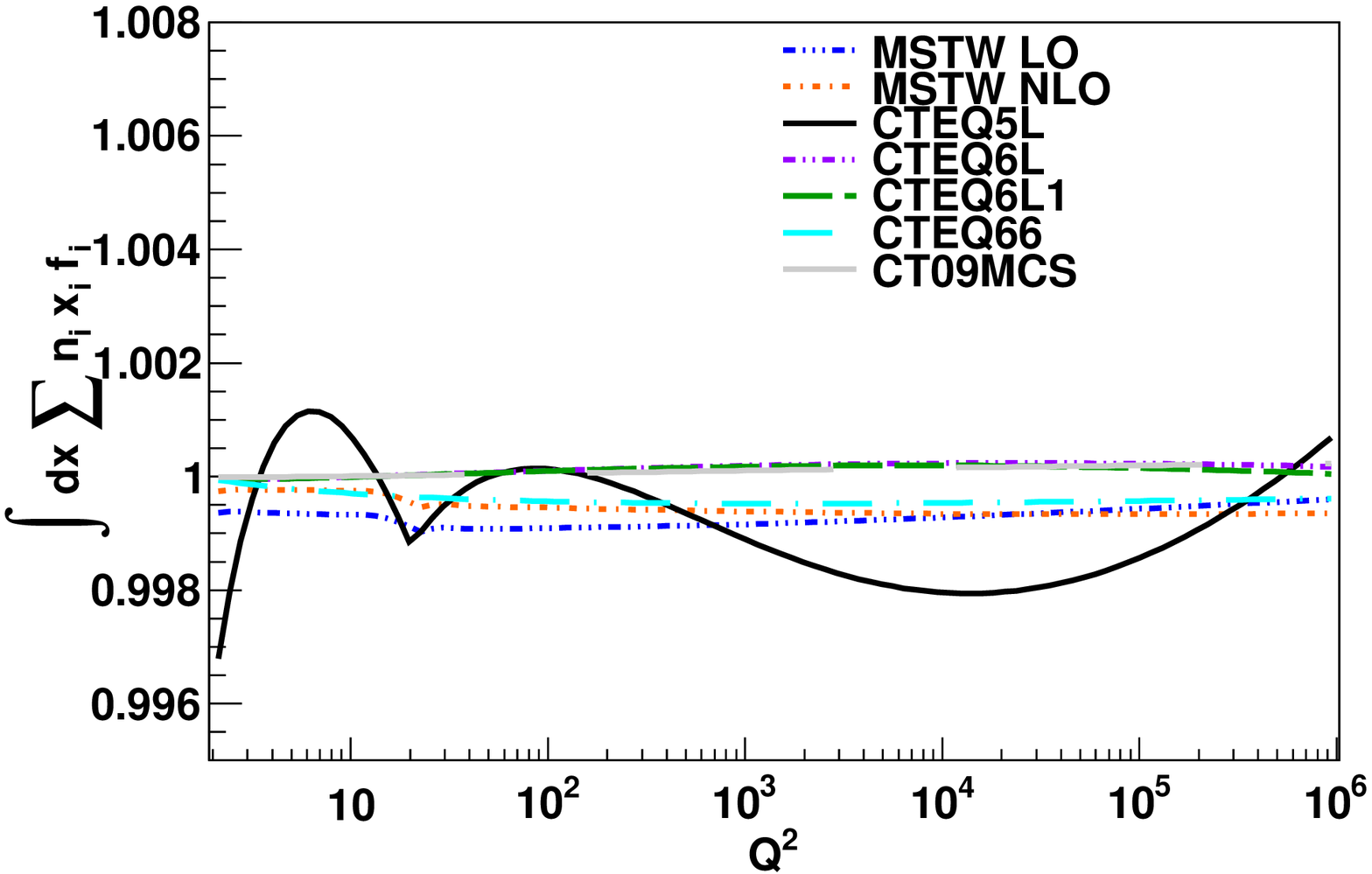}}
  \caption{The momentum fraction held by the partons in different PDFs. To
  the left are the four PDFs which break the momentum sum rule.}
  \label{fig:momsum}
\end{figure}

The PDFs are different from one another in several aspects and
Tab.~\ref{tab:momsum} show that four of them do not obey the
momentum sum rule. MC2 carries the largest momentum fraction of $1.15$ closely
followed by LO** which has the special
behavior where the fraction changes with $Q^2$ as shown in Fig~
\ref{fig:momsum1}. Although CTEQ5L also changes, see Fig~\ref{fig:momsum2},
this is unintentional, because of technical reasons, and the
scale of the changes is too small to give any noticeable effects.
\begin{table}[tp]
  \center\small
  \begin{tabular}[h]{lc}
    \toprule   
    PDF & Total Momentum Fraction \\
    \midrule
    CTEQ5L    & $1.00$ \\
    MRST LO*  & $1.12$ \\
    MRST LO** & $1.14$ \\
    MSTW LO   & $1.00$ \\
    MSTW NLO  & $1.00$ \\
    CTEQ6L    & $1.00$ \\
    CTEQ6L1   & $1.00$ \\
    CTEQ66    & $1.00$ \\
    CT09MC1   & $1.10$ \\
    CT09MC2   & $1.15$ \\
    CT09MCS   & $1.00$ \\
    \bottomrule
  \end{tabular}
  \caption{The total fraction of the protons momentum held by all the partons
    for the different PDFs evaluated at $Q^2 = 10^3$.}
  \label{tab:momsum}
\end{table}

\begin{figure}[tp]
  \centering
  \subfloat[]{\label{fig:uxQ24-1}\includegraphics[width=0.5\textwidth]
    {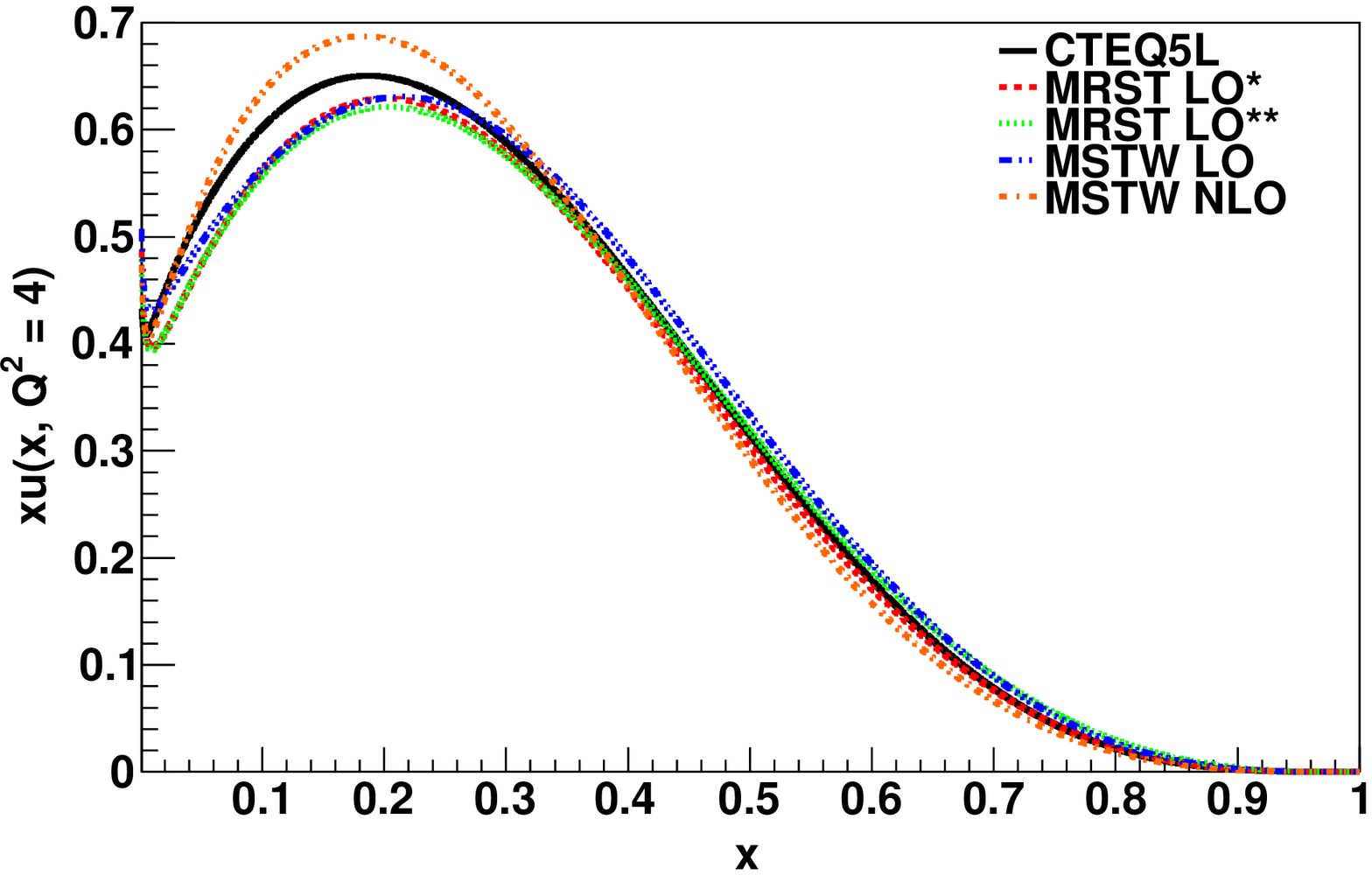}}                
  \subfloat[]{\label{fig:uxQ24-2}\includegraphics[width=0.5\textwidth]
    {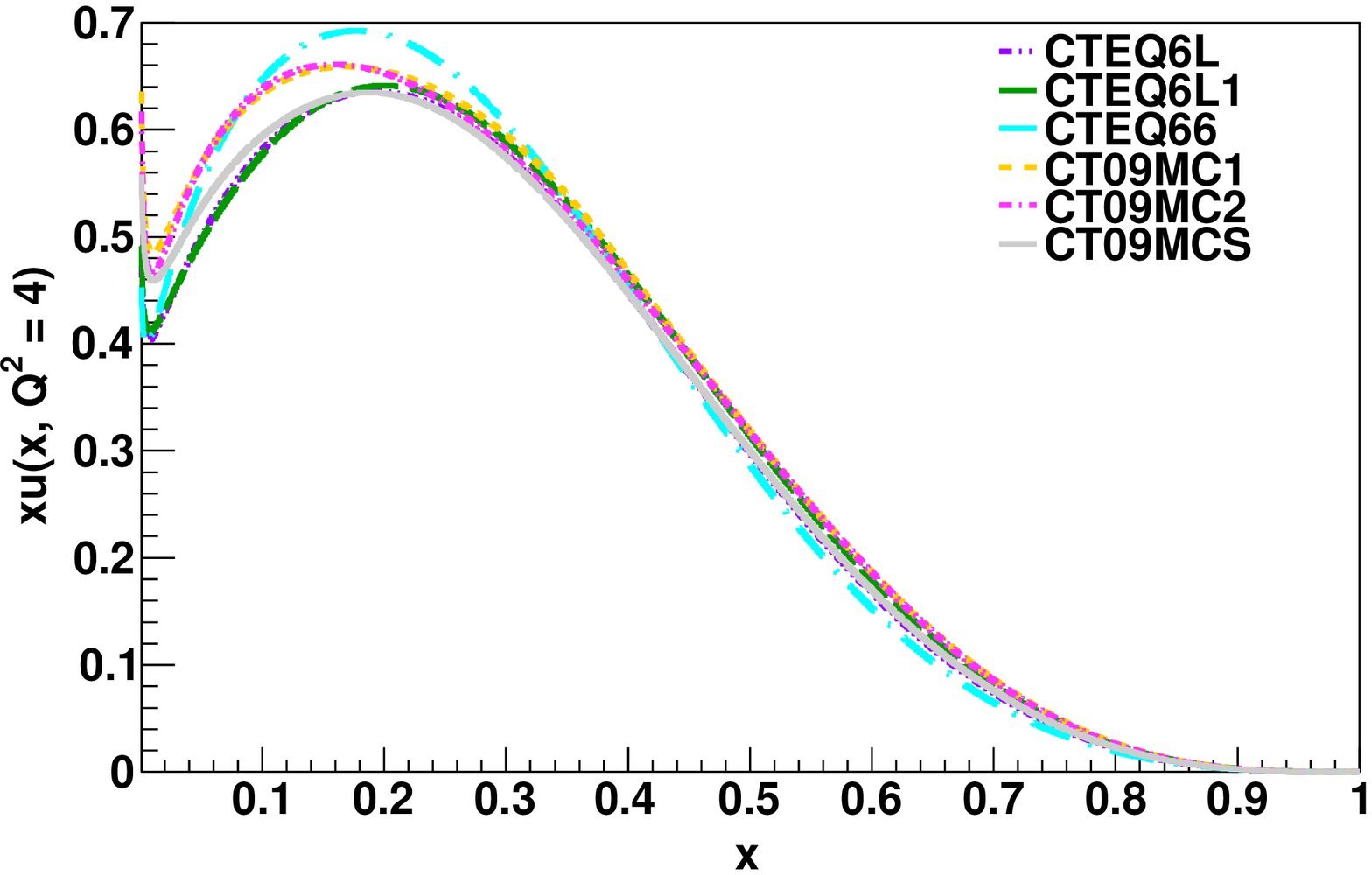}} \\
  \subfloat[]{\label{fig:gLogxQ24-1}\includegraphics[width=0.5\textwidth]
    {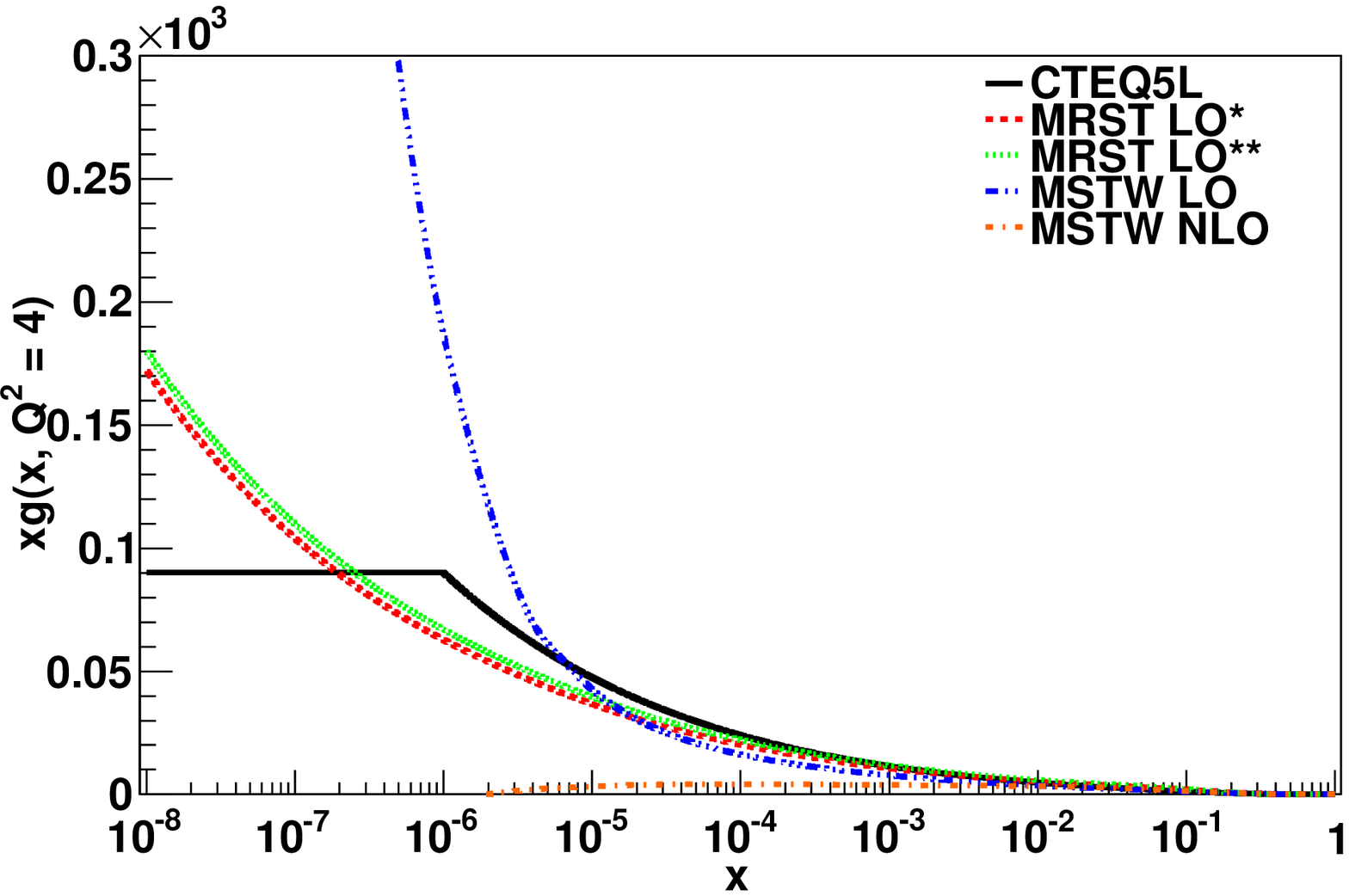}}                
  \subfloat[]{\label{fig:gLogxQ24-2}\includegraphics[width=0.5\textwidth]
    {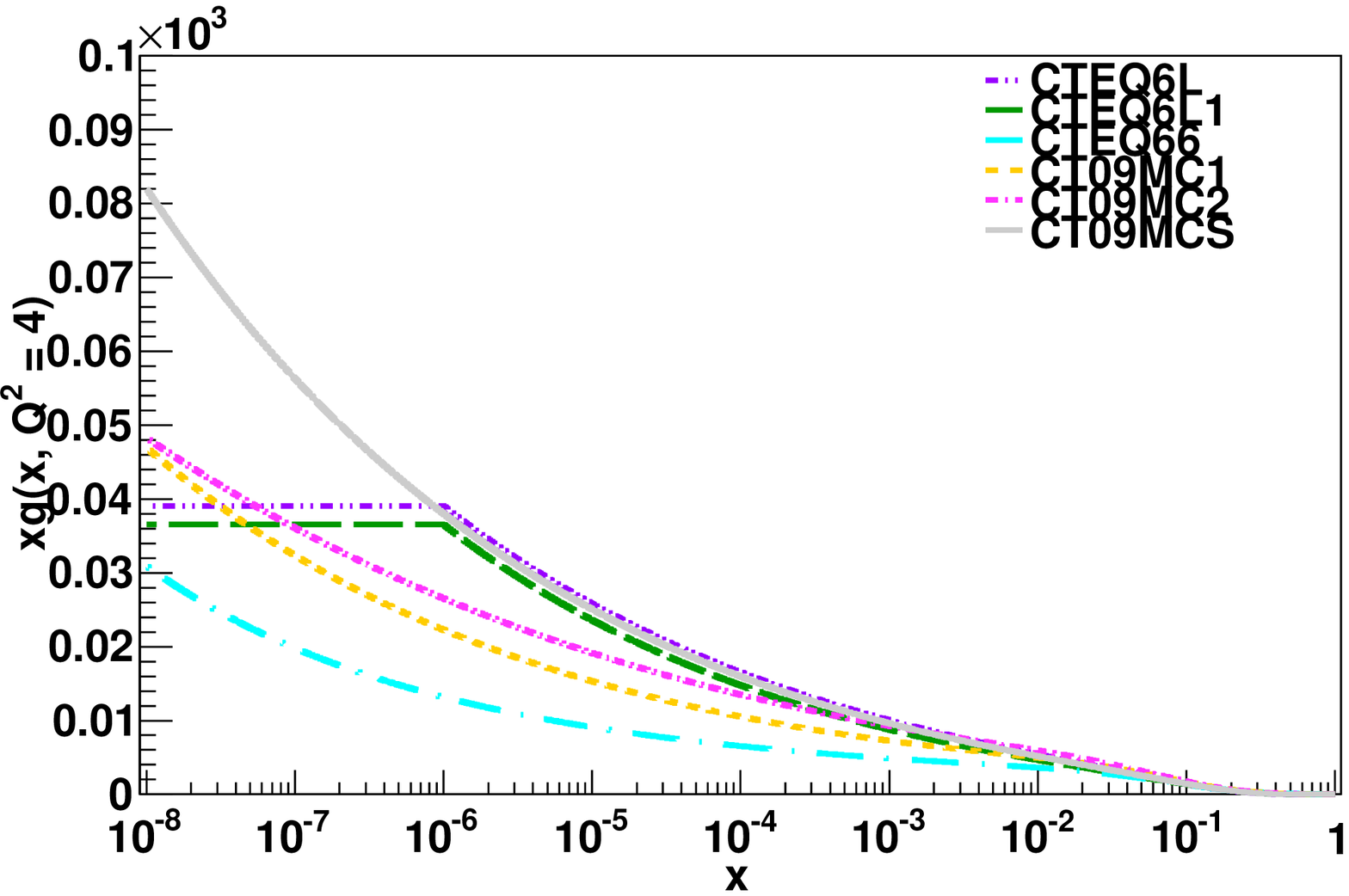}}
  \caption{Up quark (a, b) and gluon (c, d) distributions at
    $Q^2=4$~GeV$^2$. Note difference in horizontal and vertical scales.}
  \label{fig:xfQ24}
\end{figure}

Minimum bias events are sensitive to low $Q$ and a $Q^2$ around $4$~GeV$^2$ is
a typical scale for such simulations. Looking at the gluon and up
distributions  at $Q^2=4$~GeV$^2$ in Fig.~\ref{fig:xfQ24} we see that the
up distributions are all similar, with slight differences for the two NLO
PDFs and CT09 MC1 and MC2. For the gluon distribution we see large differences at
small $x$ values. MSTW LO has a much steeper rise and gets much larger than
the others. All the MRST/MSTW distributions give larger values at small $x$
than the CTEQ ones. The MC-adapted PDFs follow each other within both distributions,
except for MCS which is more similar to CTEQ6L and CTEQ6L1. The two NLO PDFs
stand apart from the rest and MSTW NLO is negative in a large region. One can also see
that CTEQ5L, CTEQ6L and CTEQ6L1 all freeze at $x=10^{-6}$. 

\begin{figure}[tp]
  \centering
  \subfloat[]{\label{fig:uxQ21e3-1}\includegraphics[width=0.5\textwidth]
    {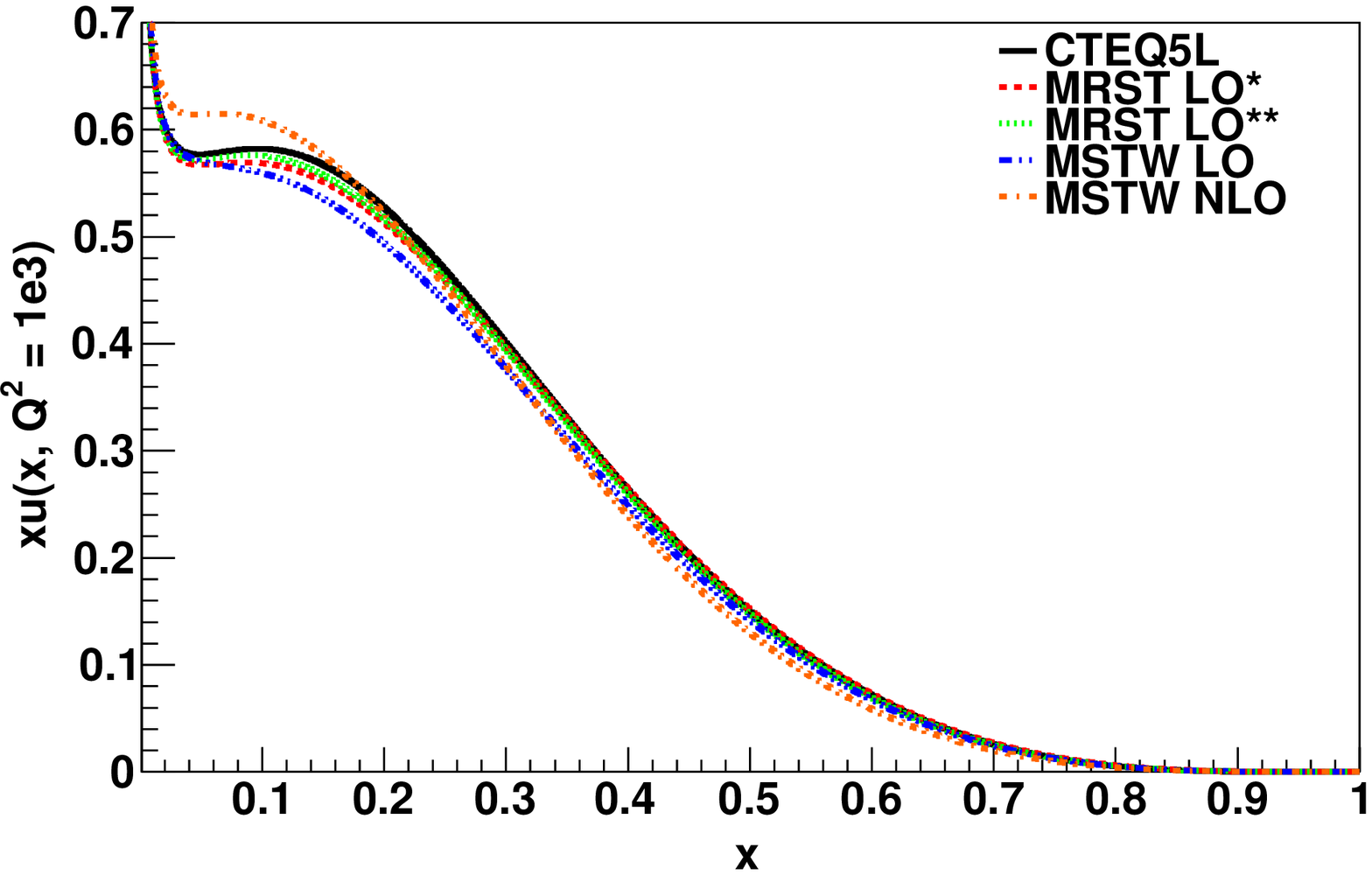}}                
  \subfloat[]{\label{fig:uxQ21e3-2}\includegraphics[width=0.5\textwidth]
    {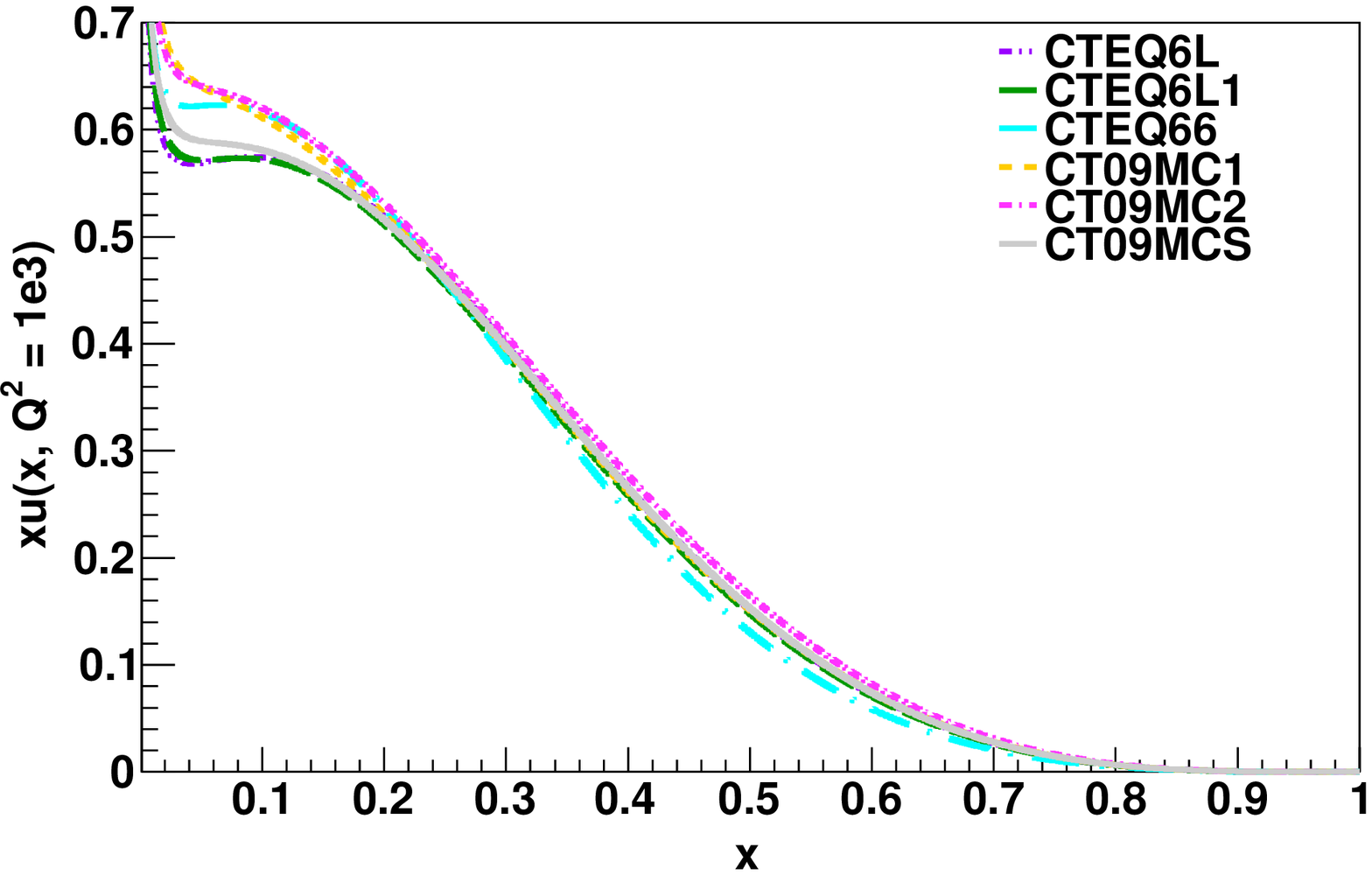}}\\
  \subfloat[]{\label{fig:gLogxQ21e3-1}\includegraphics[width=0.5\textwidth]
    {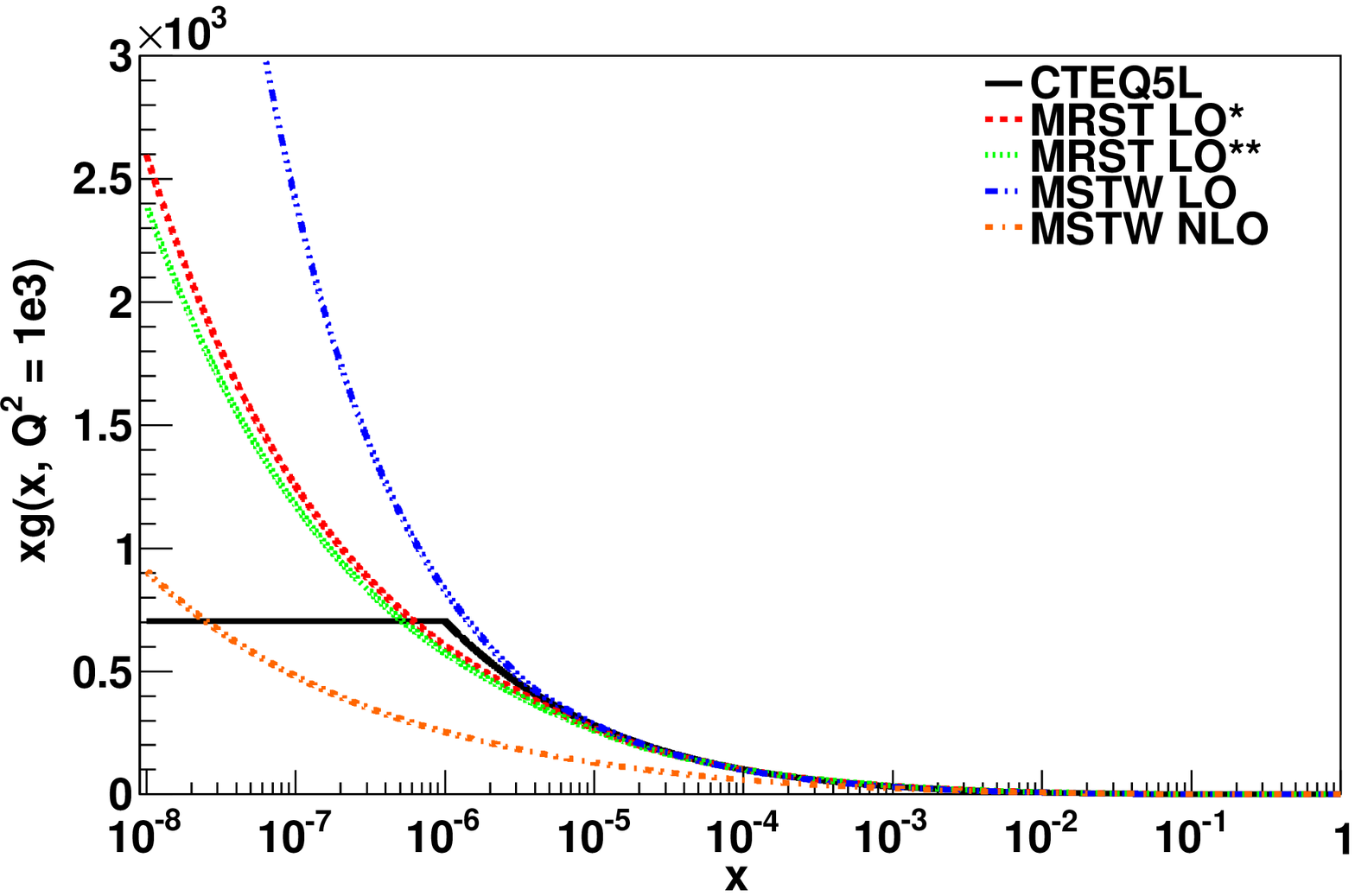}}                
  \subfloat[]{\label{fig:gLogxQ21e3-2}\includegraphics[width=0.5\textwidth]
    {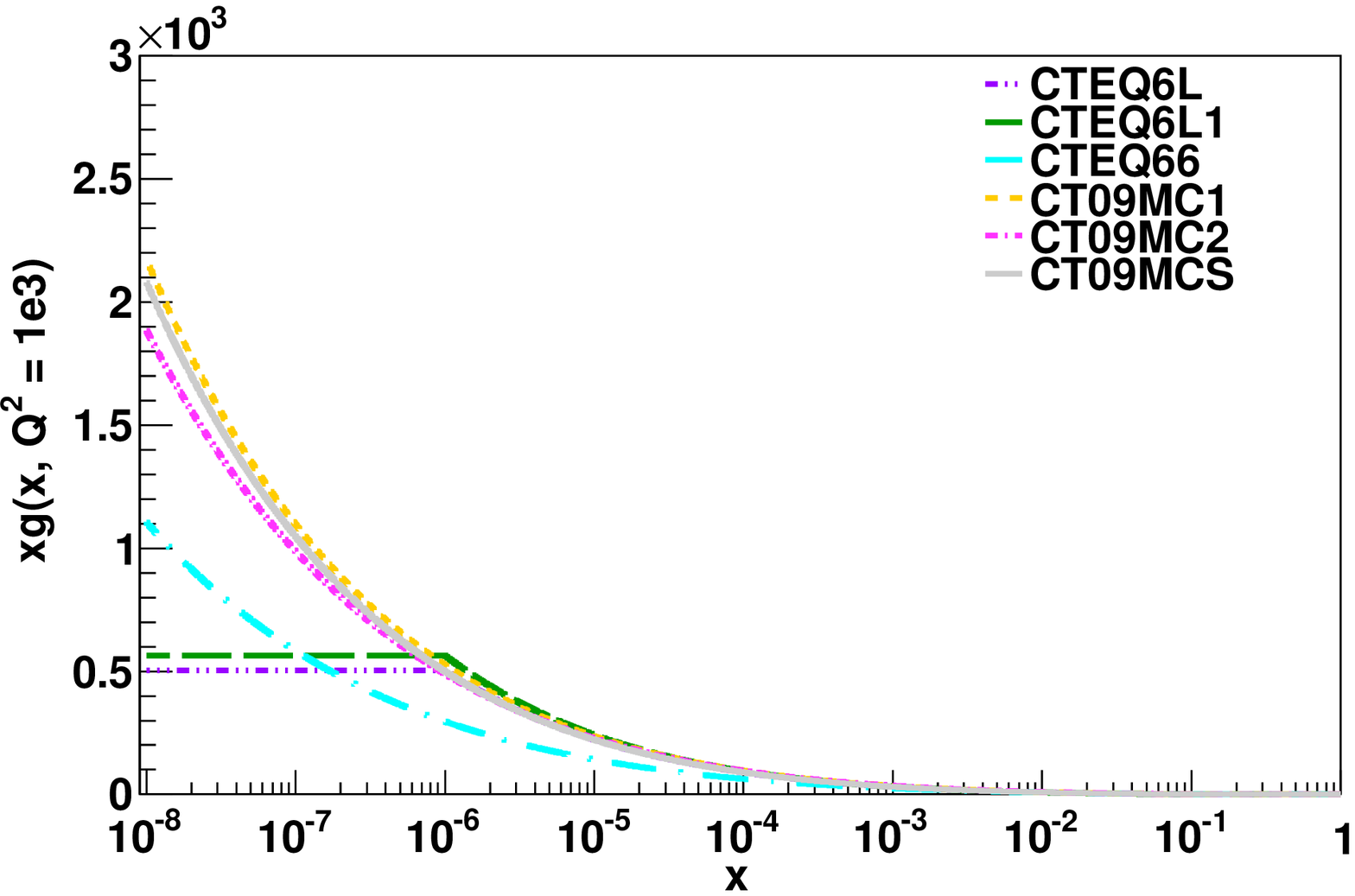}} 
  \caption{Up quark (a, b) and gluon (c, d) distributions at $Q^2=10^{3}$~GeV$^2$. Note difference in horizontal and vertical scales.}
  \label{fig:xfQ21e3}
\end{figure}

\begin{figure}[tp]
  \centering
  \subfloat[]{\label{fig:xsQ250-1}\includegraphics[width=0.5\textwidth]
    {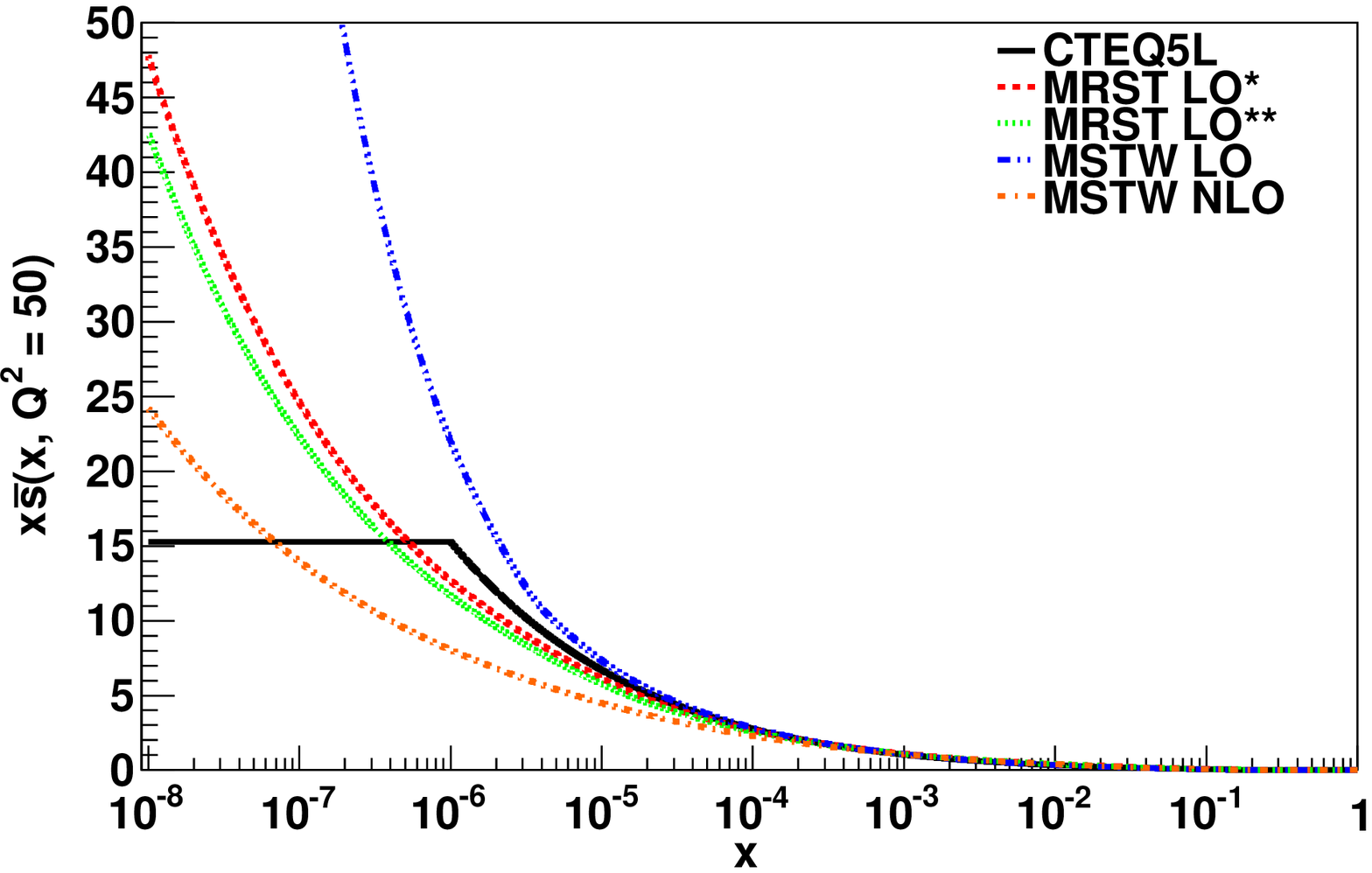}}                
  \subfloat[]{\label{fig:xsQ250-2}\includegraphics[width=0.5\textwidth]
    {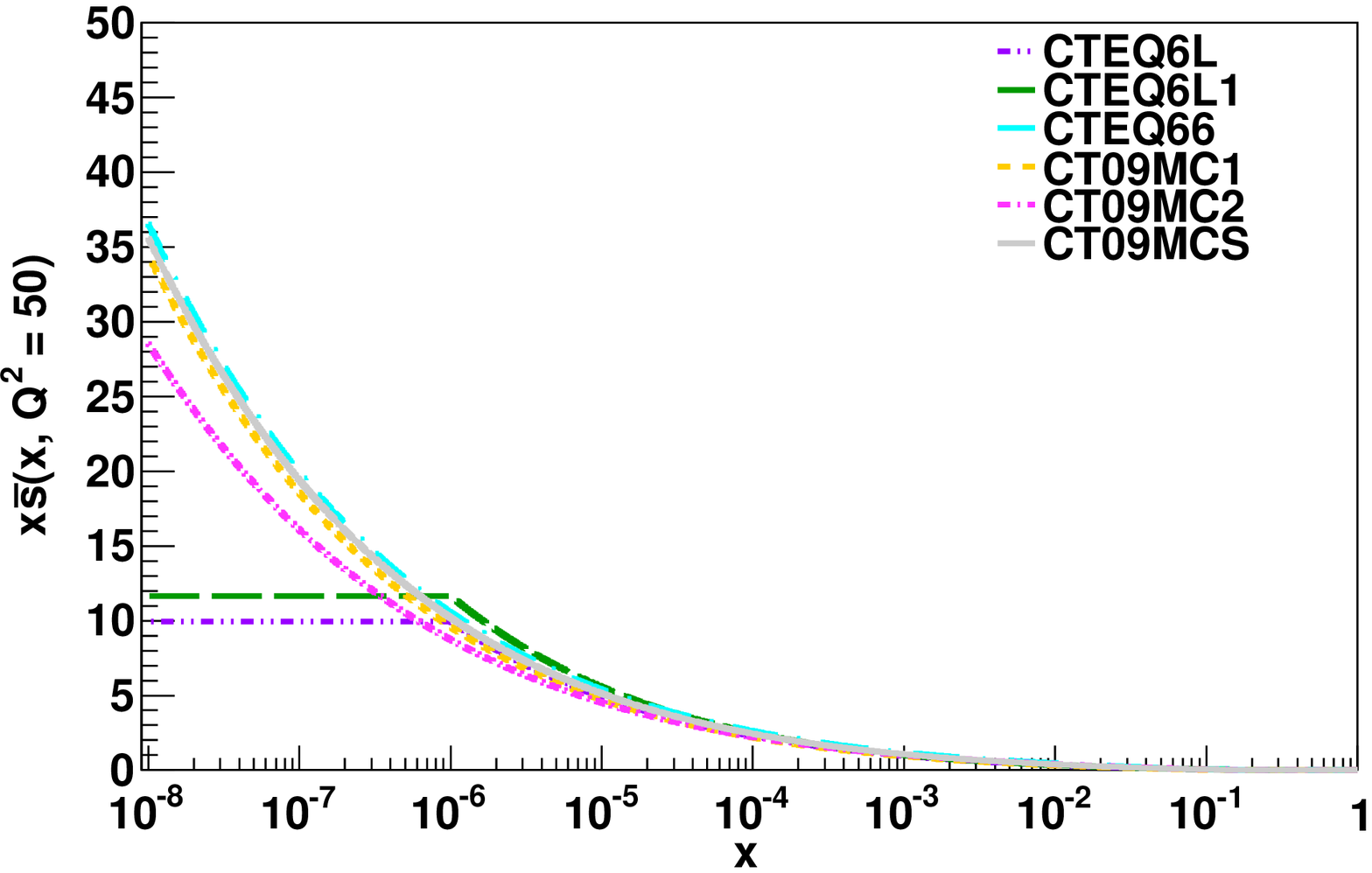}}
  \caption{$\bar{s}$ distributions at $Q^2=50$~GeV$^2$.}
  \label{fig:xsQ250}
\end{figure}

Fig.~\ref{fig:xfQ21e3} shows
the distributions at larger
$Q^2=10^3$ and here the up distributions show the same pattern. This is also
the case for the gluon distributions. However, the differences between the PDFs
are now smaller, and especially the difference between the two groups is no
longer as prominent. We can also see in Fig.~\ref{fig:gLogxQ21e3-2} that all three MC-adapted PDFs
from CTEQ are similar at this $Q^2$. Taking a look at the $\bar{s}$ in Fig.~\ref{fig:xsQ250} at an
intermediate $Q^2=50$~GeV$^2$ we see a similar pattern as we saw for the gluons.

In general the MSTW LO blows up at small $x$ values and is
much larger than all the others in this region. The MC-adapted PDFs show
strong similarities, especially within their respective collaboration, while MCS
stands out by sometimes resembling the ordinary LO PDFs. The similarity
between the two NLO PDFs is also clear and CTEQ66 looks
similar to MSTW NLO for large $Q^2$ but does not go negative at $Q^2=4$~GeV$^2$. Comparing the two groups the CTEQ
distributions have smaller distributions at small $x$ both for the PDFs that
freeze, and also for the ones with grid ranging down to $10^{-8}$.\\


\section{Minbias Events}
\label{sec:minbias}
\subsection{Introduction}
In experimental physics minimum bias events are what would be seen in a totally inclusive trigger where
everything except elastic and (most) diffractive events is accepted. In \textsc{Pythia8},
minbias events are equivalent to inelastic nondiffractive events. They are
managed by the multiple interactions machinery and is affected by the PDFs
through cross section calculations.

Minbias events tend to
have low
average transverse energy, low particle multiplicity and consist largely of soft
inelastic interactions which are interesting both in
their own right and because they constitute background when studying hard
interactions. Because of the low $p_{\perp}$ the interacting partons only need a
small portion of the momenta of the incoming hadrons, and hence minbias events probe parton distributions in the small $x$ region dominated by the gluon distribution.

We examine the rapidity, multiplicity and $p_{\perp}$ distributions from
simulations with different PDFs, both at Tevatron and LHC energies. With the aid of Rivet \cite{Rivet} we also
compare $p_{\perp}$ and $\sum E_{\perp}$ particle spectra as well as average
$p_{\perp}$ evolution with
multiplicity to real data taken by the CDF experiment at Tevatron Run 2 \cite{Aaltonen2}.

\subsection{Multiplicity and Tuning}
The larger momentum carried by the partons in LO*, LO**, MC1 and MC2
yield a larger activity and hence a larger multiplicity than with the
ordinary leading-order PDFs. Furthermore the NLO PDFs give less activity, so
before comparing the simulations we first tune \textsc{Pythia8} so that all PDFs 
have the same average charge particle multiplicity as CTEQ5L. We
choose CTEQ5L as reference because it is the default PDF in \textsc{Pythia8} and is
most commonly used in \textsc{Pythia8} simulations. We are not making a complete tune
and only intend to get a first impression of relative differences, under
comparable conditions. The tuning is accomplished by tweaking the
$p_{\perp 0}^{\stext{Ref}}$ parameter in \textsc{Pythia8}
\begin{equation}
  p_{\perp 0}=p_{\perp 0}^{\stext{Ref}}\left( \frac{E_{\stext{CM}}}{E_{\stext{CM}}^{\stext{Ref}}}\right)^p
\end{equation}
where $E_{\stext{CM}}^{\stext{Ref}} = 1800$~GeV and $p = 0.24$. $p_{\perp 0}$ is used for the regularization of the divergence of the QCD
cross section as $p_{\perp}\rightarrow 0$, eqn.~\ref{eqn:pt0}, and a smaller $p_{\perp0}$ cause the
regularization to kick in at a lower $p_{\perp}$ increasing the charged
particle multiplicity, $n_{ch}$. The simulations were done with 100 000 events and
$p_{\perp 0}^{\stext{Ref}}$ tuned until the relative difference
($\frac{\langle n_{PDF} \rangle - \langle n_{5L} \rangle}{\langle n_{PDF} \rangle + \langle n_{5L} \rangle}$) was less than 1\%. The tuning
was done for the $\alpha_S$ value and leading-order running which is default
in \textsc{Pythia8}, as well as with $\alpha_S$ determined individually by the PDFs. Results are shown in Tab.~\ref{charmult1}~and~\ref{charmult2}
respectively. With the \textsc{Pythia8} default $\alpha_S$ the largest multiplicity is
obtained with LO**, followed by MC2 and LO* which also are the PDFs with
most momentum, but 6L manages to squeeze in before MC1 follows. This changes for the $\alpha_S$
determined by the PDFs since MC1 and LO experience a large increase in multiplicity. 

For MSTW NLO the integrated interaction cross
section is smaller than the nondiffractive inelastic one and therefore \textsc{Pythia8} automatically lowers $p_{T0}$. This occurs for both
NLO PDFs when we use the $\alpha_S$ specific to the individual PDFs and,
although this does not cause any trouble, it is a reminder of the danger of using next-to-leading-order PDFs together with leading
order MC generators.

\begin{table}[tp]
  \center\small
  \begin{tabular}[h]{lcc}
    \toprule   
      & \multicolumn{1}{c}{Charged Particle} \\
    PDF & Multiplicity & $p_{\perp 0}^{\stext{Ref}}$ \\
    \midrule
    CTEQ5L    & 54.48 & 2.25 \\
    MRST LO*  & 59.74 & 2.50 \\
    MRST LO** & 63.52 & 2.63 \\
    MSTW LO   & 49.10 & 2.06 \\
    MSTW NLO  & 48.02 & 1.56 \\
    CTEQ6L    & 54.92 & 2.25 \\
    CTEQ6L1   & 51.71 & 2.13 \\
    CTEQ66    & 42.85 & 1.75 \\
    CT09MC1   & 53.92 & 2.25 \\
    CT09MC2   & 60.37 & 2.50 \\
    CT09MCS   & 54.87 & 2.25 \\
    \bottomrule
  \end{tabular}
  \caption{Average charged particle multiplicity for the different PDFs with the default value of
    $p_{\perp 0}^{\stext{Ref}} = 2.25$ and also the $p_{\perp 0}^{\stext{Ref}}$ required to tune the charge
    multiplicity equal to the value for CTEQ5L.}
  \label{charmult1}
\end{table}

\begin{table}[tp]
  \center\small
  \begin{tabular}[h]{lccc}
    \toprule   
      & \multicolumn{1}{c}{Charged Particle} \\
    PDF & Multiplicity & $p_{\perp 0}^{\stext{Ref}}$ \\
    \midrule
    CTEQ5L    & 54.48 & 2.25 \\
    MRST LO*  & 57.43 & 2.38 \\
    MRST LO** & 56.03 & 2.31 \\
    MSTW LO   & 60.58 & 2.50 \\
    MSTW NLO  & 46.17 & 1.31 \\
    CTEQ6L    & 51.11 & 2.13 \\
    CTEQ6L1   & 54.18 & 2.25 \\
    CTEQ66    & 43.89 & 1.63 \\
    CT09MC1   & 62.35 & 2.50 \\
    CT09MC2   & 55.78 & 2.31 \\
    CT09MCS   & 50.83 & 2.13 \\
    \bottomrule
  \end{tabular}
  \caption{Average charged particle multiplicity for the different PDFs with the default value of
    $p_{\perp 0}^{\stext{Ref}} = 2.25$ and also the $p_{\perp 0}^{\stext{Ref}}$ required to tune the charge
    multiplicity equal to the value for CTEQ5L. With $\alpha_S$ value and
    running set individually.}
  \label{charmult2}
\end{table}

\subsection{Results}
\begin{figure}[tp]
  \subfloat[]{\label{fig:CRapid-allstrip1}\includegraphics[width=0.5\textwidth]
    {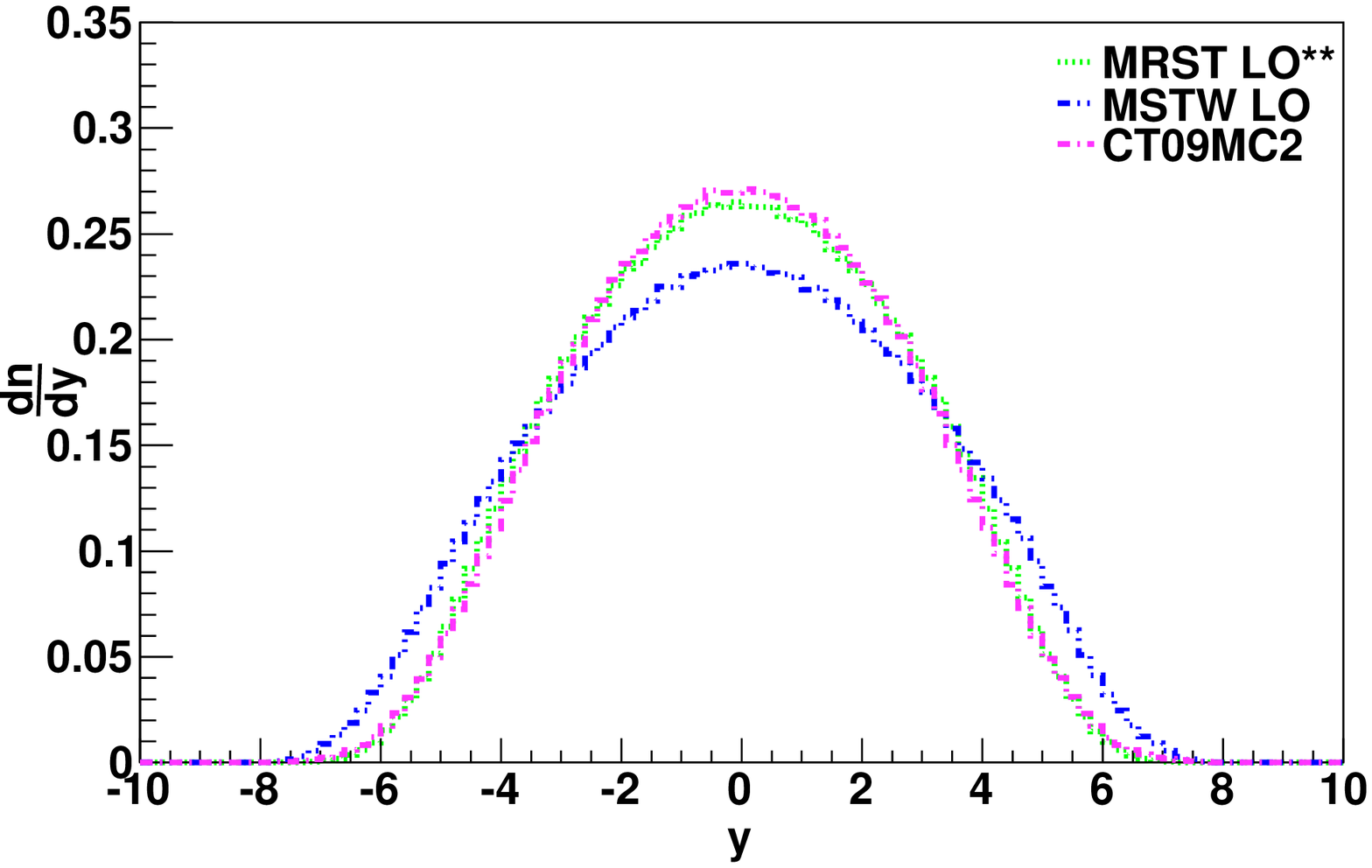}}
   \subfloat[]{\label{fig:CRapid-allstrip2}\includegraphics[width=0.5\textwidth]
    {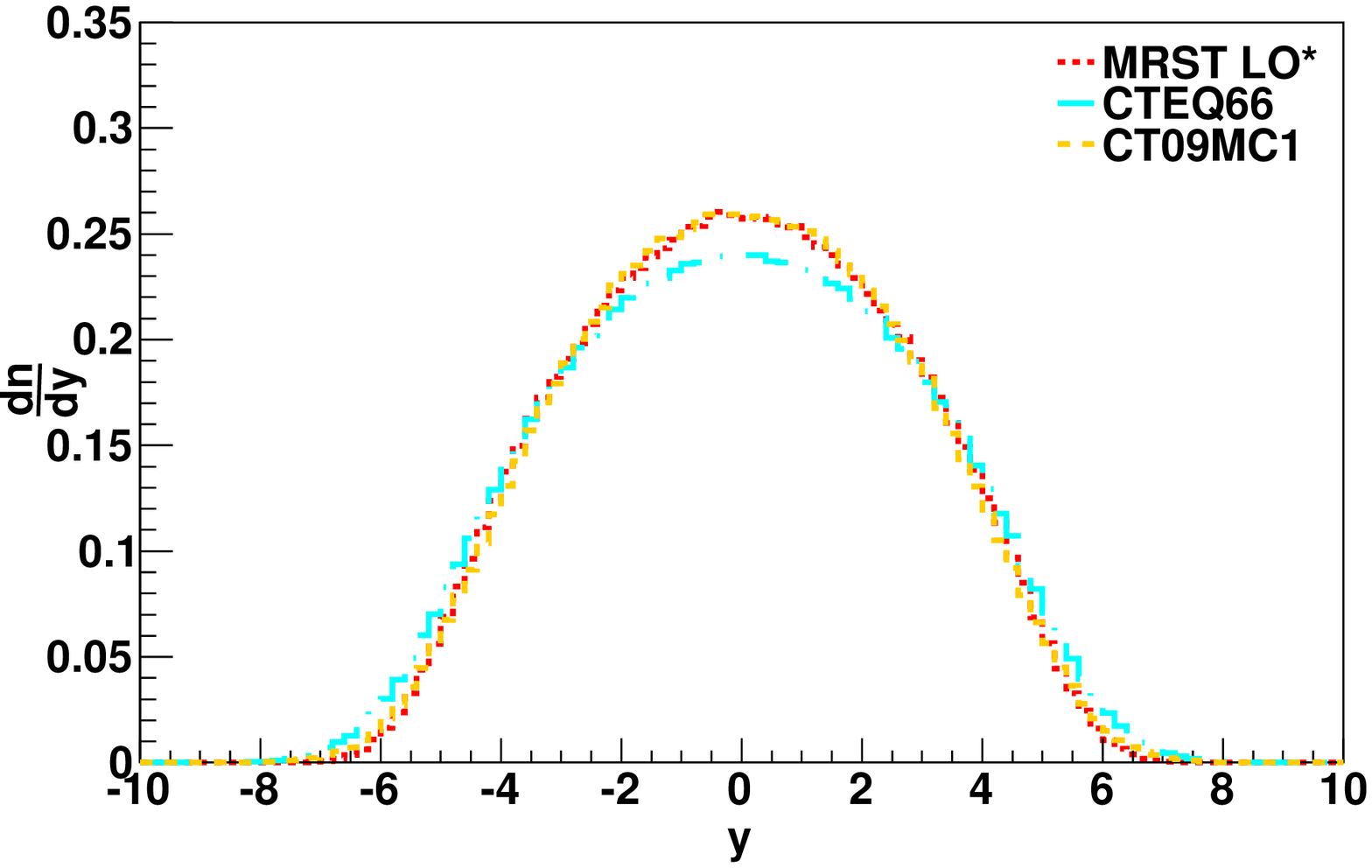}}
  \caption{Rapidity distributions of the partons created in the $2 \rightarrow 2$ sub-process.}
  \label{fig:Rapid-strip}
\end{figure}

In this section all simulations are done at the CM-energy of 1960~GeV, if not
explicitly stated otherwise. So as not to cram the pictures, not all sets are
shown all the time. We have
tried to choose a selection of PDFs to show in each plot what represents both
the extremes and the middle way. The
rapidity distributions of the outgoing particles at the parton level when only
the $2 \rightarrow 2$ sub-process is considered are presented in
Fig.~\ref{fig:Rapid-strip}. MSTW LO has a broader distribution than the rest of the PDFs with more
particles at larger rapidities as an effect of the large gluon
distribution at the small $x$. We can also see that LO** closely resembles
MC2, as does LO* and MC1 while the two NLO distributions are lower in the central
rapidity region. The remaining leading-order
distributions all show similarities to the MC-adapted ones.

\begin{figure}[tp]
  \subfloat[]{\label{fig:CRapid1}\includegraphics[width=0.5\textwidth]
    {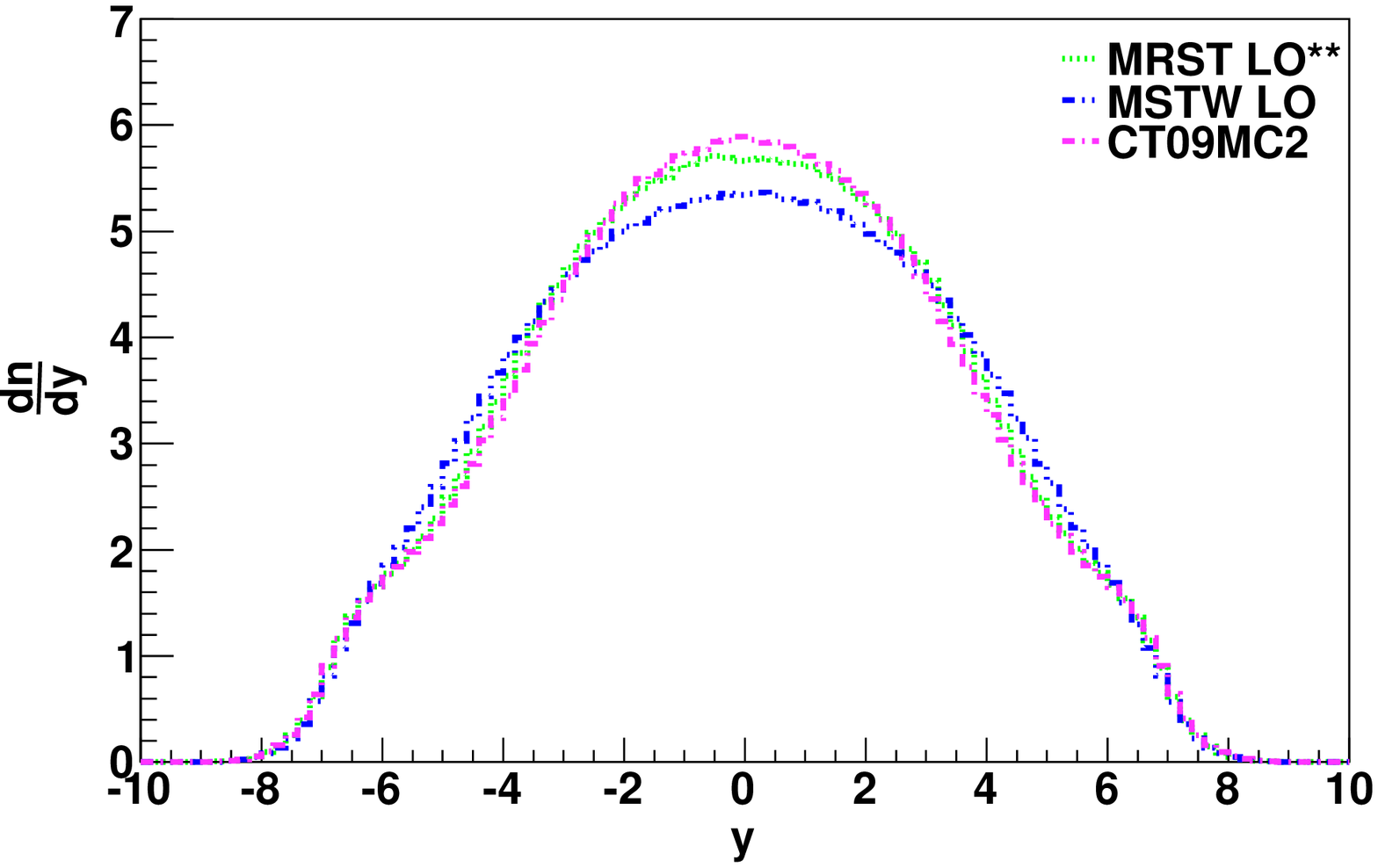}}
   \subfloat[]{\label{fig:CRapid2}\includegraphics[width=0.5\textwidth]
    {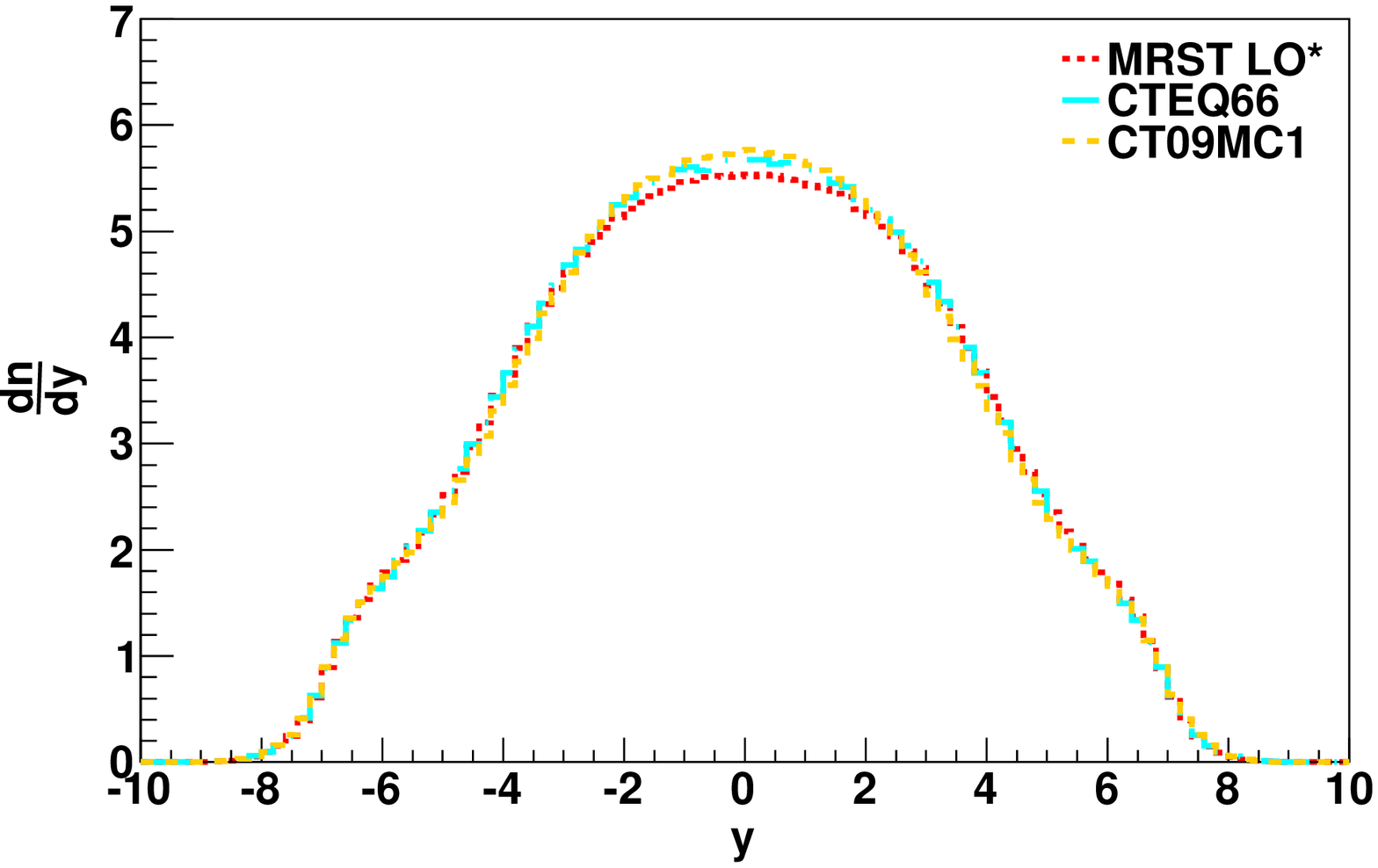}}
  \caption{Rapidity distributions of charged particles after hadronization at
    minbias simulations with $s=1960$~GeV$^2$.}
  \label{fig:Rapid}
\end{figure}

Turning on the
rest of the \textsc{Pythia8} machinery and looking at the distribution of charged
particles after hadronization, shown in Fig.~\ref{fig:Rapid}, the distribution
changes its shape. There are now more particles at larger rapidities as a
result of fragmenting color field strings stretched out to the beam remnants and most of the differences
between the PDFs get blurred. Some differences still remain and MSTW LO is
still smaller for central rapidities, and as a remnant of the wider
distribution lack the inward dents that all other PDFs have at rapidities
around $\pm5$. The peaks of LO** and MC2 are a bit sharper than for LO*
and MC1 but the trace of the lower value of the CTEQ66 at central rapidity is
gone.

\begin{figure}[tp]
  \subfloat[]{\includegraphics[width=0.5\textwidth]
    {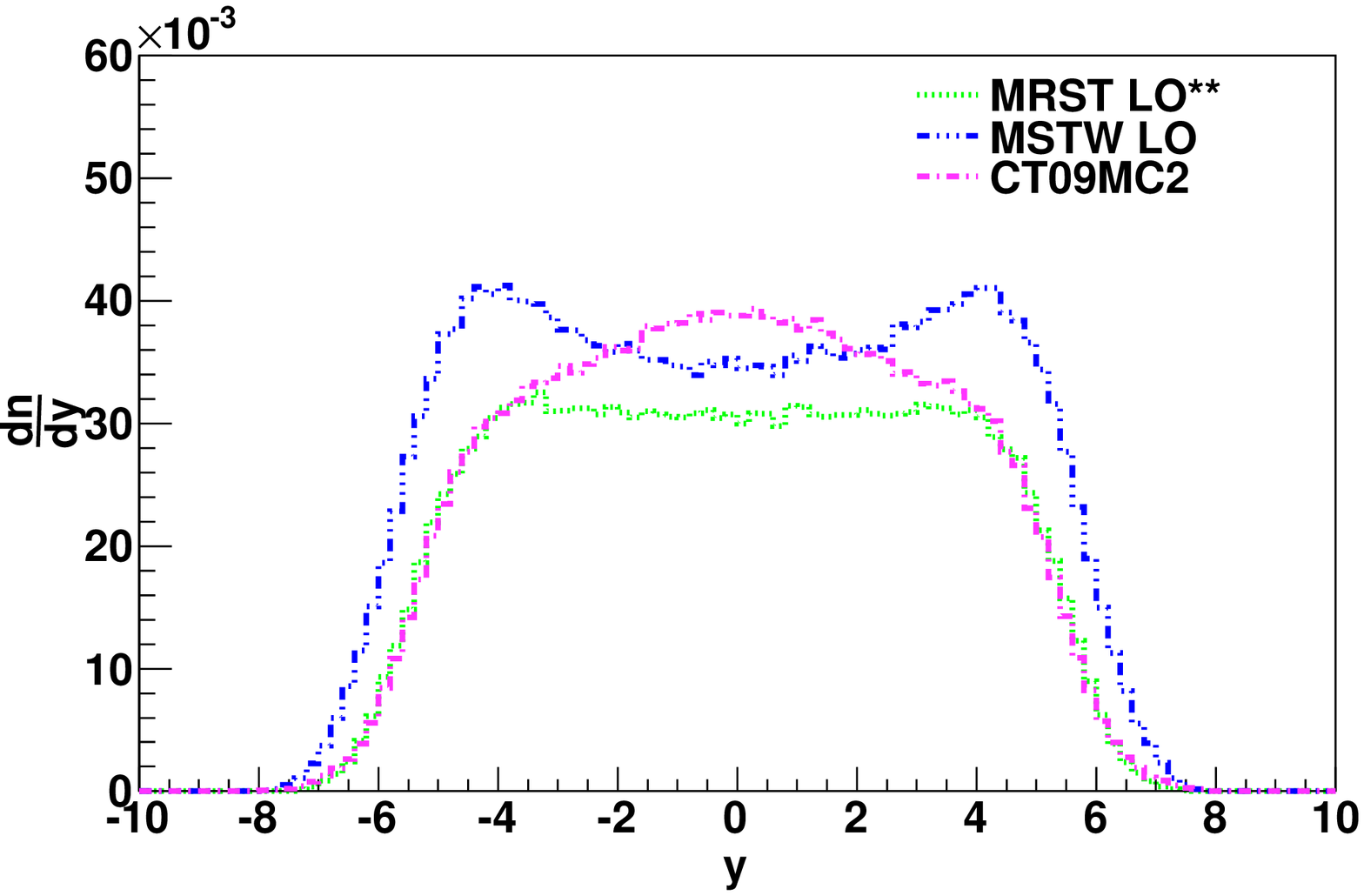}}
   \subfloat[]{\includegraphics[width=0.5\textwidth]
    {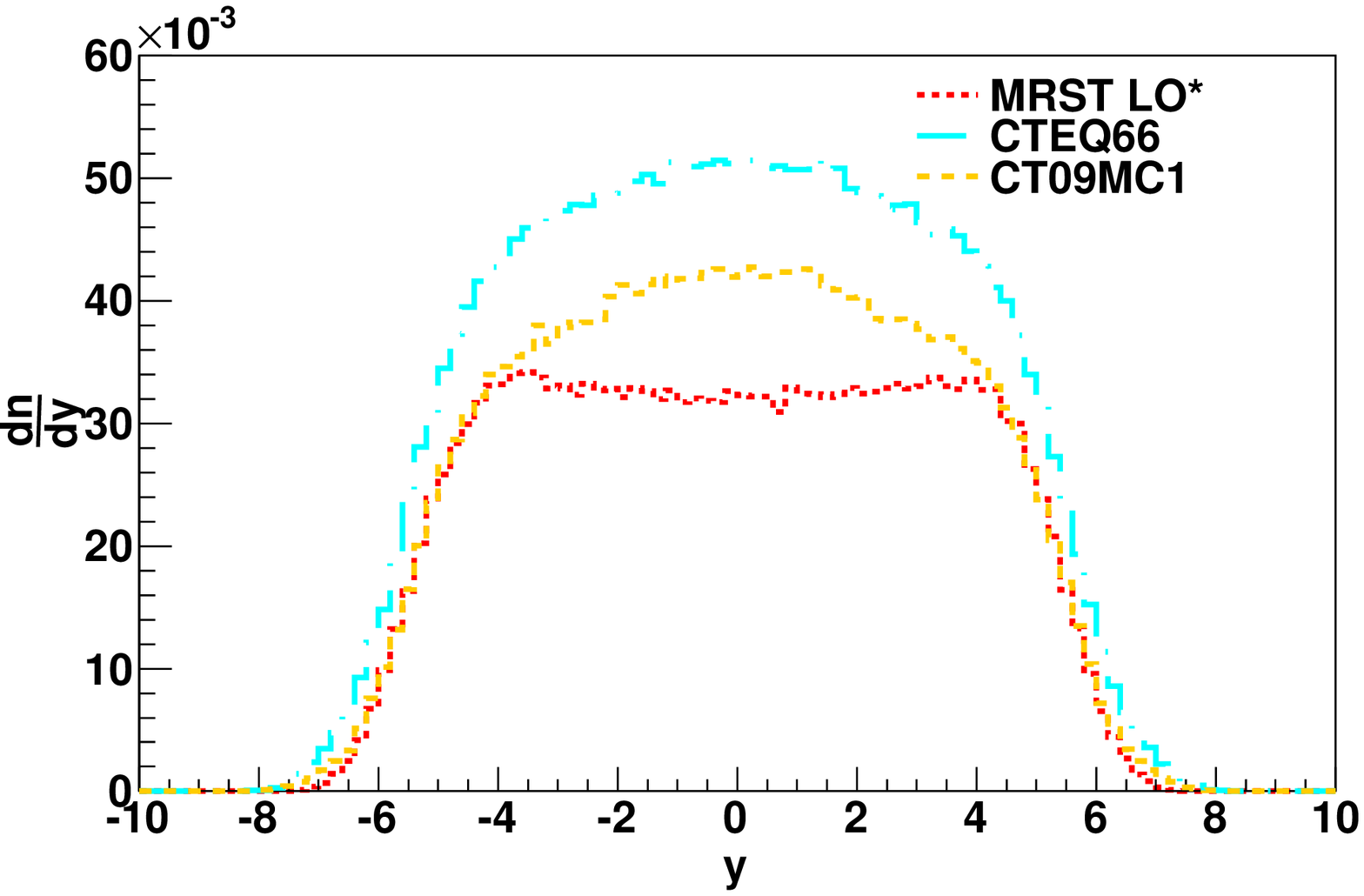}}
  \caption{Rapidity distribution for the outgoing quarks with the different PDFs. Only the
    $2 \rightarrow 2$ sub-process.}
  \label{fig:Rapid-quarks}
\end{figure}

 Minimum bias events are dominated by $gg\rightarrow gg$ interactions but
if we go back to the partons in Fig.~\ref{fig:Rapid-strip} and select
only the outgoing quarks, i.e. mainly from $qg\rightarrow qg$, large underlying differences are revealed, see
Fig.~\ref{fig:Rapid-quarks}. The CT09 and CTEQ66 distributions show smooth shapes leading up to a peak at zero
rapidity, while LO*/** have a very flat distribution at central
rapidities. The outlier is once again the MSTW LO distribution which shows
peaks at large rapidities. This is once again caused by the large gluon
distribution at small $x$ and we believe a reason, that it is so much more
prominent for the quarks, is that MSTW LO has a relatively small gluon
distribution at large $x$ which in asymmetric $gg$ interactions somewhat compensates,
while the up-quark distribution is similar to the other PDFs. The normal LO distributions from CTEQ
show similar distributions as the MC-adapted ones from MSTW. Actually CTEQ5L show similar peaks as MSTW LO, and looking at the
gluon distribution at low energies we can see that before it freezes as $x=10^{-6}$ it is
larger than in all other PDFs. There is also a clear difference in the amount of quarks produced
where CTEQ66 has a much larger amount and thereby produces much more quark
initiated jets than the rest of the PDFs.

\begin{figure}[tp]
  \subfloat[]{\includegraphics[width=0.5\textwidth]
    {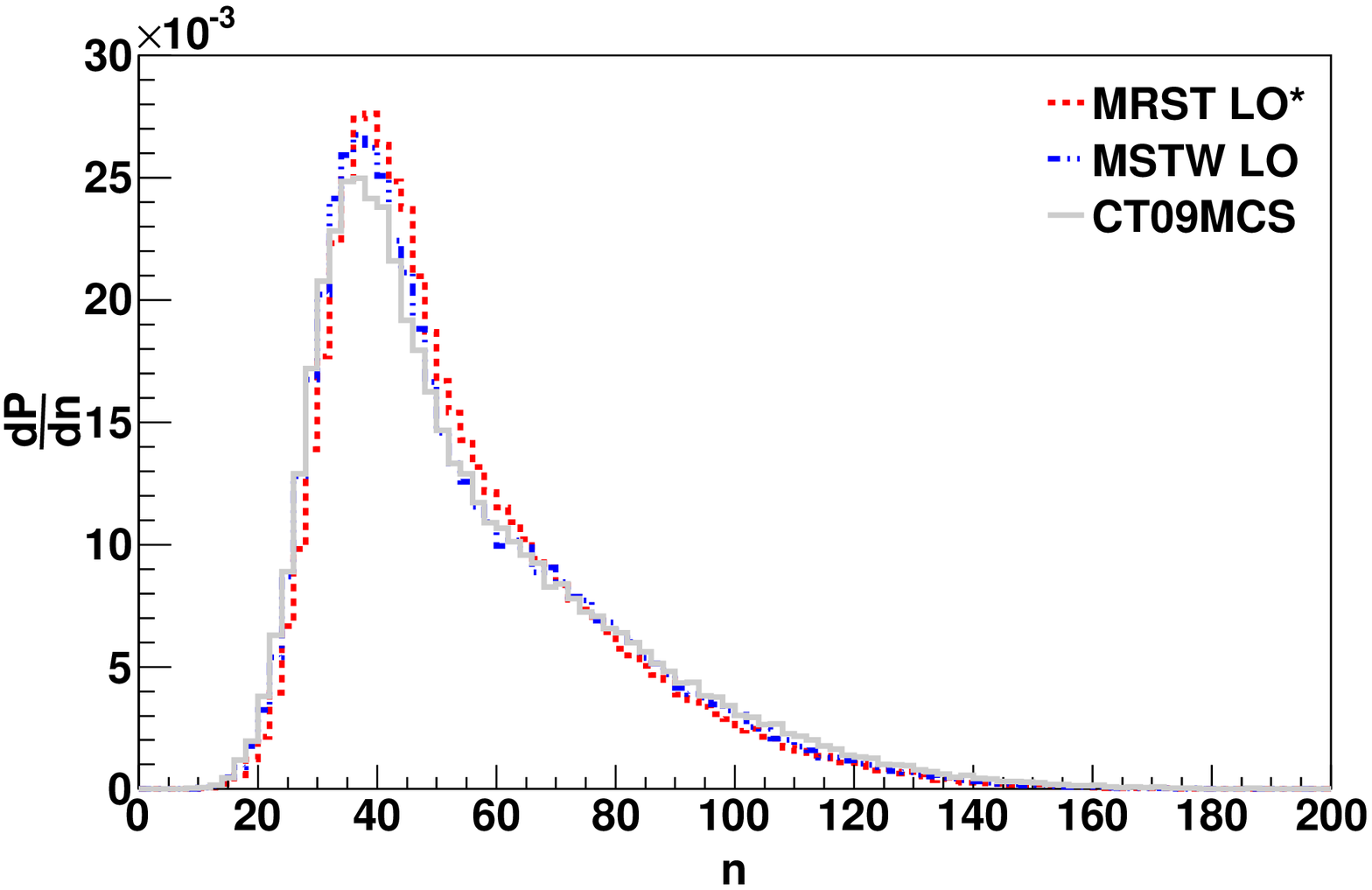}}
   \subfloat[]{\includegraphics[width=0.5\textwidth]
    {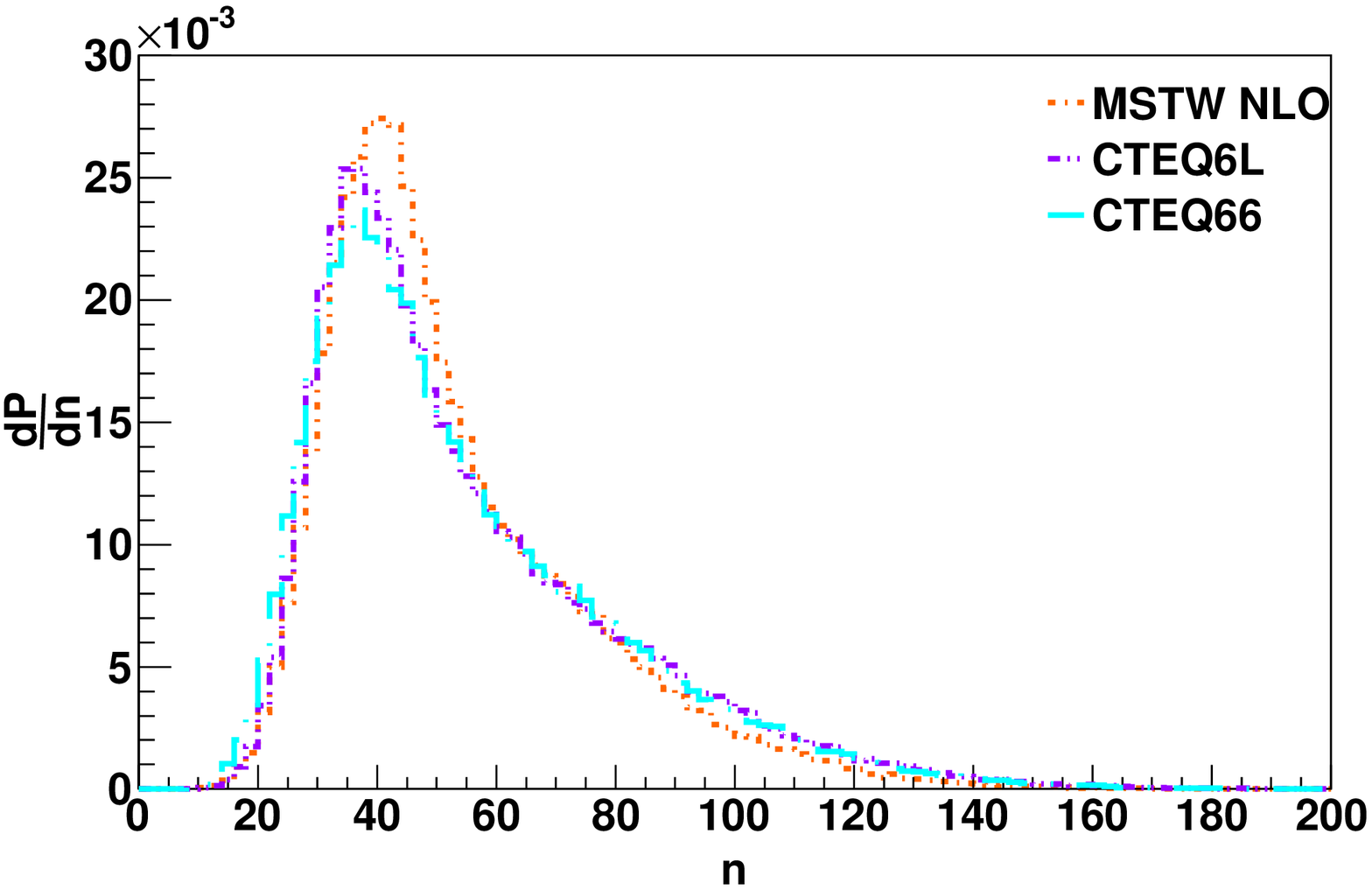}}
  \caption{Charged particle multiplicity distributions.}
  \label{fig:Multdist}
\end{figure}

The multiplicity distributions are similar for most
PDFs. The two NLO distributions stand out as two extremes in different
directions where MSTW NLO has the highest peak and the shortest tail as shown
in Fig.~\ref{fig:Multdist}. All MC-adapted PDFs except MCS
have a peak slightly shifted to larger multiplicities but are different in
height where the two from MRST have a larger peak value. The three normal
leading-order distributions are all similar and we only show the CTEQ6L.

\begin{figure}[tp]
  \subfloat[]{\includegraphics[width=0.5\textwidth]
    {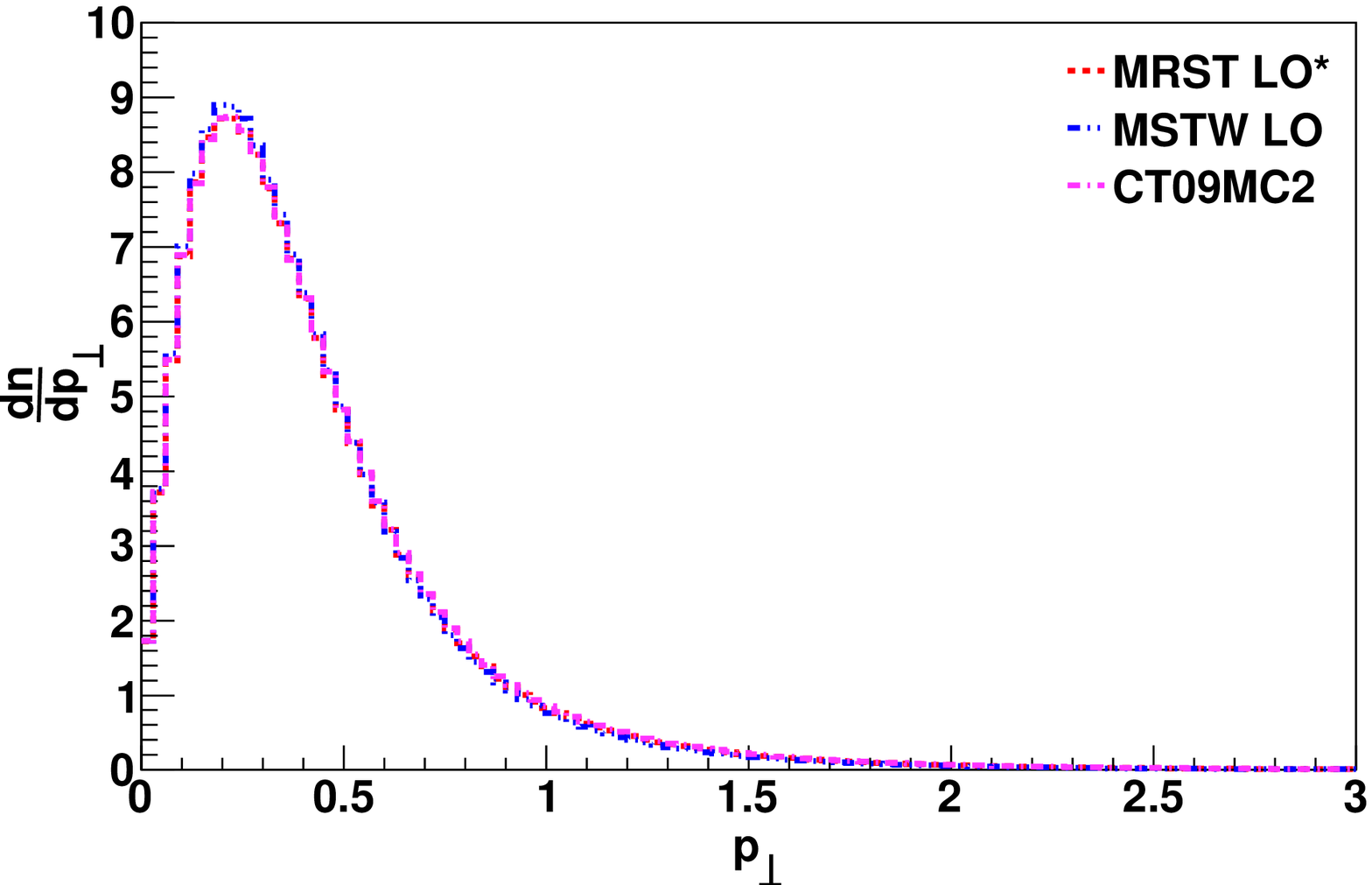}}
   \subfloat[]{\includegraphics[width=0.5\textwidth]
    {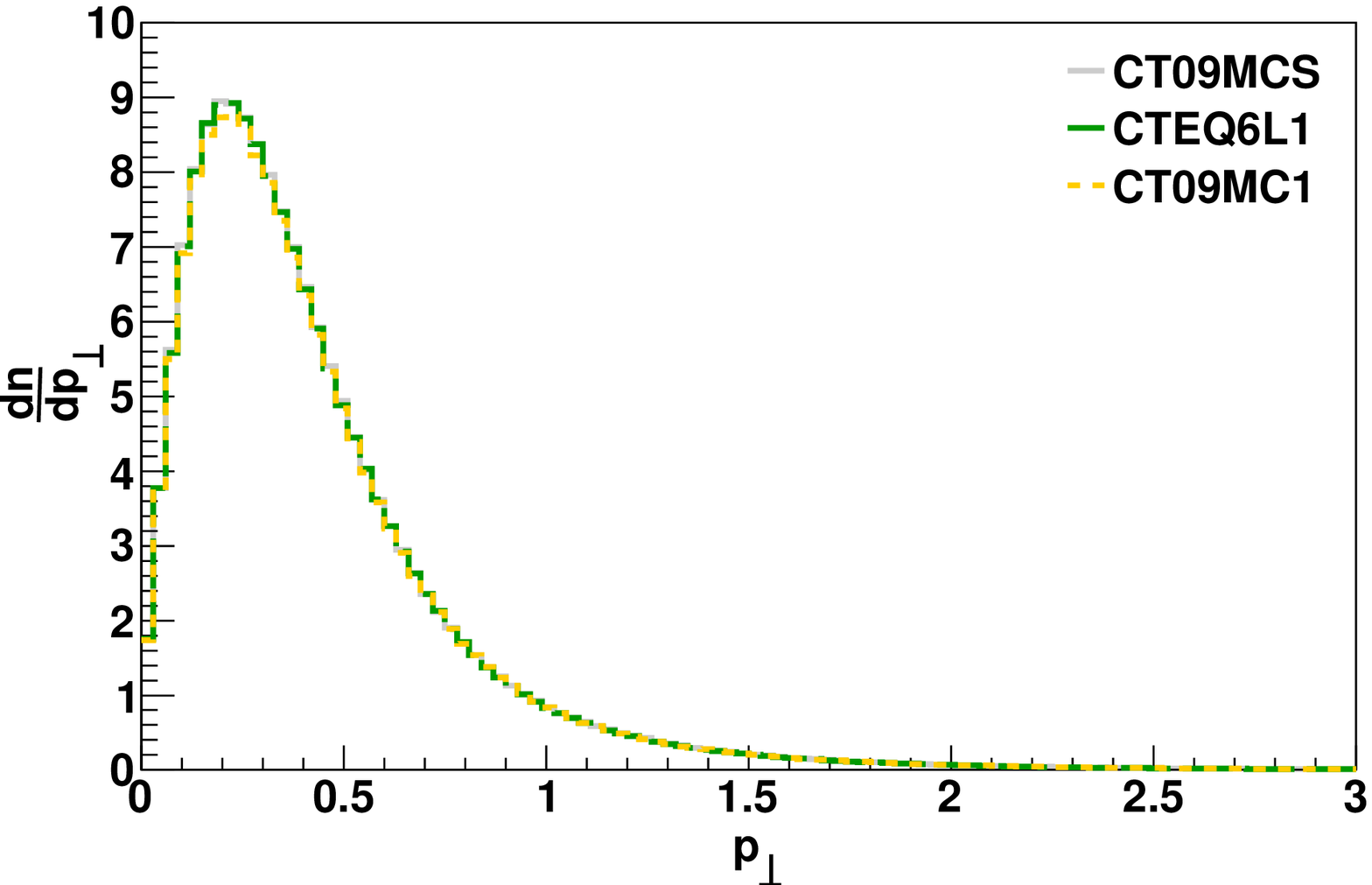}}
  \caption{$p_{\perp}$ distributions of charged particles.}
  \label{fig:pTdist}
\end{figure}

Fig.~\ref{fig:pTdist} show the $p_{\perp}$ distribution which is almost completely
independent of PDF and the overlap of the three distributions in each figure
makes them impossible to distinguish.

\begin{figure}[tp]
   \subfloat[]{\label{fig:CRapid-Comp4}\includegraphics[width=0.5\textwidth]
    {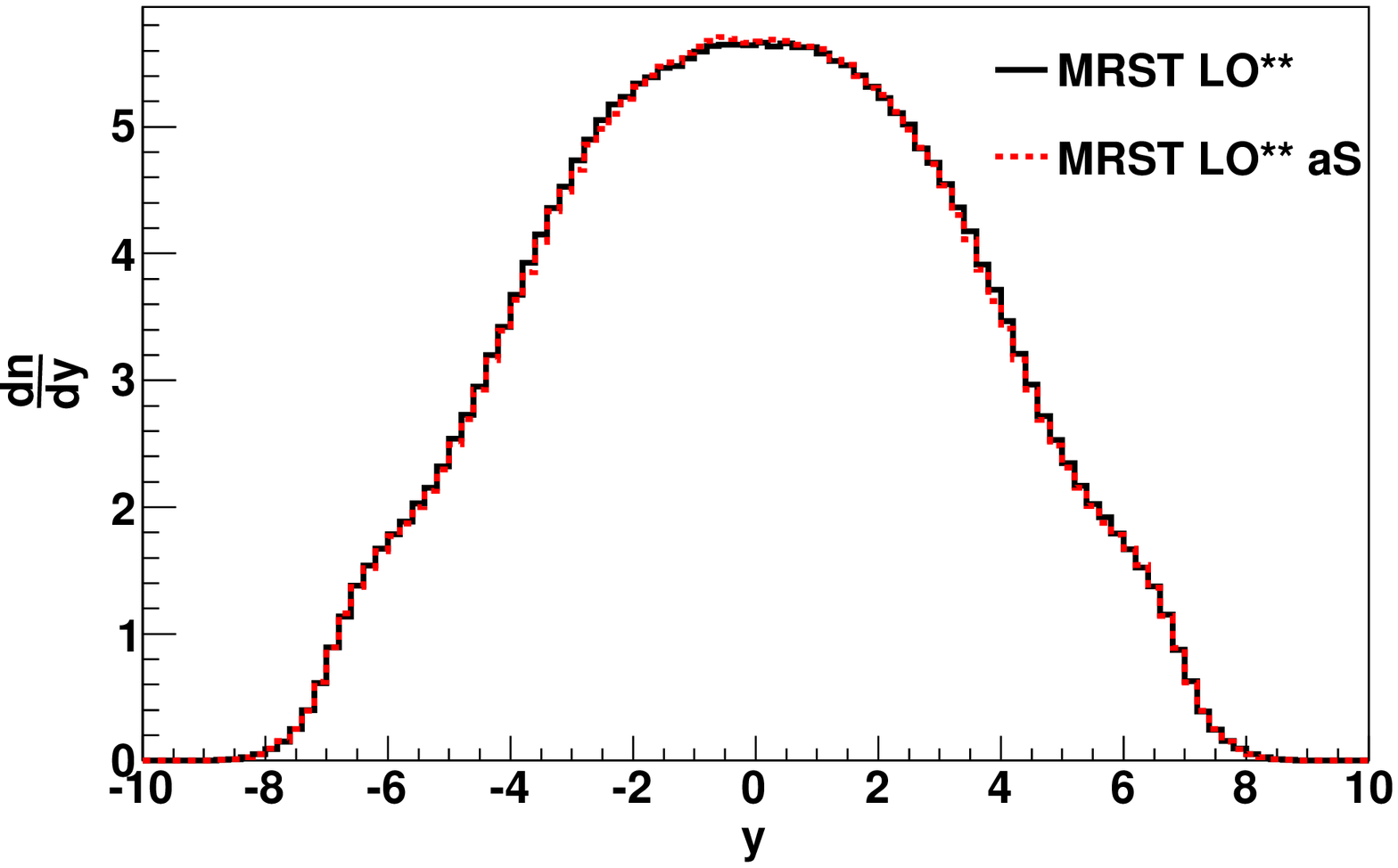}}
  \subfloat[]{\label{fig:CMult-Comp4}\includegraphics[width=0.5\textwidth]
    {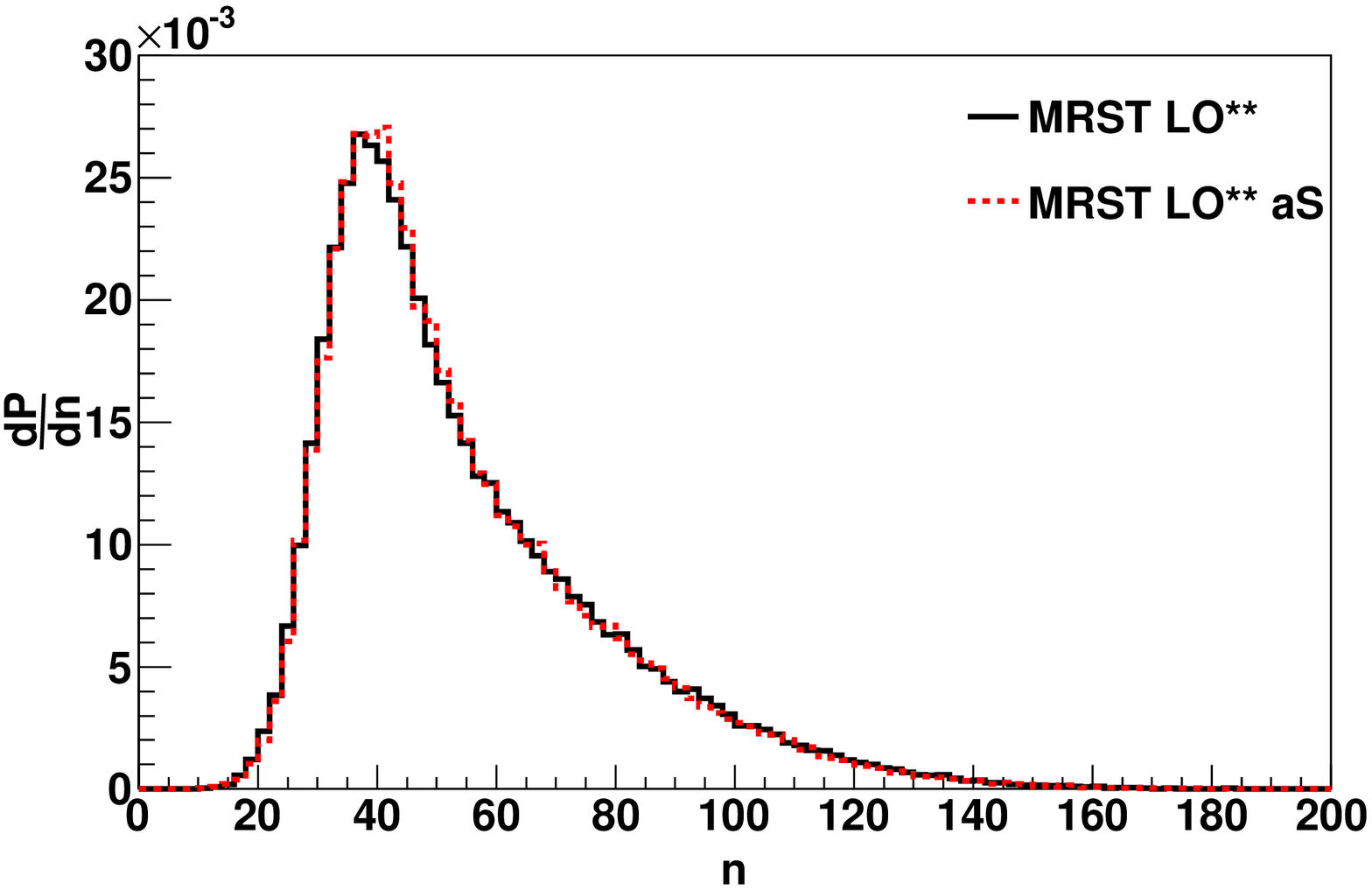}} 
  \caption{MRST LO** comparison between simulations with $\alpha_S$ determined
    by \textsc{Pythia8} and from the PDF. Rapidity distribution to the left and charge
    multiplicity distribution to the right. The two distributions overlap in all four figures.}
  \label{fig:Comp}
\end{figure}

Repeating the simulations but now with $\alpha_S$ set differently for the
different PDFs, i.e. the value at $M_Z$ and the order of the running is
determined by the PDFs, does not significantly change anything, once the multiplicity
has been retuned. As an illustration of this point, the results with the two
different $\alpha_S$ for the LO** PDF are shown in Fig~\ref{fig:Comp}. CT09MCS
uses a varying running of $\alpha_S$ which is not implemented in \textsc{Pythia8}
and the simulations with this PDF use a NLO running $\alpha_S$ instead. This
should not change the results much since the varying $\alpha_S$ is fine tuning
\cite{Huston}.

\begin{figure}[tp]
  \subfloat[]{\includegraphics[width=0.5\textwidth]
    {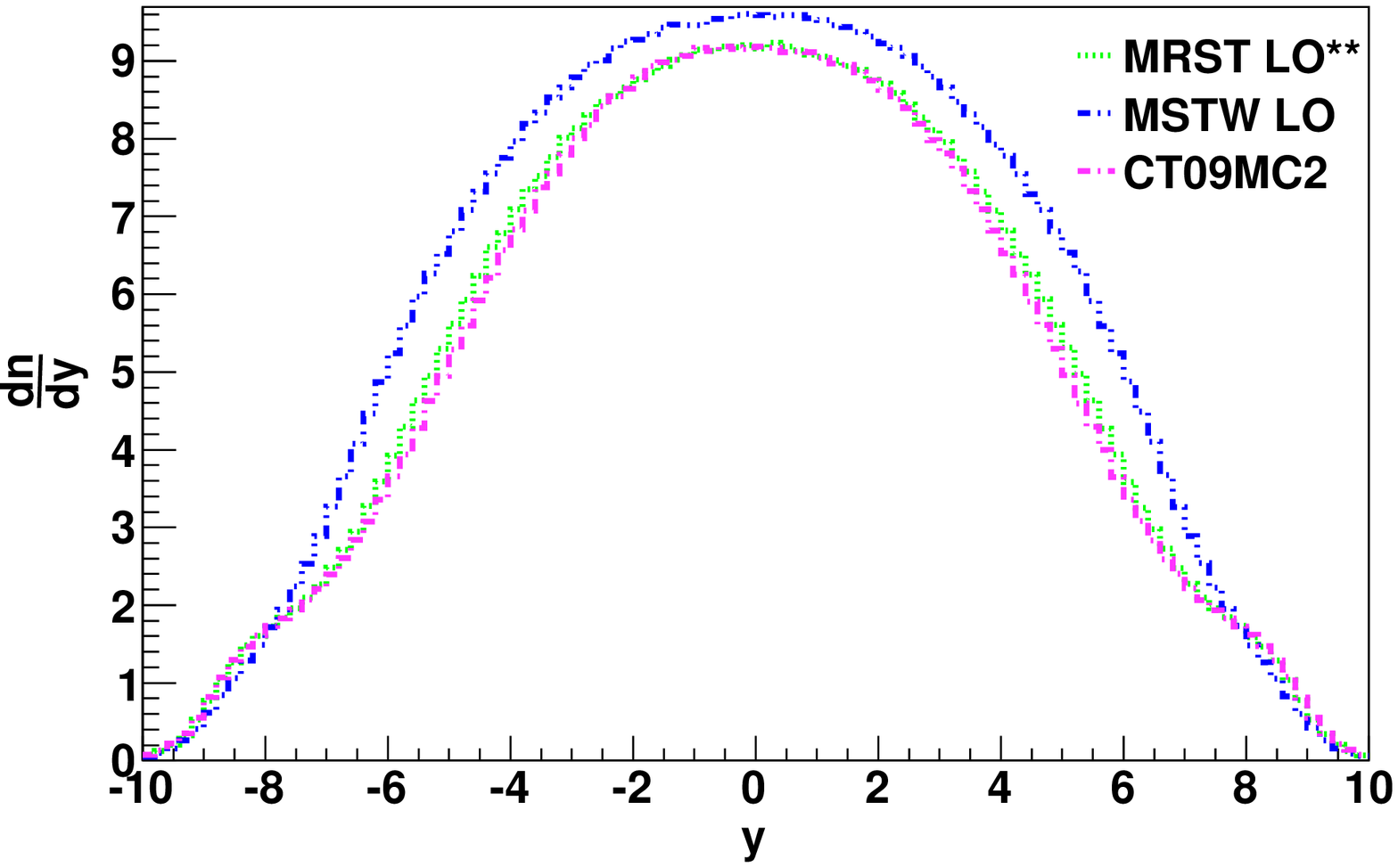}}
   \subfloat[]{\includegraphics[width=0.5\textwidth]
    {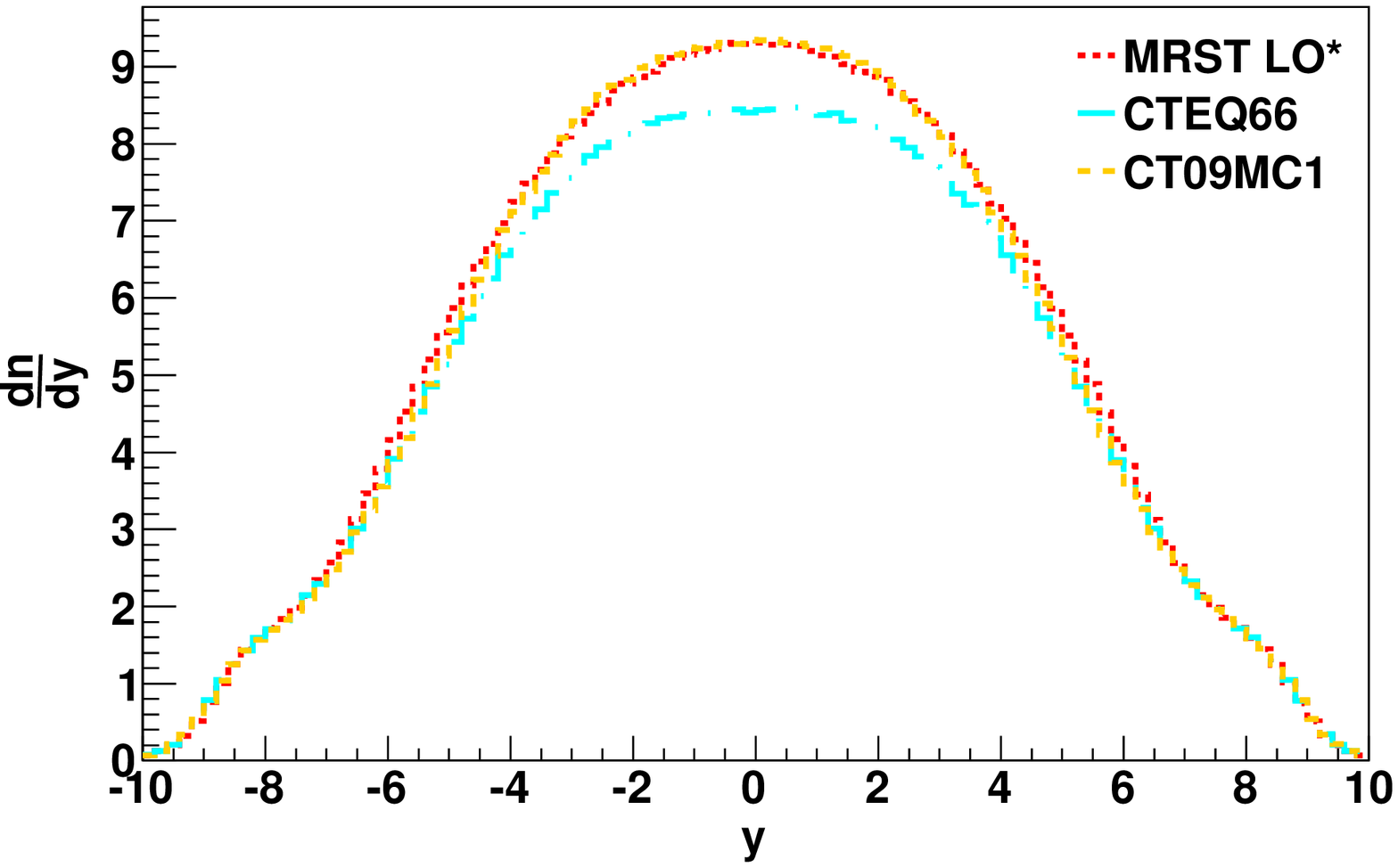}}
  \caption{Rapidity distributions at LHC ($E_{CM}=14$~TeV).}
  \label{fig:Rapid-LHC}
\end{figure}

Increasing the energy
to the level of a fully operational LHC enhances the differences
seen at Tevatron energy, especially for MSTW LO and the two NLO PDFs. The
multiplicity of these three evolve with energy in a different way than for the
other PDFs. The rapidity distribution, shown in Fig.~\ref{fig:Rapid-LHC},
naturally extends to larger rapidities and the total
charged particle
multiplicity increases since the energy available is larger. MSTW LO here gives
a much broader distribution and also has a much higher total charged particle
multiplicity. This is because as the energy increases even lower values of $x$
come into play, so that the effect of the gluon distribution in this region
has larger impact on the results. The two NLO PDFs have a flatter peak
than the MC-adapted PDFs and are similar in shape to MSTW LO but have much
smaller multiplicity. The rest of the PDFs evolve in a fashion similar to the
MC-adapted PDFs shown in the figure but with some more variation. MC1 and LO* are a little bit larger at central rapidities than MC2
and LO**.

\begin{figure}[tp]
  \subfloat[]{\includegraphics[width=0.5\textwidth]
    {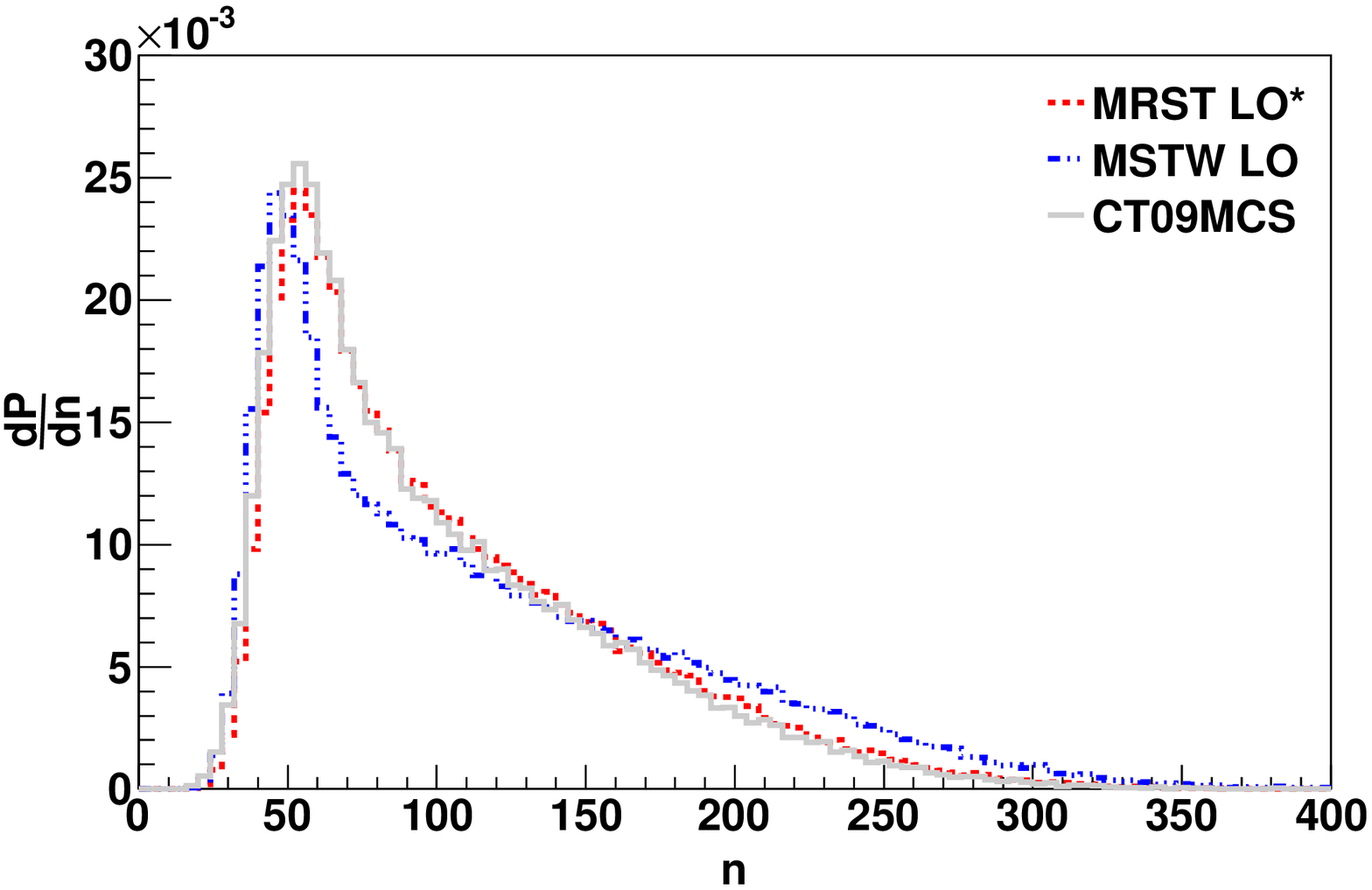}}
   \subfloat[]{\includegraphics[width=0.5\textwidth]
    {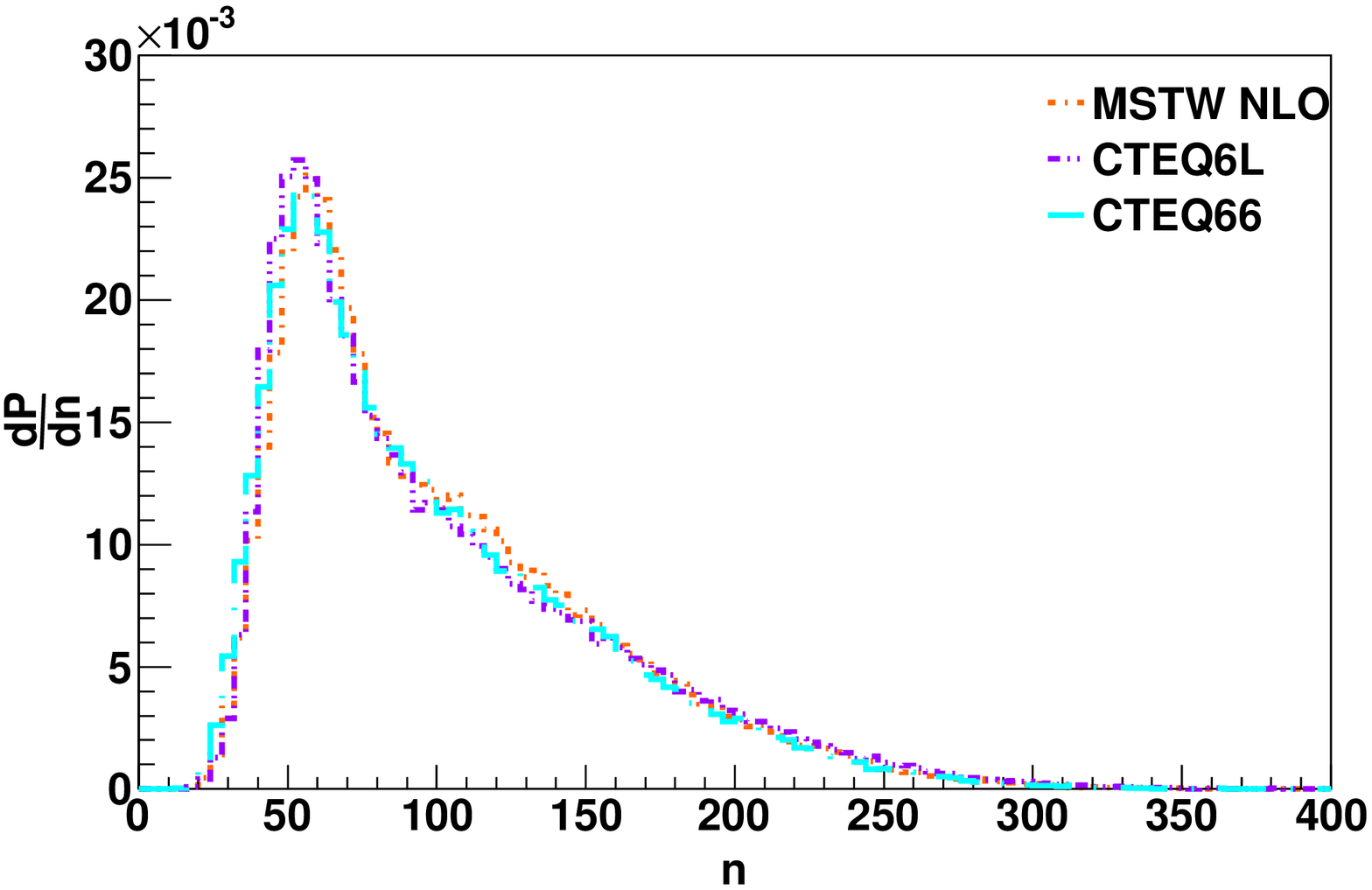}}
  \caption{Charged particle multiplicity distributions at LHC.}
  \label{fig:Mult-LHC}
\end{figure}

The multiplicity distributions in Fig.~\ref{fig:Mult-LHC} also show increased differences except for the
two NLO PDFs which converge at this energy. Not only is there a larger mean multiplicity
and peak at a higher value, but the same distributions that stand out from the
rest with the rapidity also do so with their charge particle multiplicity
distribution.

\begin{figure}[tp]
  \subfloat[]{\includegraphics[width=0.5\textwidth]
    {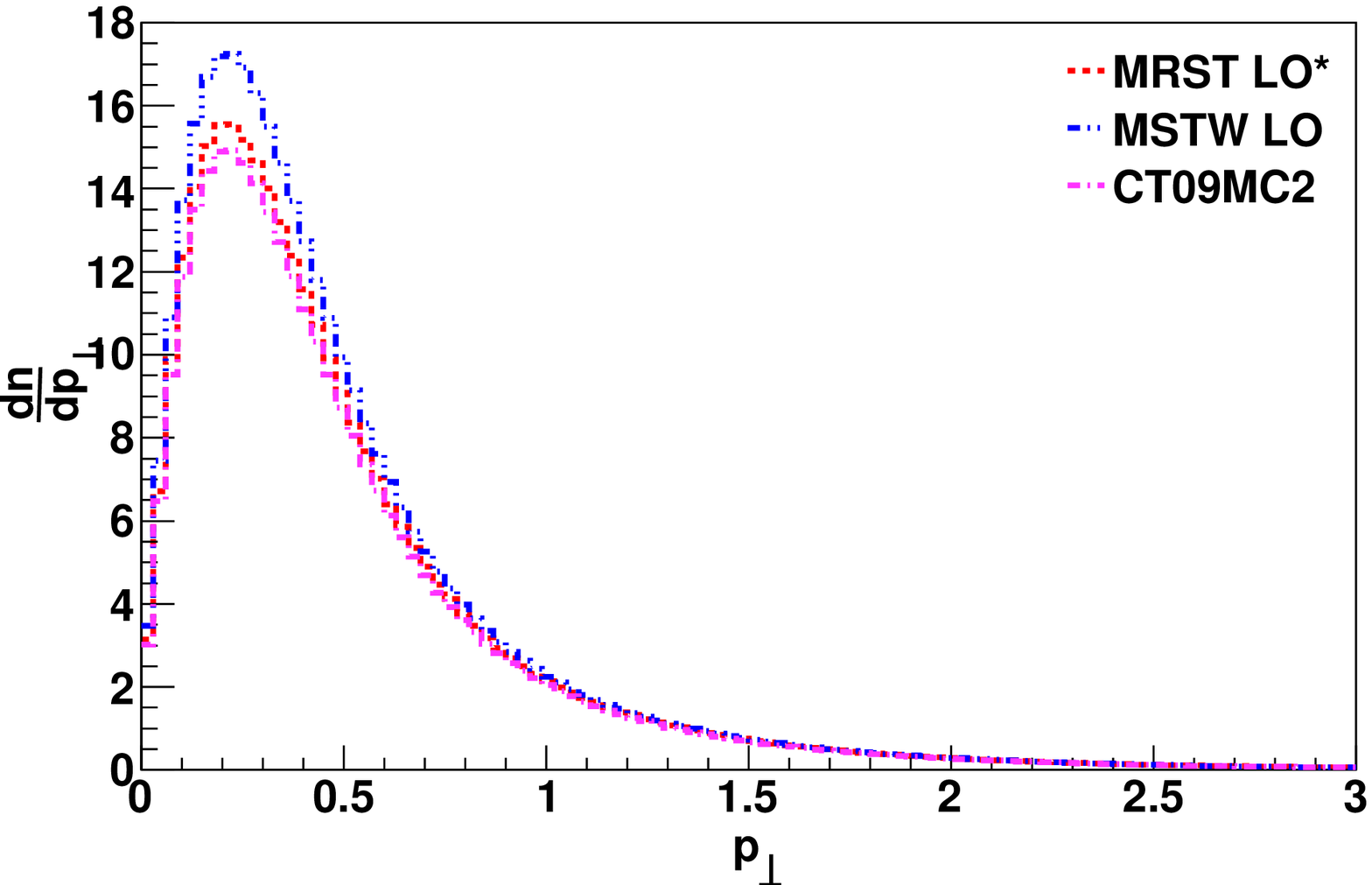}}
   \subfloat[]{\includegraphics[width=0.5\textwidth]
    {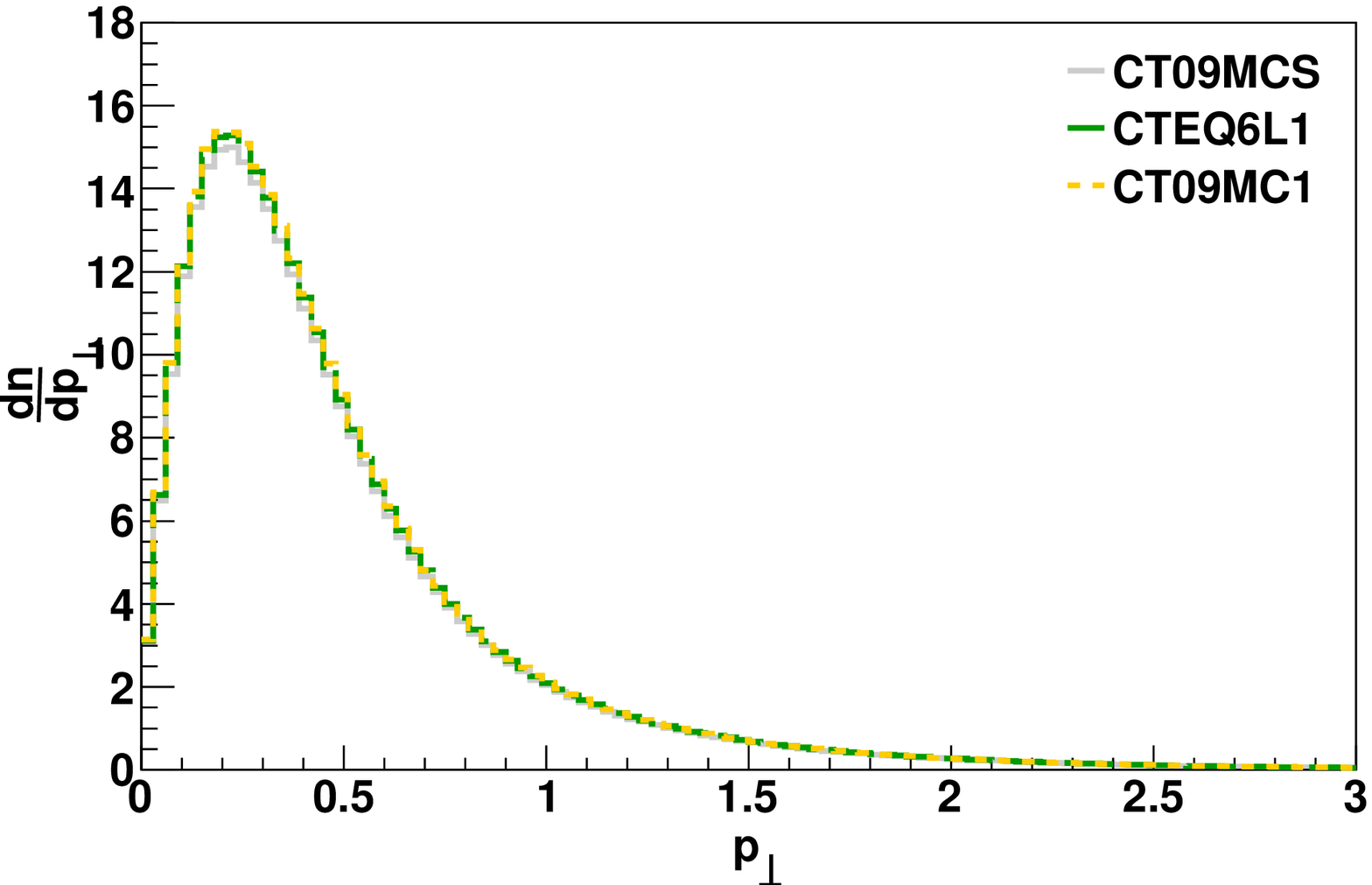}}
  \caption{$p_{\perp}$ distributions for the different PDFs at LHC.}
  \label{fig:pT-LHC}
\end{figure}

This is the case also for the $p_{\perp}$ distributions but these are still
very similar, excluding the MSTW LO with its large multiplicity, Fig.~\ref{fig:pT-LHC}.

\subsubsection{Comparison to CDF Run 2 data}
\begin{figure}[tp]
  \centering
  \subfloat[]{\includegraphics[width=0.5\textwidth]
    {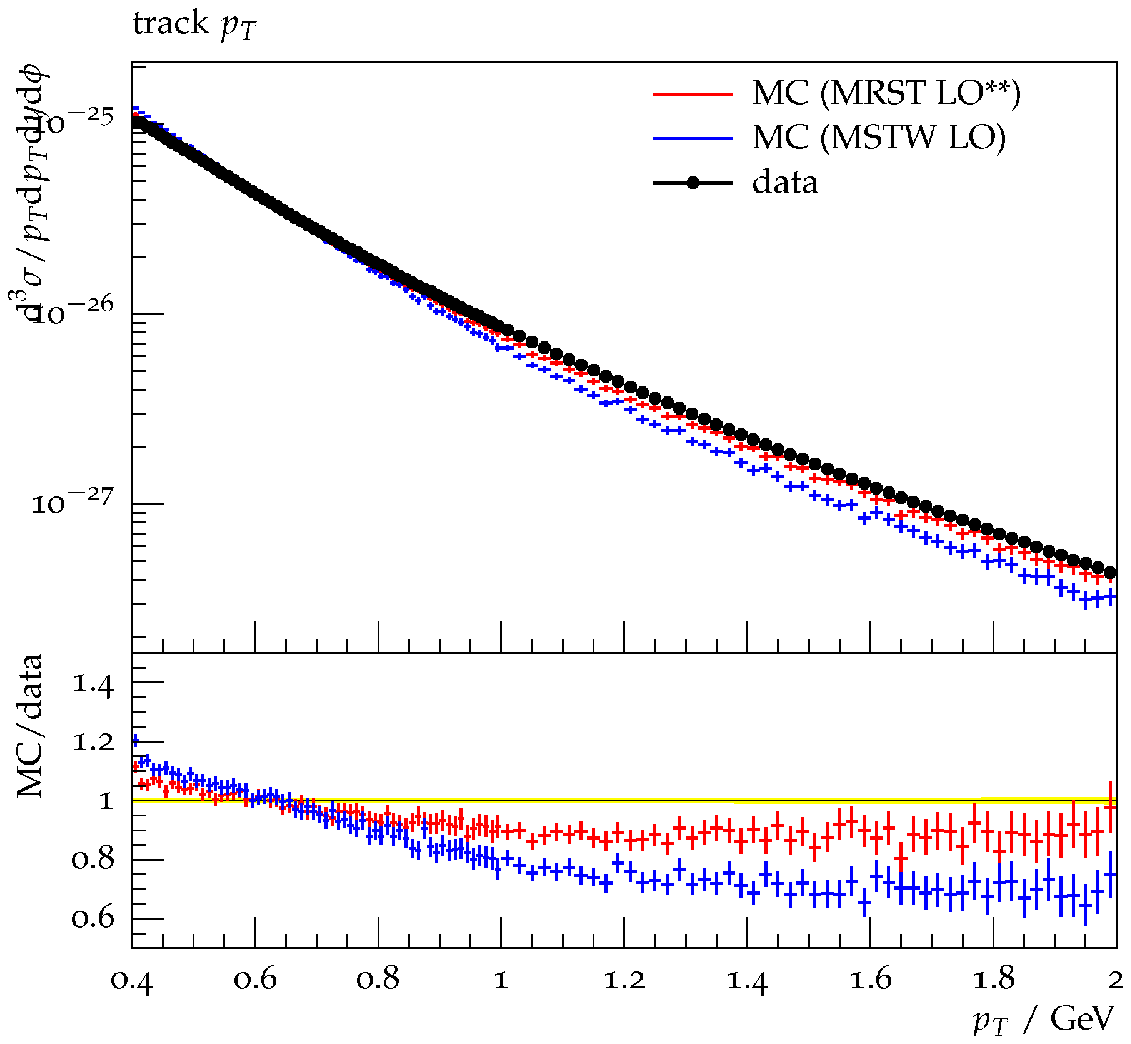}} 
  \subfloat[]{\includegraphics[width=0.5\textwidth]
    {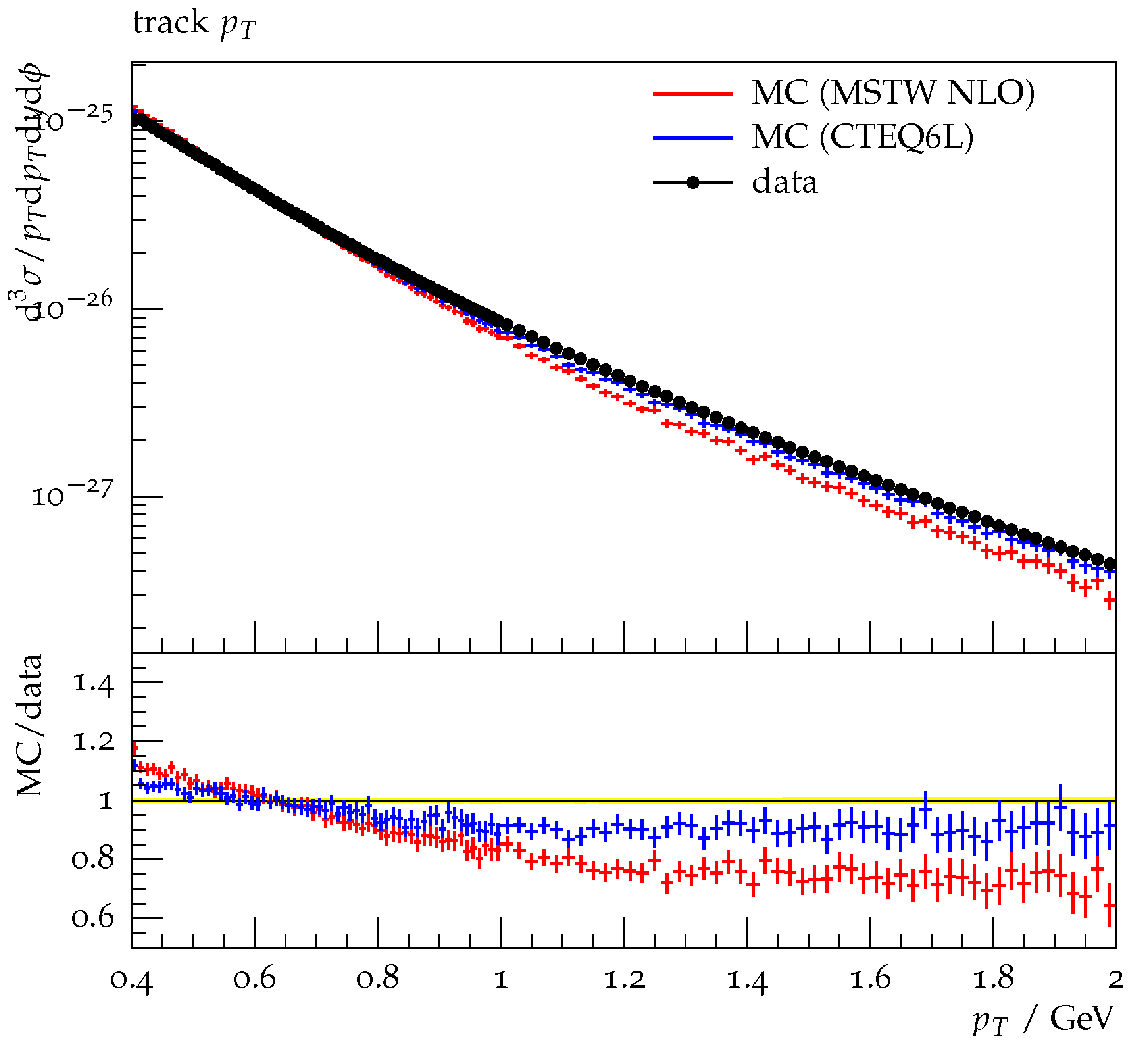}} \\
  \subfloat[]{\includegraphics[width=0.5\textwidth]
    {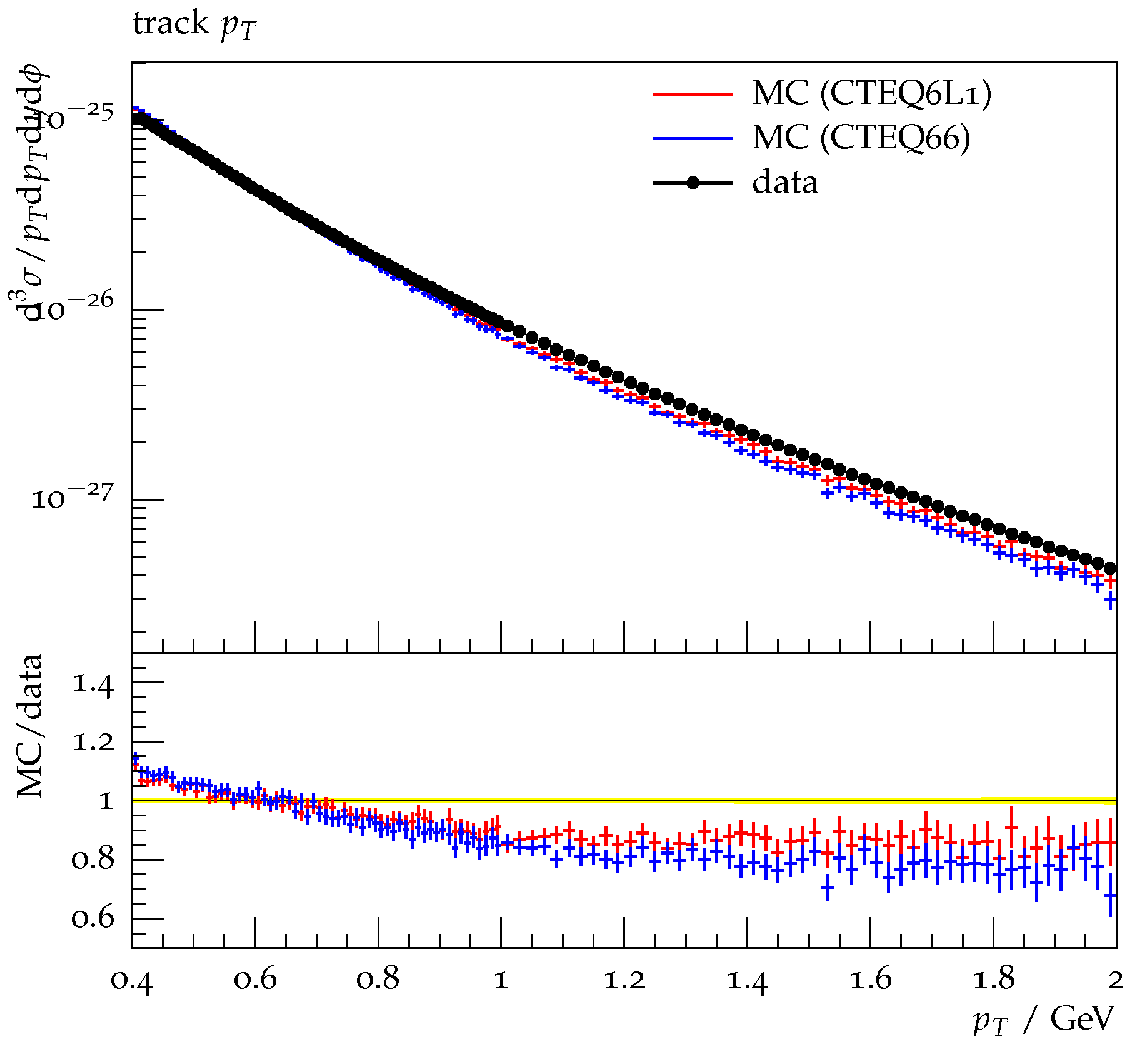}} 
  \subfloat[]{\includegraphics[width=0.5\textwidth]
    {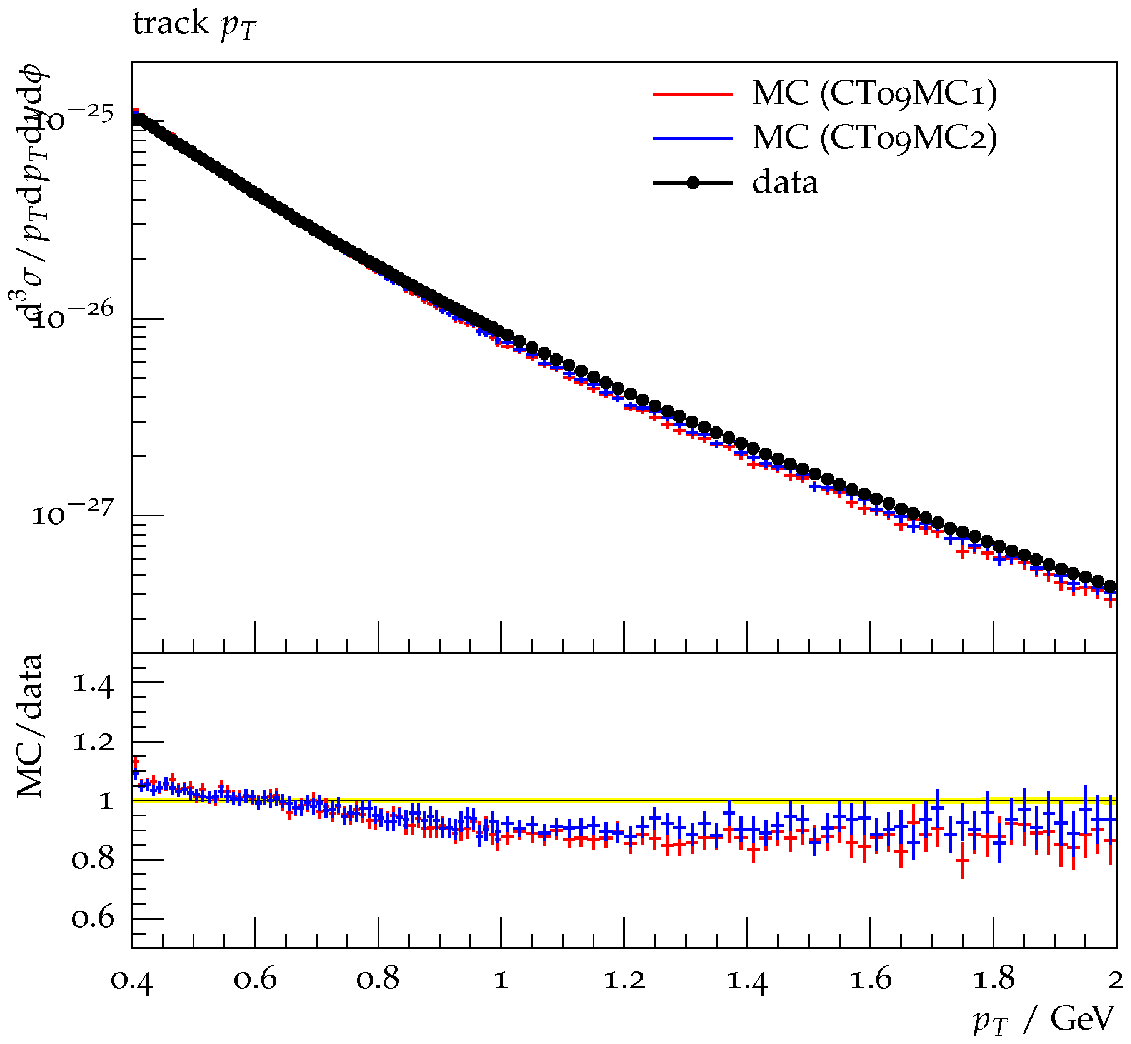}}
  \caption{$p_{\perp}$ spectra of charged particles from the CDF Run 2 experiment compared to
    simulations with different PDFs.}
  \label{fig:SigmapT-Riv}
\end{figure}

The analyses in Rivet ensure that the comparisons with data have the same cuts
and corrections as the original experiment. Therefore only the central
pseudorapidity region is used and also cuts in transverse
momentum \cite{Aaltonen2}. 
$p_{\perp}$ spectra of charged particles in Fig.~\ref{fig:SigmapT-Riv} show the same
relative shape for all PDFs, which gives too large values at the low
$p_{\perp}$ end,
then decreases compared to data and gives too small differential cross sections
at the high end. The slope shows some differences depending on
the choice of PDF. MC-adapted PDFs and the CTEQ6L give results that are the
closest to data, while MSTW LO and NLO are further away than the rest.

The $\sum E_{\perp}$ spectrum of particles, neutral particles included, shows
larger dependence on the PDFs but it is the same distributions that result
in the values closest to data. Since we have not done a complete tune, this is
to indicate the importance of the PDFs and results far away from data are not
necessarily the fault of the PDFs. However the $\sum E_{\perp}$ distribution is less dependent on
details of the MC and therefore easier for PDF developers to consider in tunes. The MC-adapted PDFs from CTEQ as well as the
CTEQ6L reproduce data well, while MSTW LO goes down to less than half the cross
section of data at the larger energy end. MSTW NLO peaks at higher energies
than the rest. All PDFs give a too large value at the peak, but then decrease
too fast and differ the most from data at the high energy end.

\begin{figure}[tp]
  \centering
  \subfloat[]{\includegraphics[width=0.5\textwidth]
    {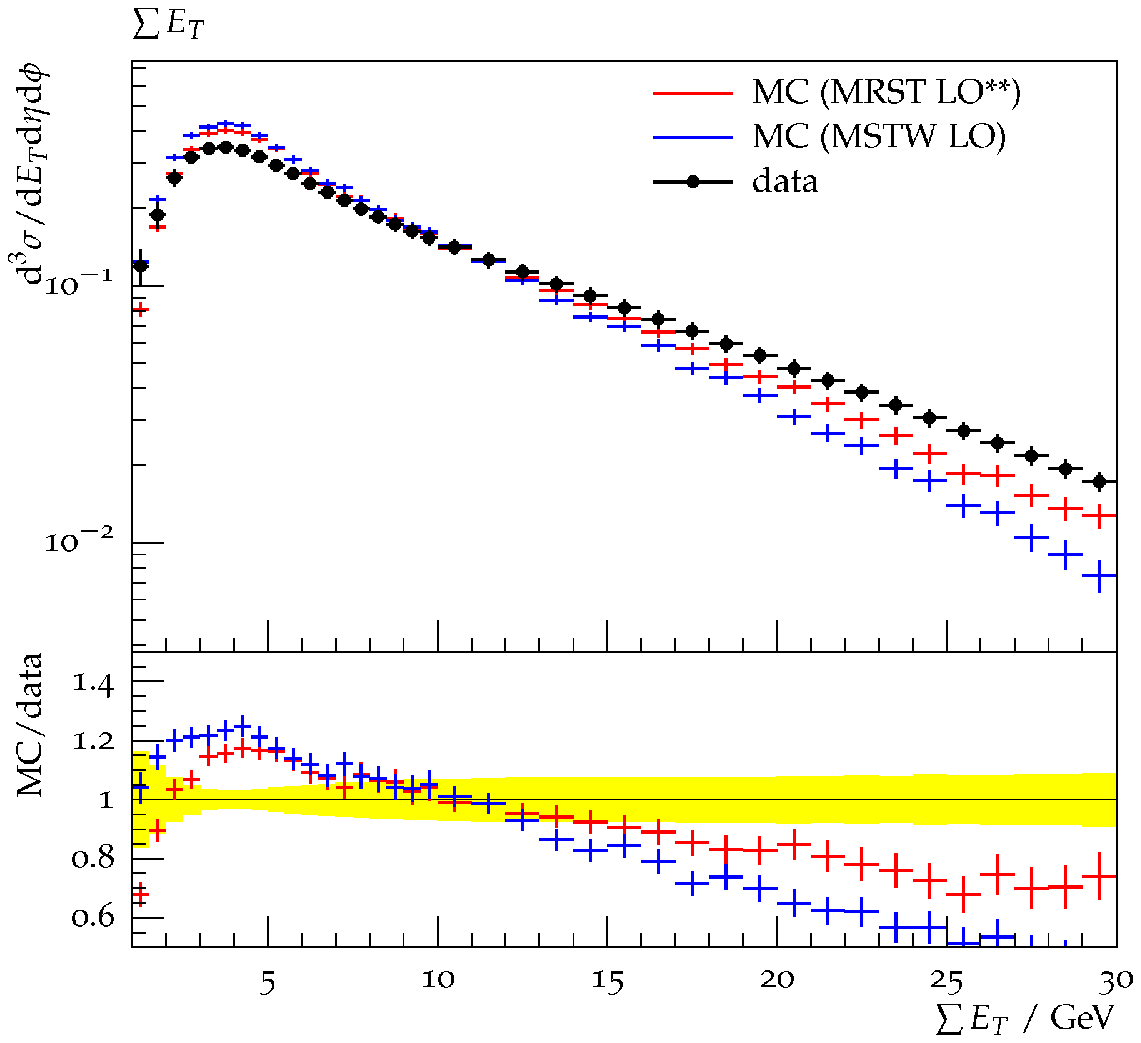}} 
  \subfloat[]{\includegraphics[width=0.5\textwidth]
    {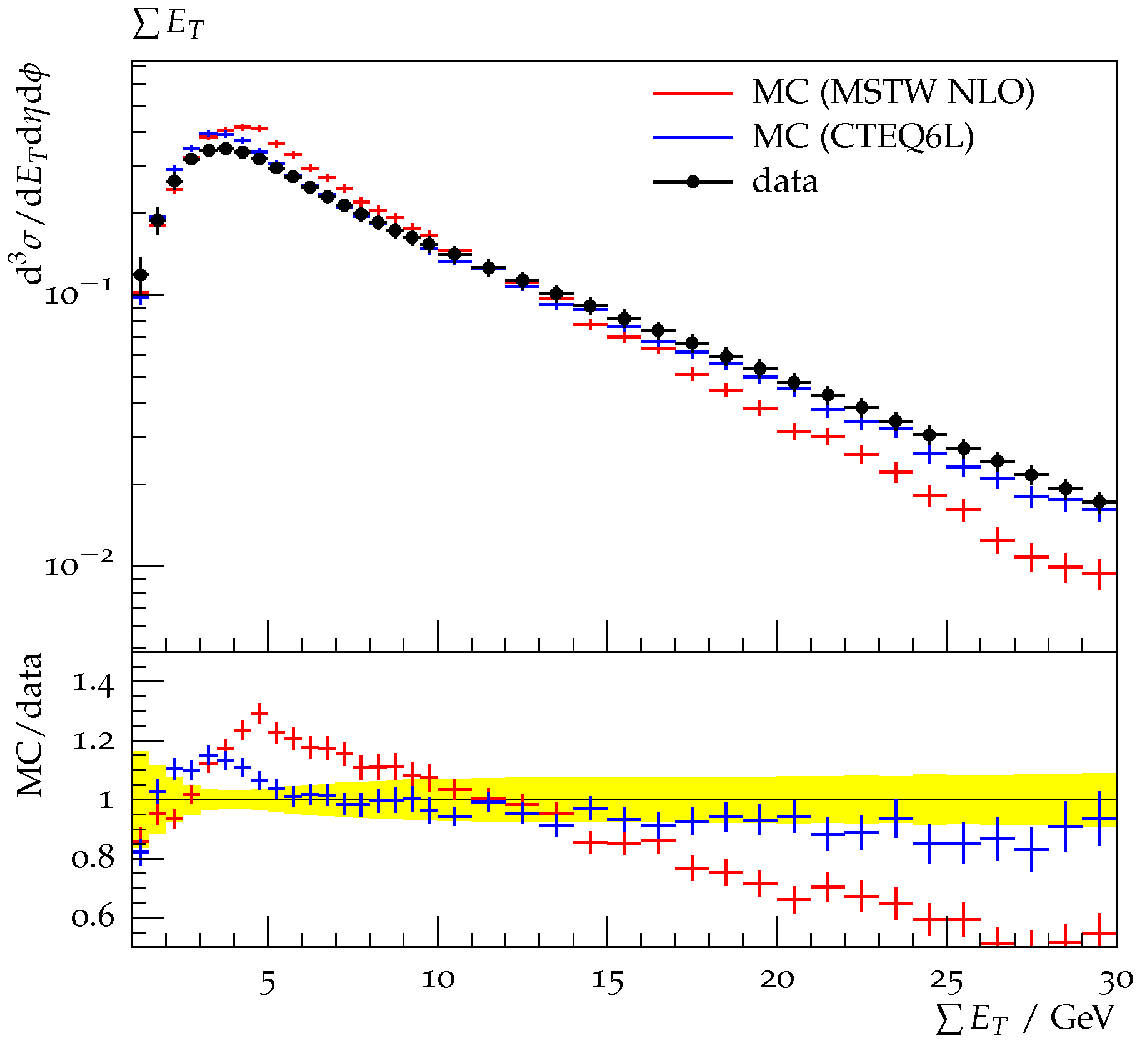}} \\
  \subfloat[]{\includegraphics[width=0.5\textwidth]
    {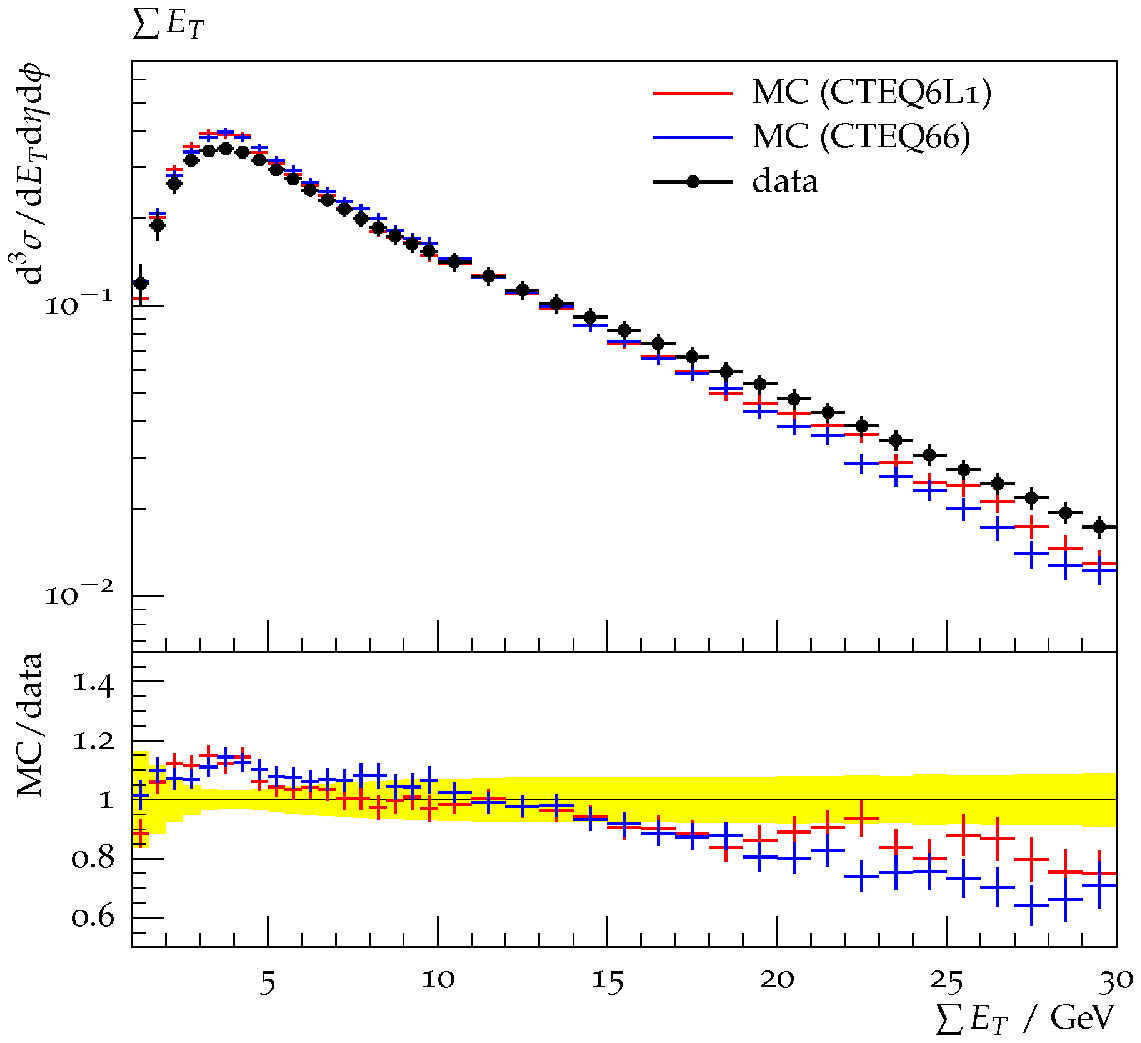}} 
  \subfloat[]{\includegraphics[width=0.5\textwidth]
    {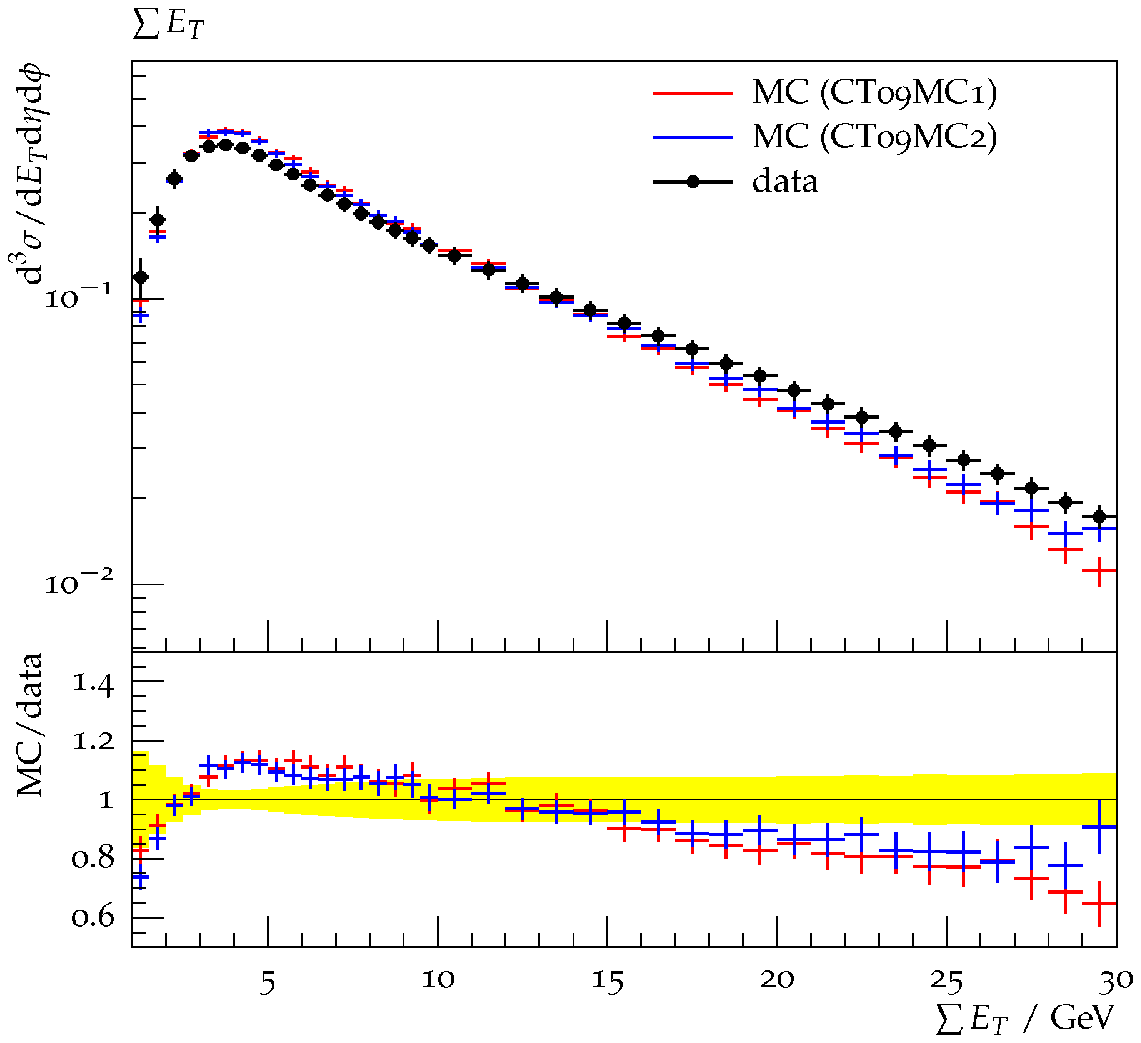}}
\vspace{-0.2cm}
  \caption{$\sum E_{\perp}$ spectra from the CDF Run 2 experiment compared to
    simulations with different PDFs.}
  \label{fig:SigmaET-Riv}
\vspace{-0.4cm}
\end{figure}

\begin{figure}[tp]
  \centering
  \subfloat[]{\label{fig:pTCMult-Riv1}\includegraphics[width=0.5\textwidth]
    {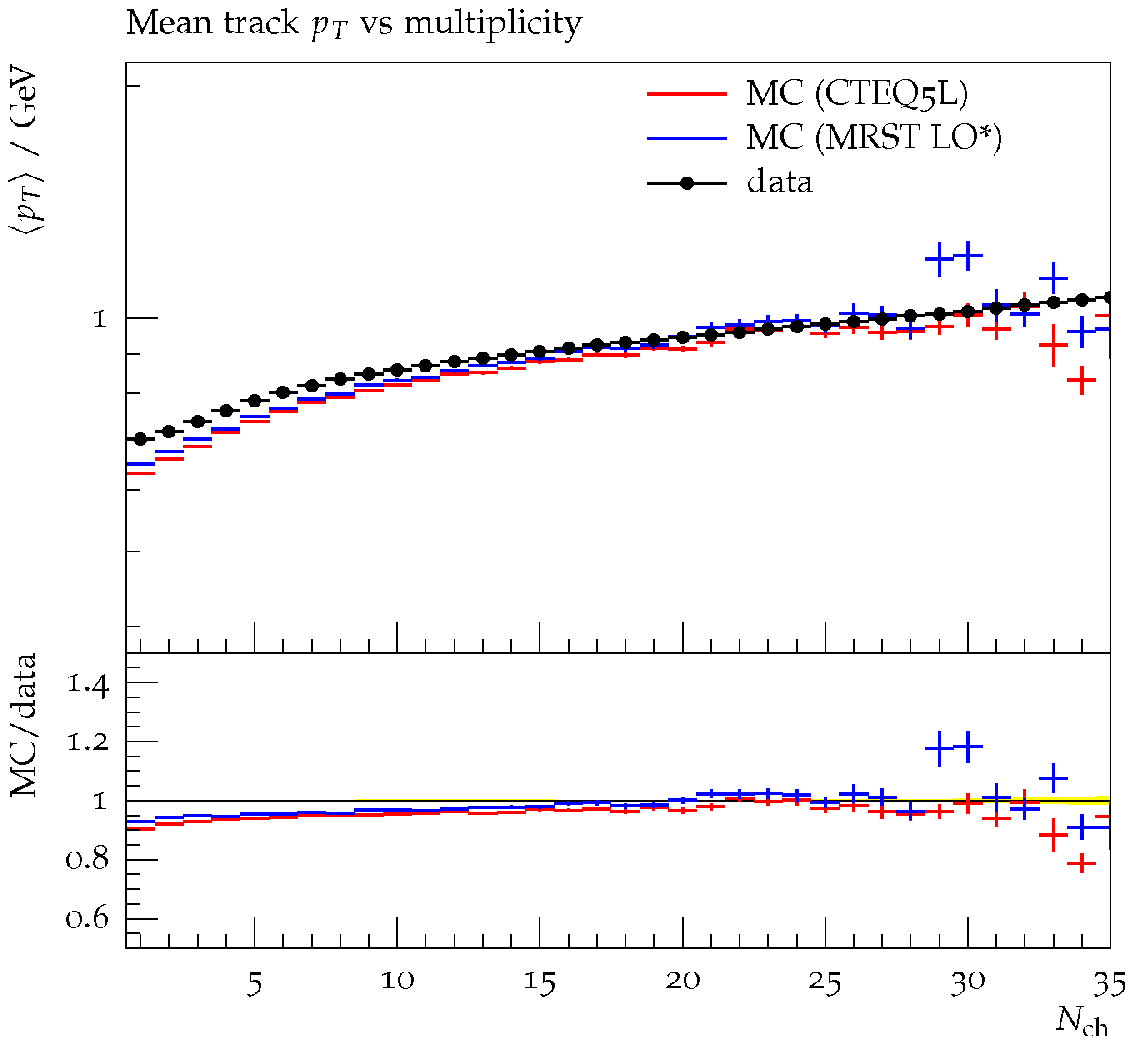}} 
  \subfloat[]{\label{fig:pTCMult-Riv2}\includegraphics[width=0.5\textwidth]
    {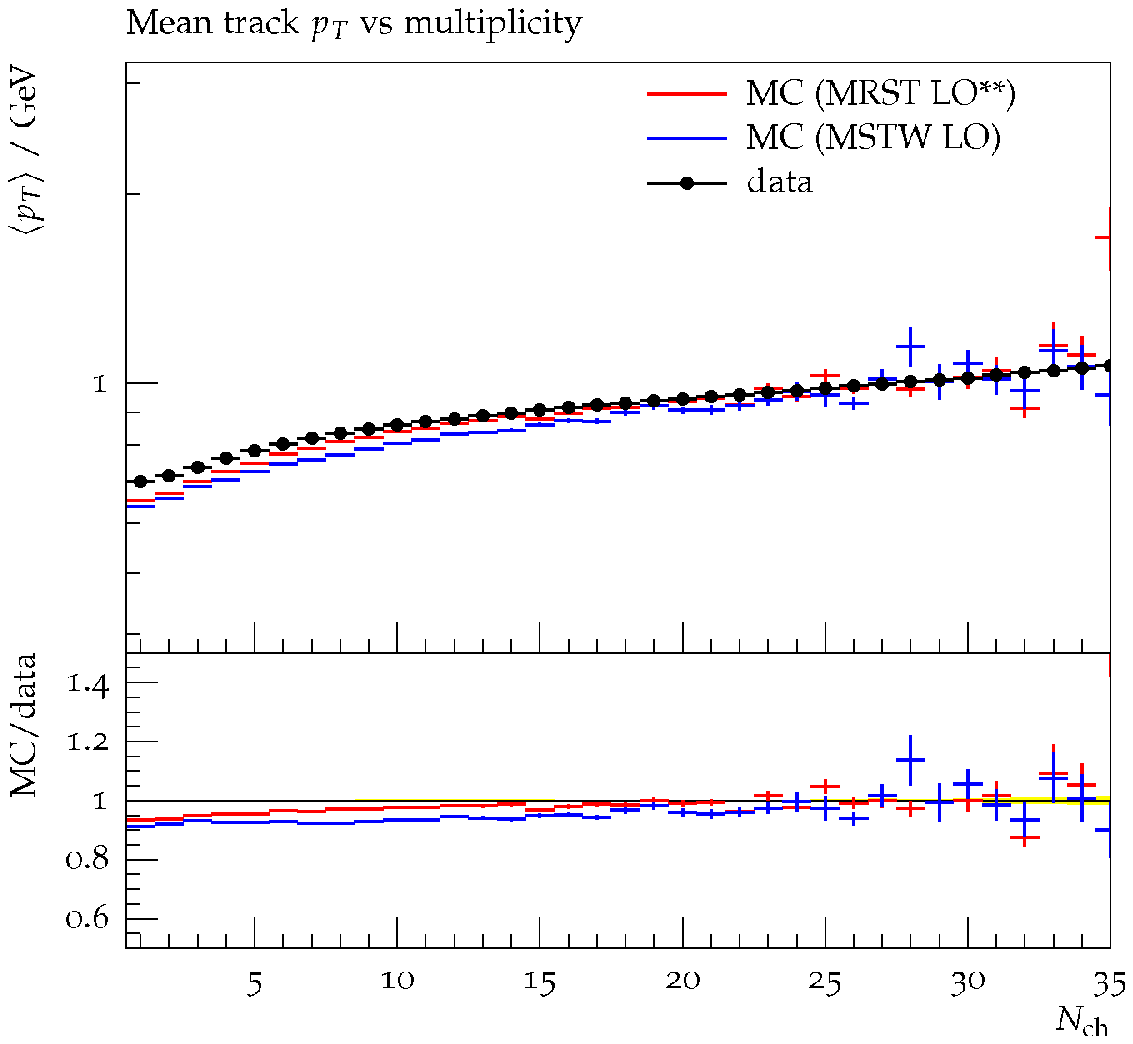}} 
  \caption{Evolution of the average transverse momentum, $<p_T>$, with charge
    multiplicity, $N_{ch}$, from the CDF Run 2 experiment compared to
    simulations with different PDFs.}
  \label{fig:pTCMult-Riv}
\end{figure}

Comparing the evolution of the average transverse momentum with charge
multiplicity from our simulations to data (which used to be one of the distribution
the MC generators struggled the most with), show that all PDFs reproduce
the data fairly well and also behave
in a similar fashion to one another. They all give too low $\langle p_{\perp}\rangle$ at low
multiplicity and then increase relative to data so that they get closer as the
multiplicity increases, Fig.~\ref{fig:pTCMult-Riv}. The only PDF that is
slightly different than the rest is MSTW LO.

\section{Inclusive Jet Cross Section}
\subsection{Introduction}
Quarks and gluons produced in collisions fragment because of the
color confinement and produce jets of color neutral hadrons. The definition
of what is a jet is far from trivial and identifying jets from data is even
more difficult. This task is performed by different jet algorithms relying on
closeness in either direction or momentum space. 

We compare the inclusive jet cross sections from our MC simulations with data, 
collected by the CDF experiment at Tevatron Run 2 \cite{Aaltonen}, over five
pseudorapidity intervals ranging up to $\eta \leq 2.1$. In the experimental
analysis the jets are identified with the midpoint cone algorithm and also
compared to results with the $k_T$ algorithm  \cite{Salam}. We are interested
mainly in the low $p_{\perp}$ region, in order to examine whether simulations can
be improved by introducing a correction ($K$-) factor, as done, for example, in $Z$
production. A $K$-factor is a factor that multiplies the cross section
in order to indirectly include known corrections from next-to-leading-order
calculations, and by that increase the total production rate. We also examine the rapidity, multiplicity and transverse momentum
distributions for the individual hadrons. The simulations were done with a
$p_{\perp}$ cut at $40$~GeV.

\subsection{Results}
\begin{figure}[tp]
  \centering
  \subfloat[]{\includegraphics[width=0.5\textwidth]
    {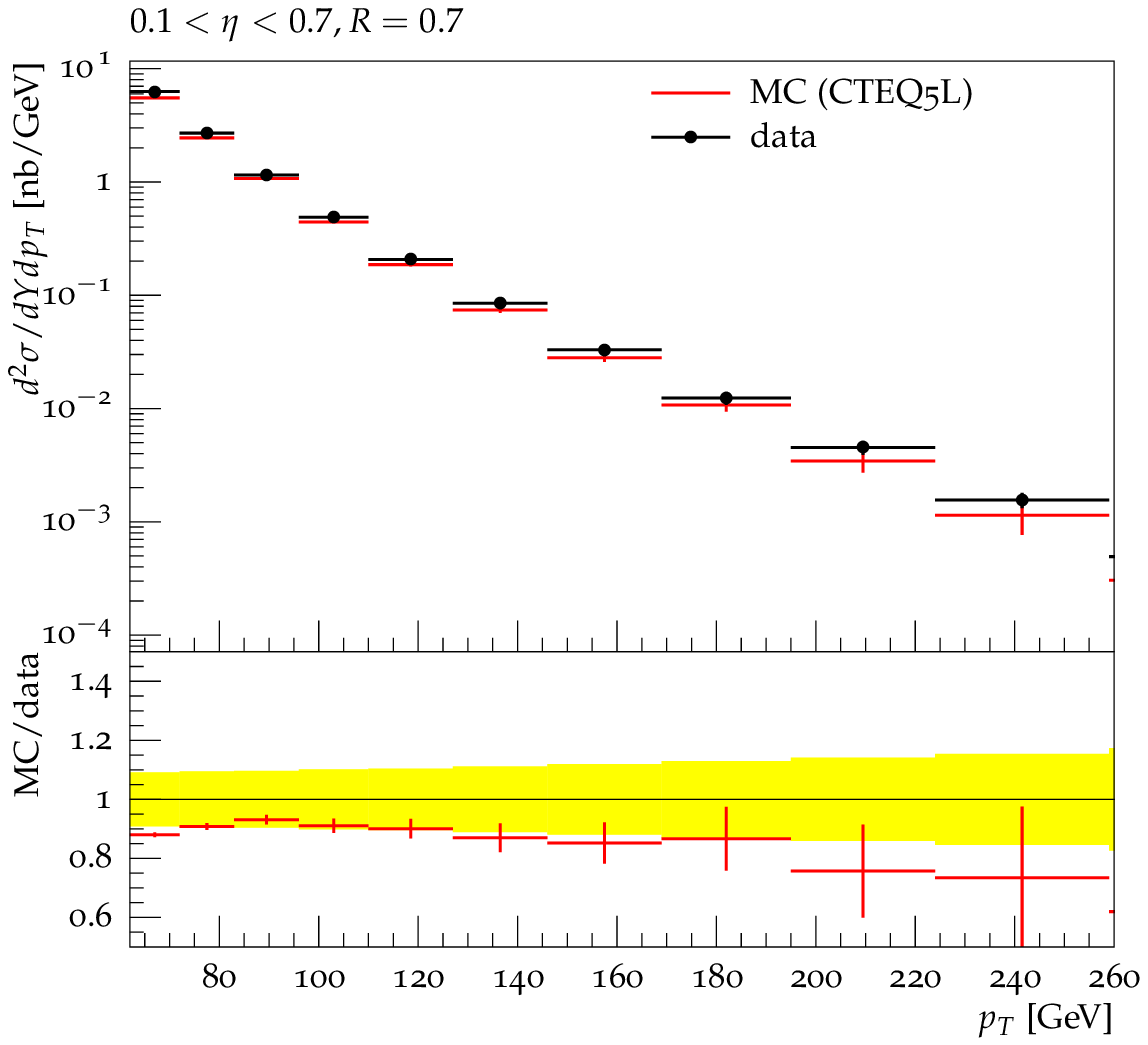}} 
  \subfloat[]{\includegraphics[width=0.5\textwidth]
    {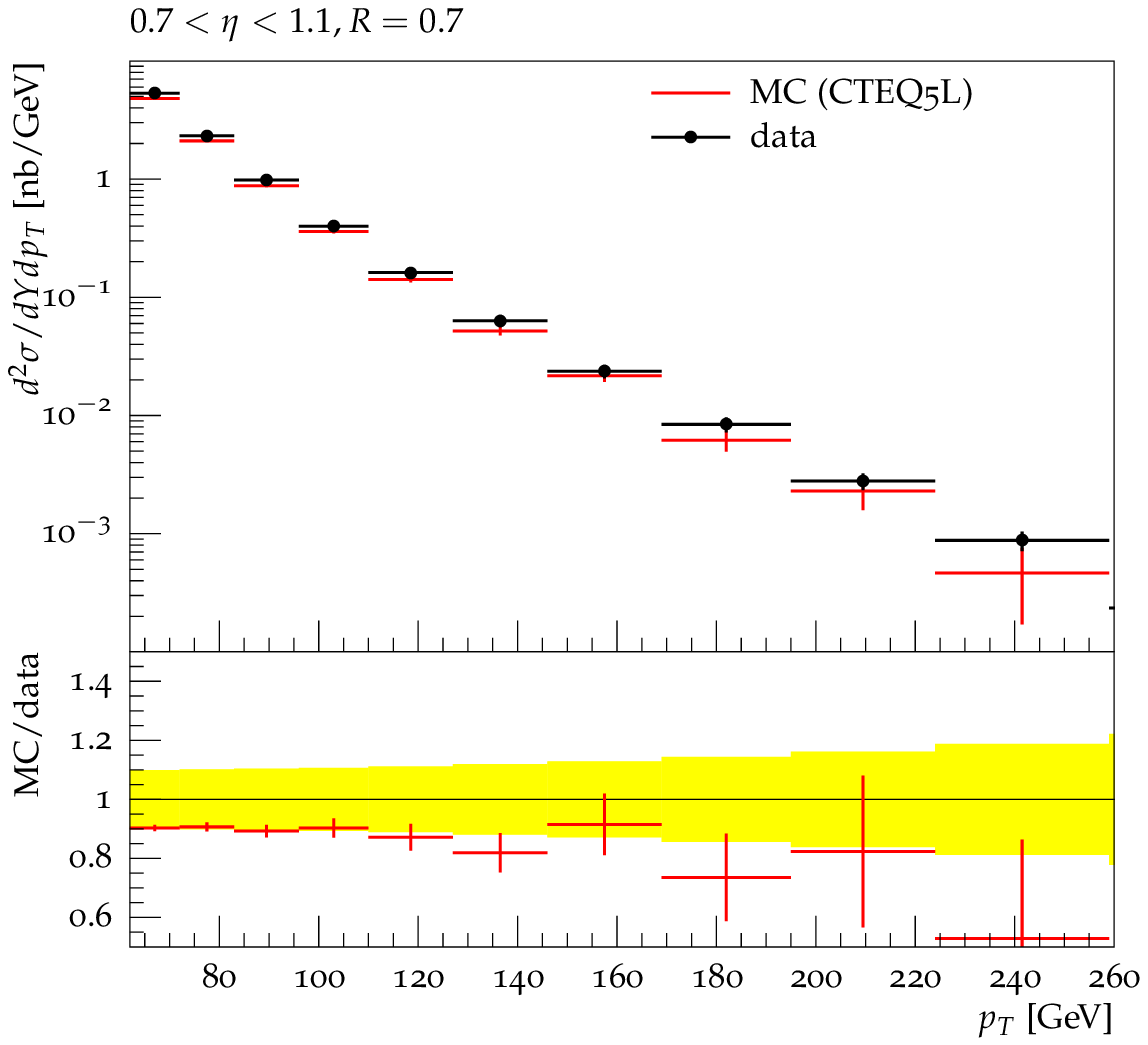}} \\
  \subfloat[]{\includegraphics[width=0.5\textwidth]
    {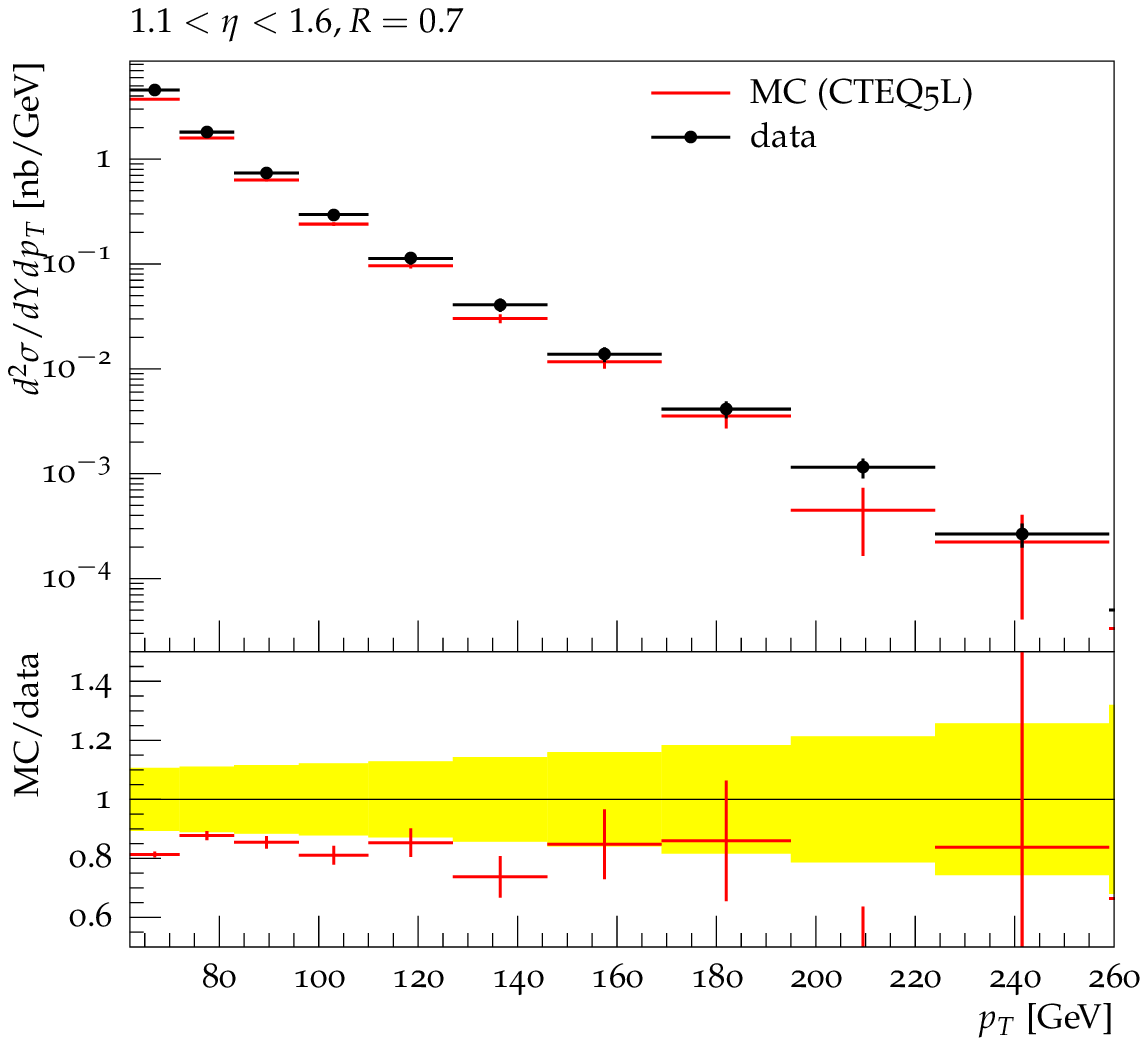}} 
  \subfloat[]{\includegraphics[width=0.5\textwidth]
    {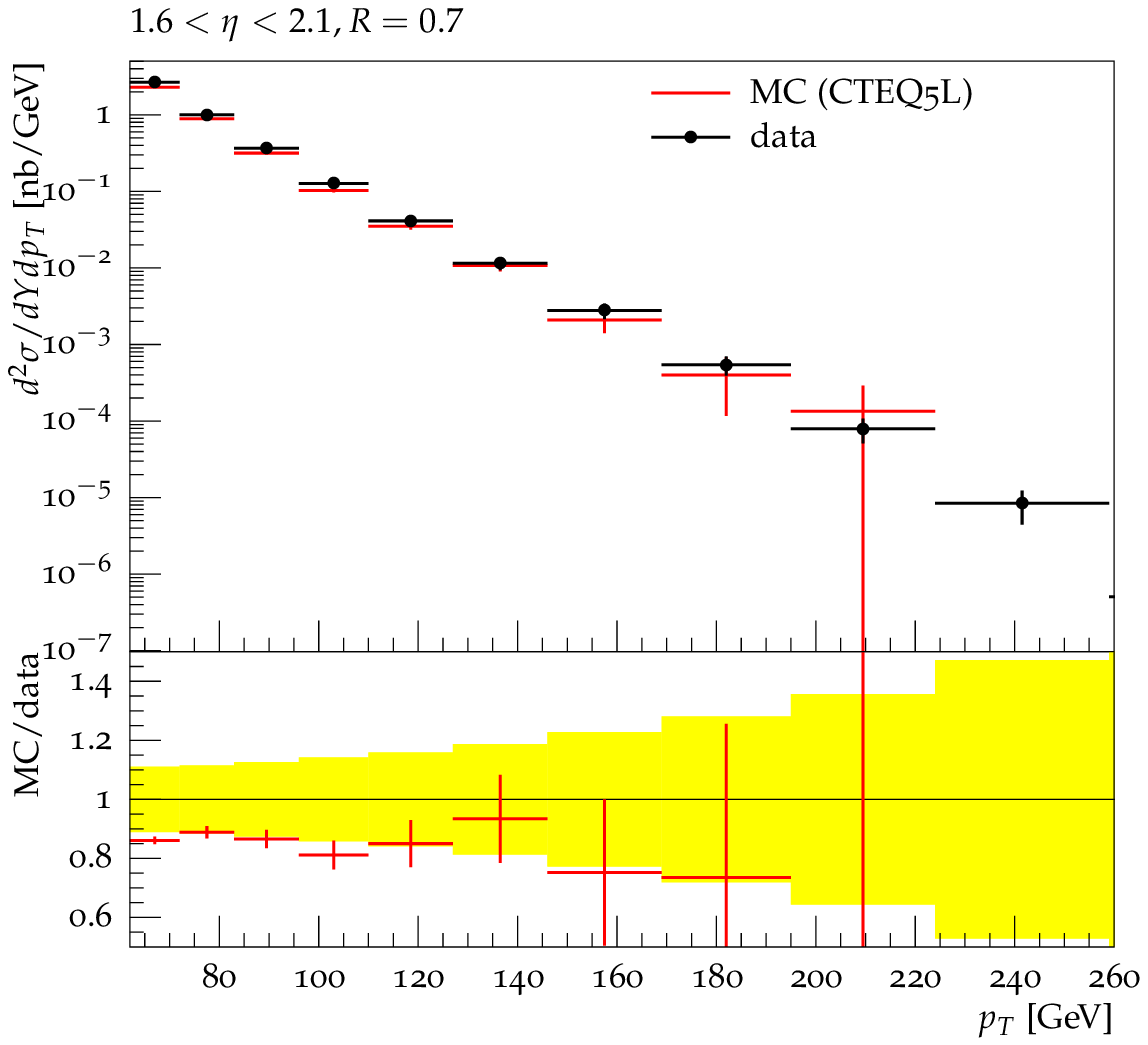}} 
  \caption{Inclusive jet cross section from CDF data compared to MC simulation
    with the CTEQ5L PDF.}
  \label{fig:jetCrossData}
\end{figure}

\begin{figure}[tp]
  \centering
  \subfloat[]{\label{fig:jetCross2b}\includegraphics[width=0.5\textwidth]
    {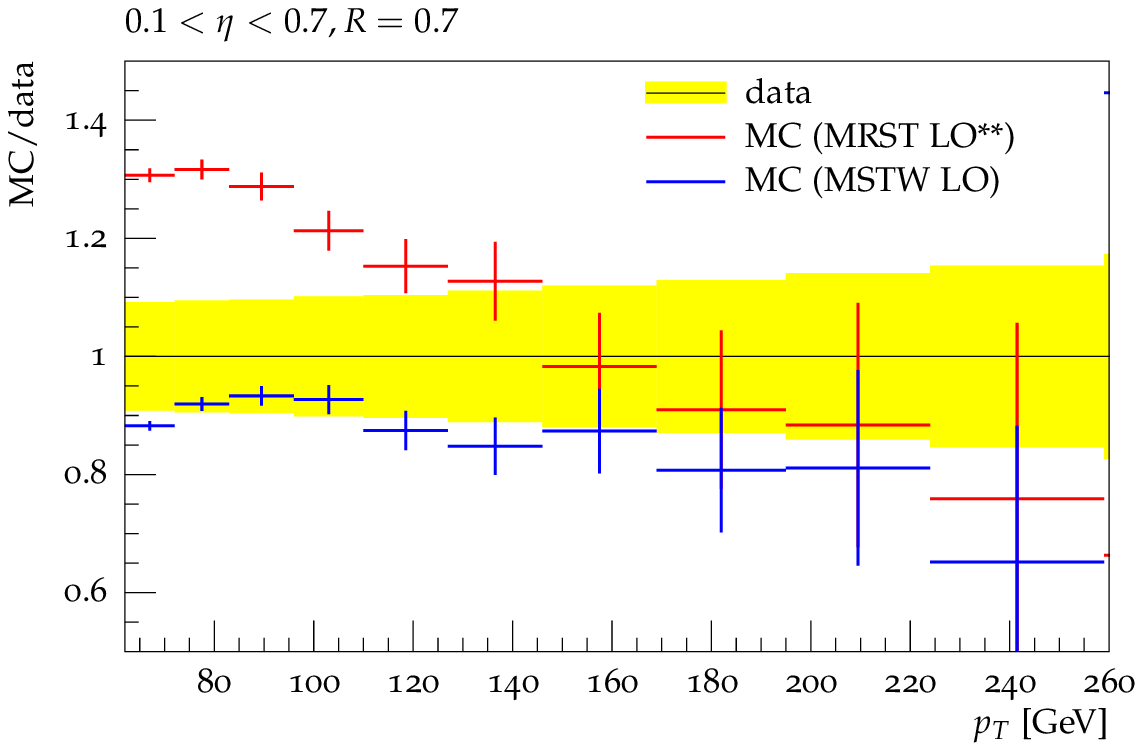}} 
  \subfloat[]{\label{fig:jetCross2c}\includegraphics[width=0.5\textwidth]
    {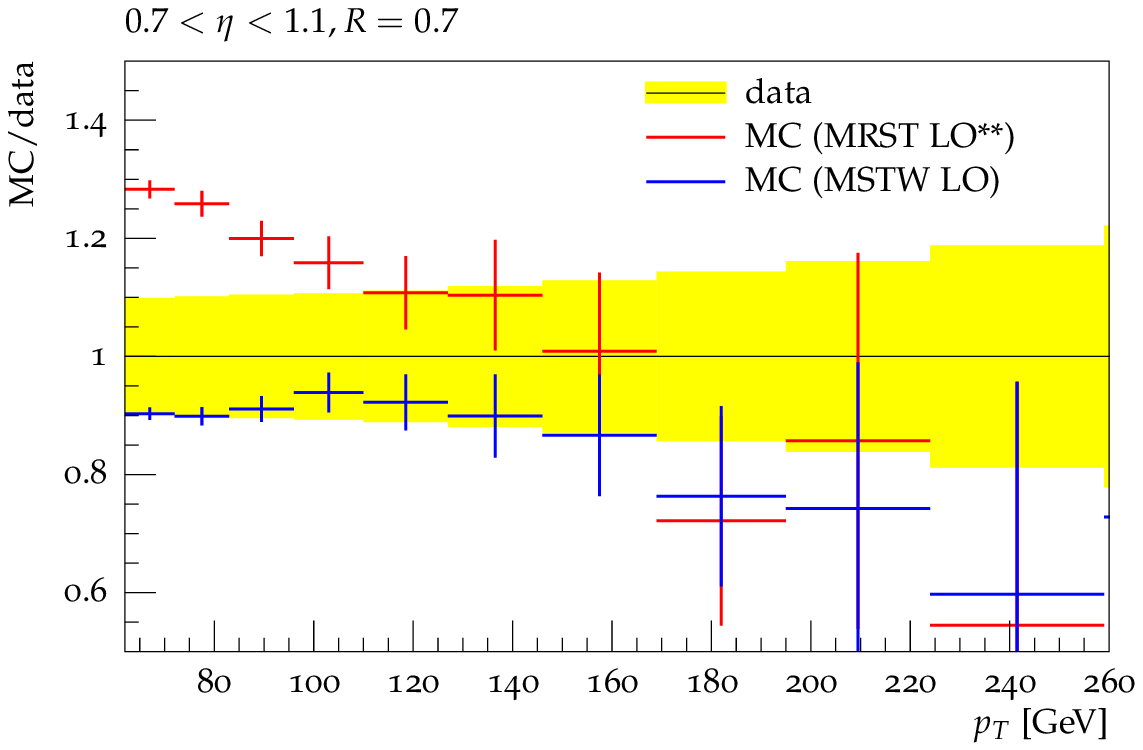}} \\
  \subfloat[]{\label{fig:jetCross2d}\includegraphics[width=0.5\textwidth]
    {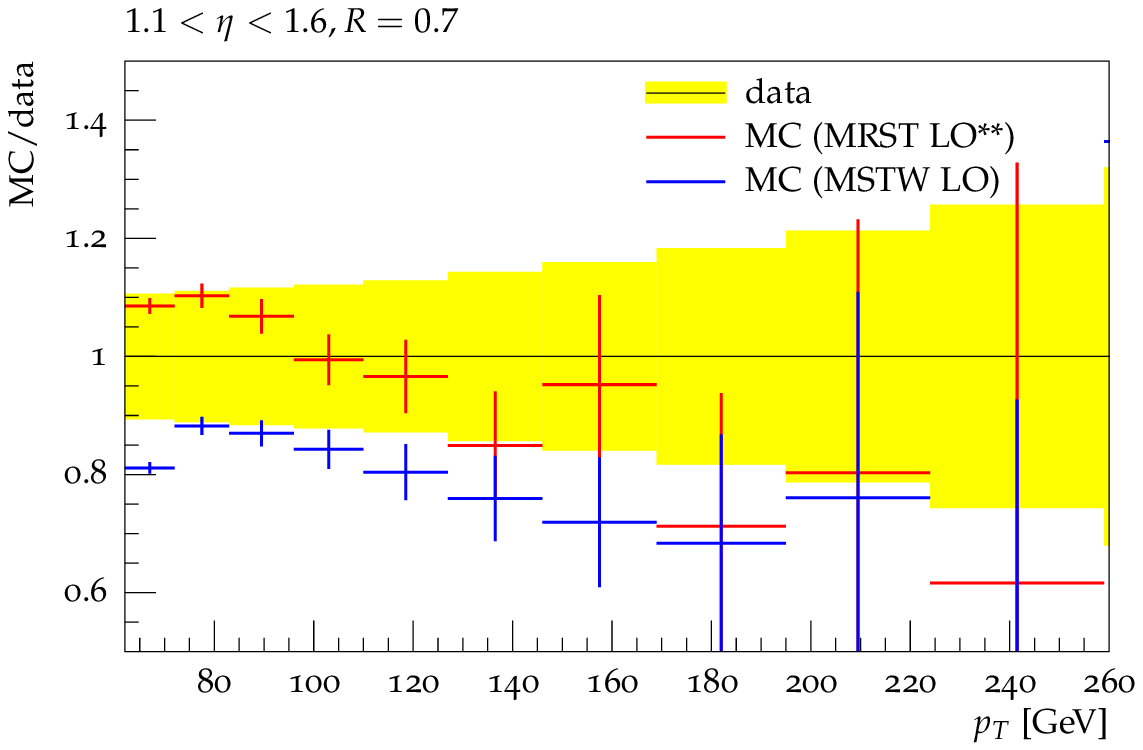}} 
  \subfloat[]{\label{fig:jetCross2e}\includegraphics[width=0.5\textwidth]
    {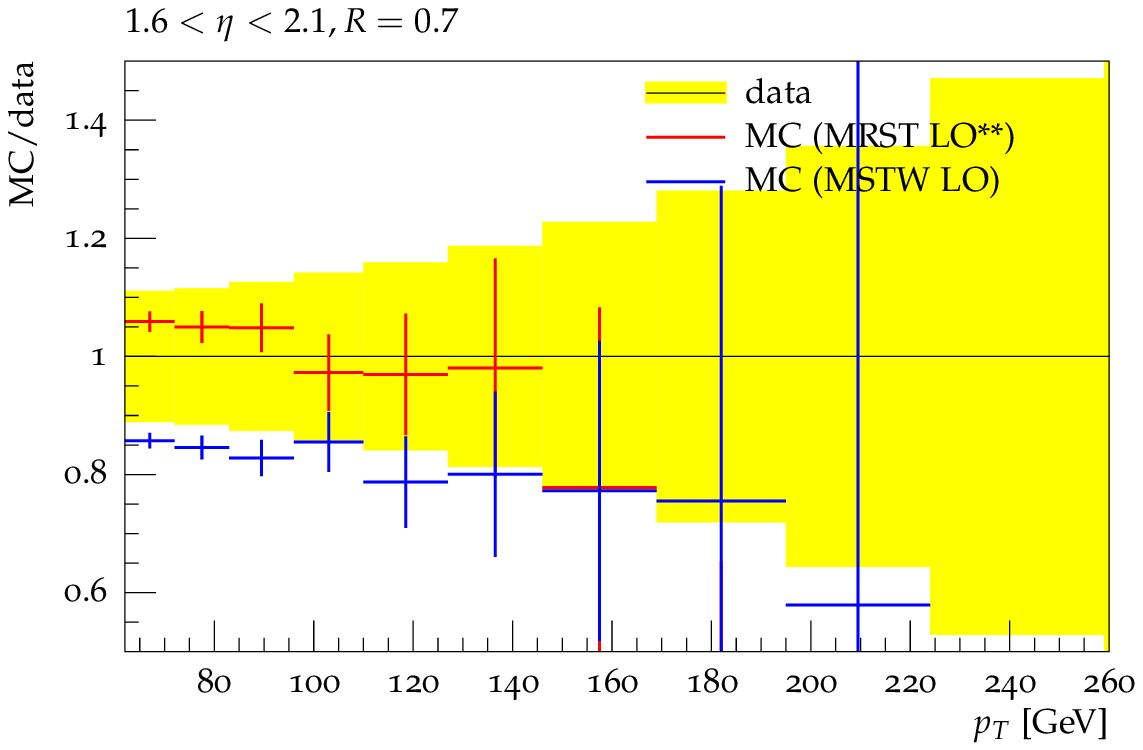}} 
  \caption{Ratio of the inclusive jet cross section, MC/Data.}
  \label{fig:jetCross2}
\end{figure}

\begin{figure}[tp]
  \centering
  \subfloat[]{\label{fig:jetCross4b}\includegraphics[width=0.5\textwidth]
    {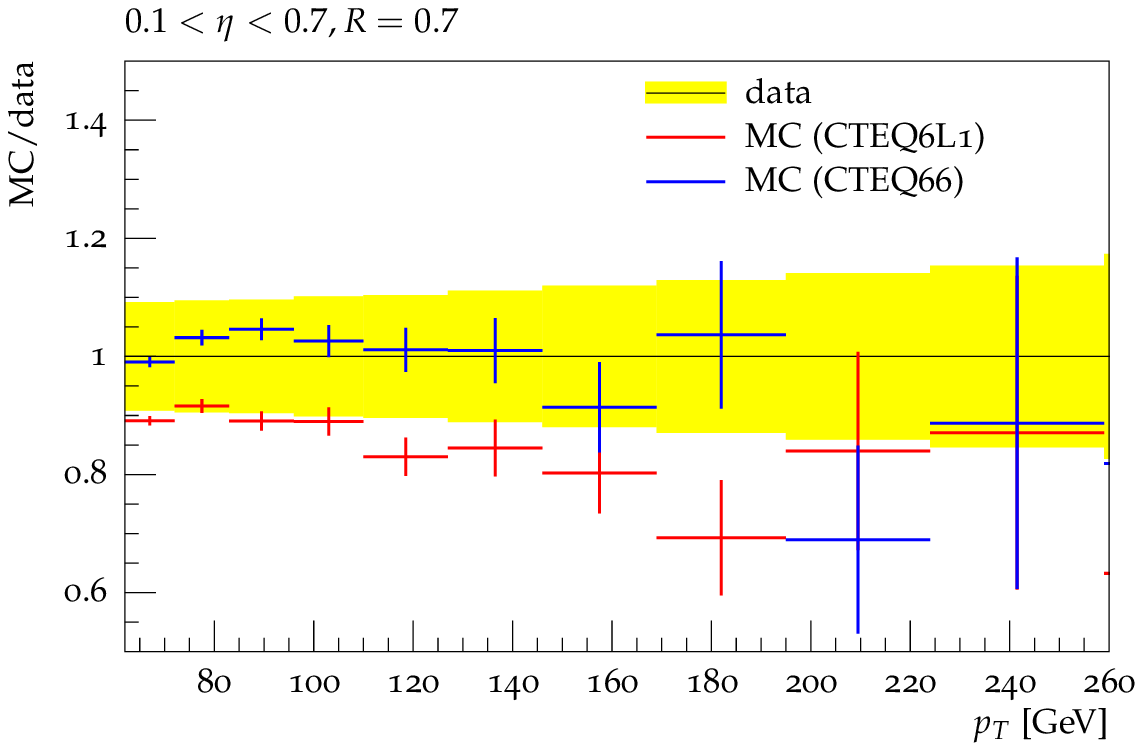}} 
  \subfloat[]{\label{fig:jetCross4c}\includegraphics[width=0.5\textwidth]
    {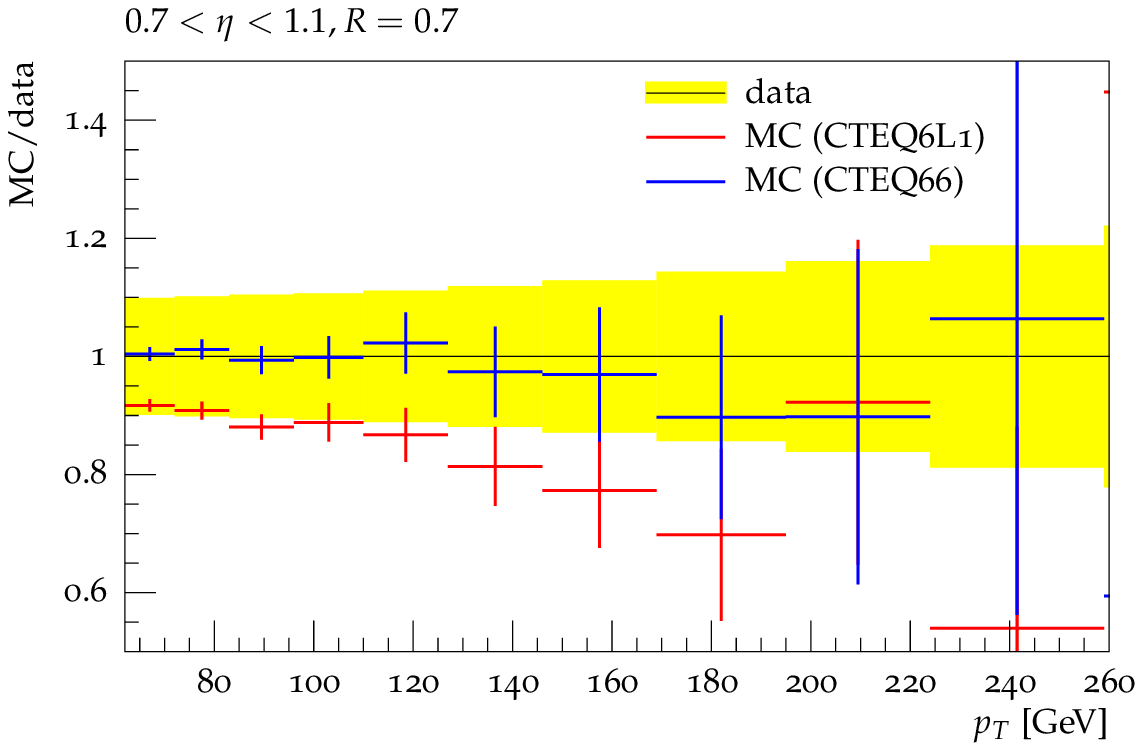}} \\
  \subfloat[]{\label{fig:jetCross4d}\includegraphics[width=0.5\textwidth]
    {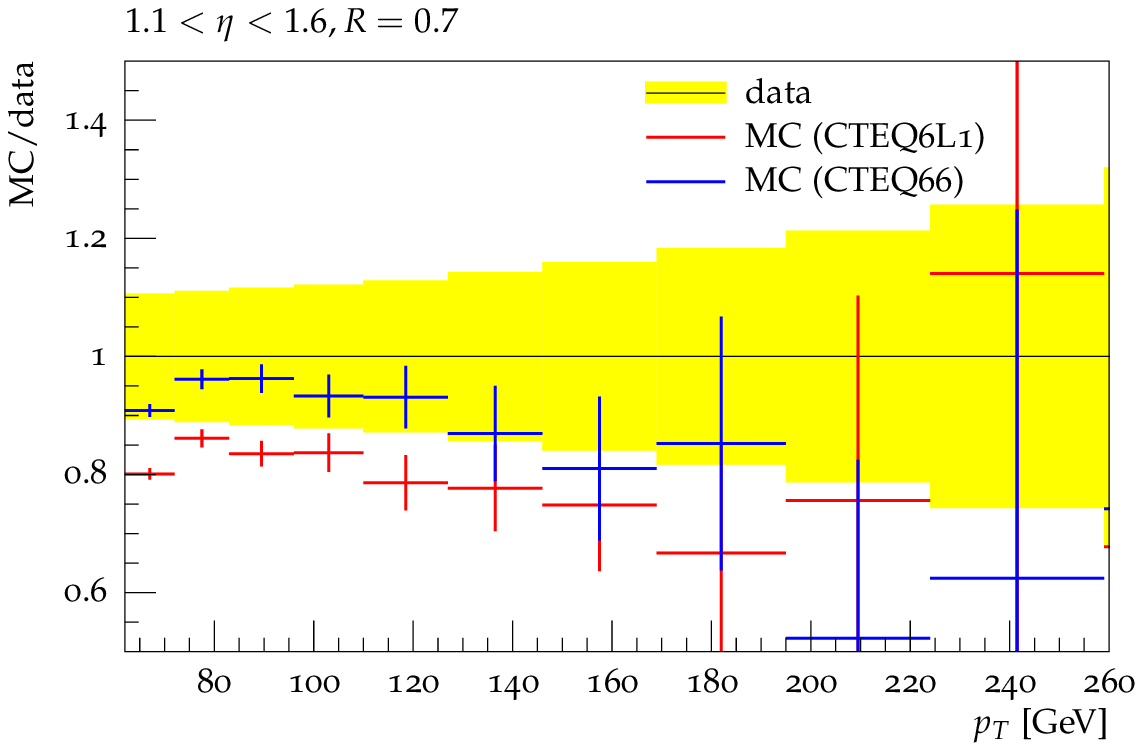}} 
  \subfloat[]{\label{fig:jetCross4e}\includegraphics[width=0.5\textwidth]
    {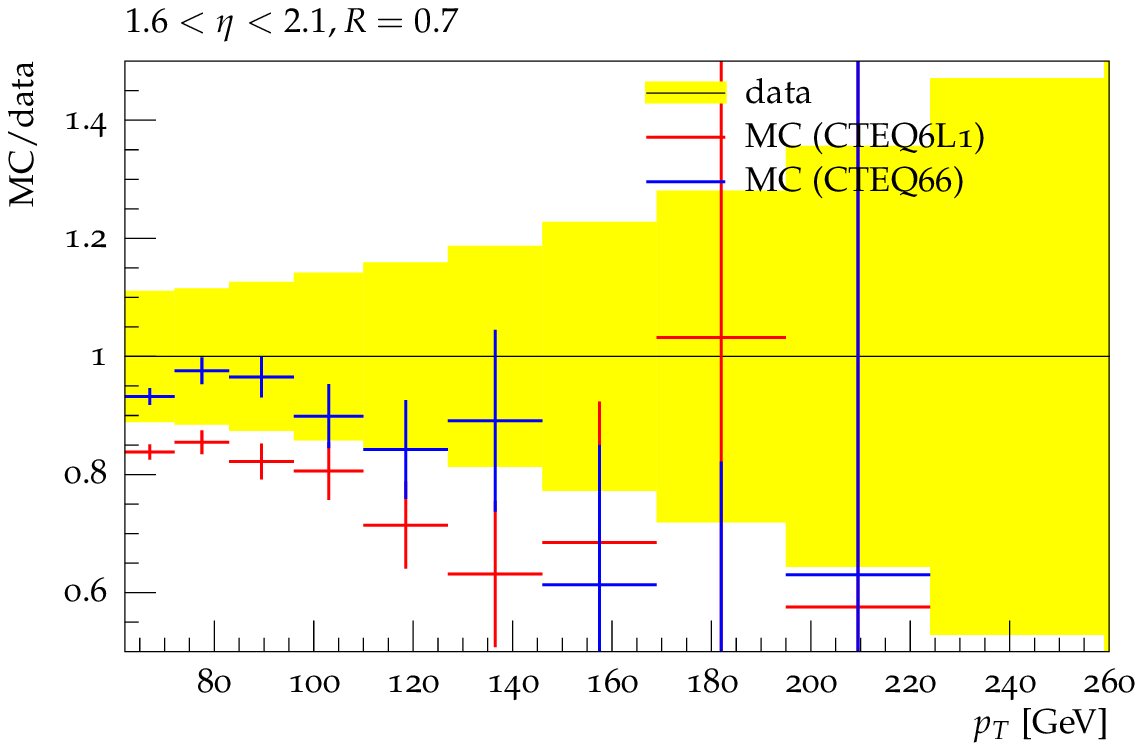}} 
  \caption{Ratio of the inclusive jet cross section, MC/Data.}
  \label{fig:jetCross4}
\end{figure}

\begin{figure}[tp]
  \centering
  \subfloat[]{\label{fig:jetCross5b}\includegraphics[width=0.5\textwidth]
    {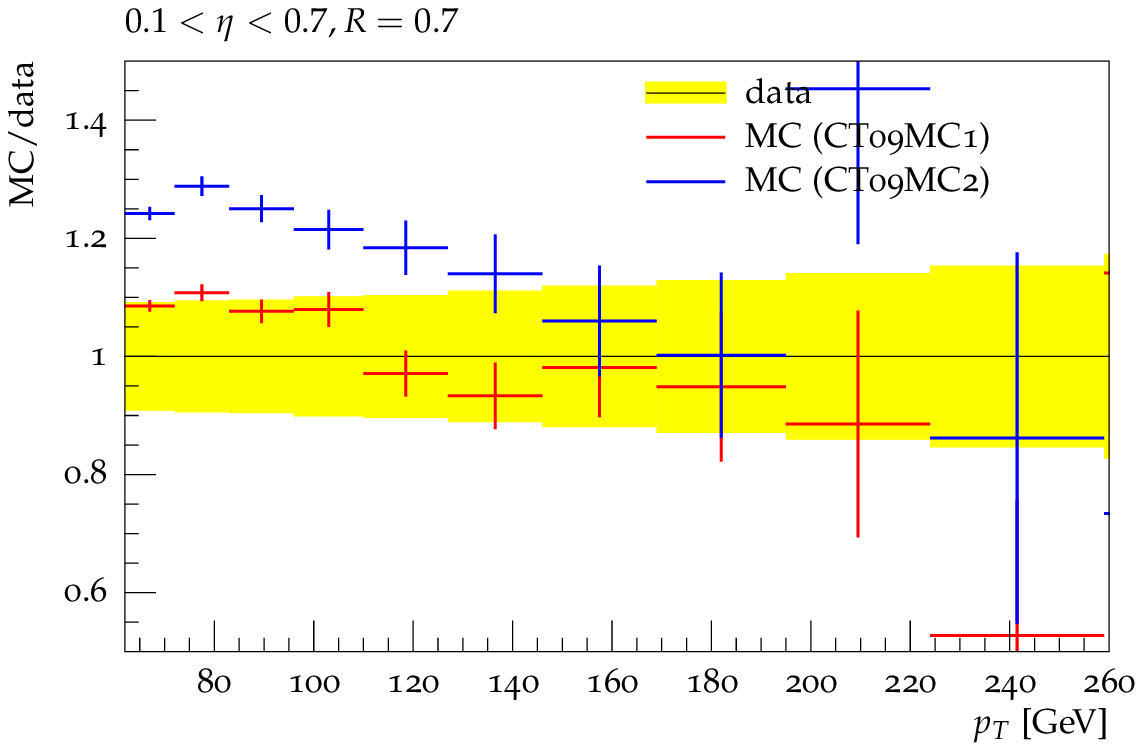}} 
  \subfloat[]{\label{fig:jetCross5c}\includegraphics[width=0.5\textwidth]
    {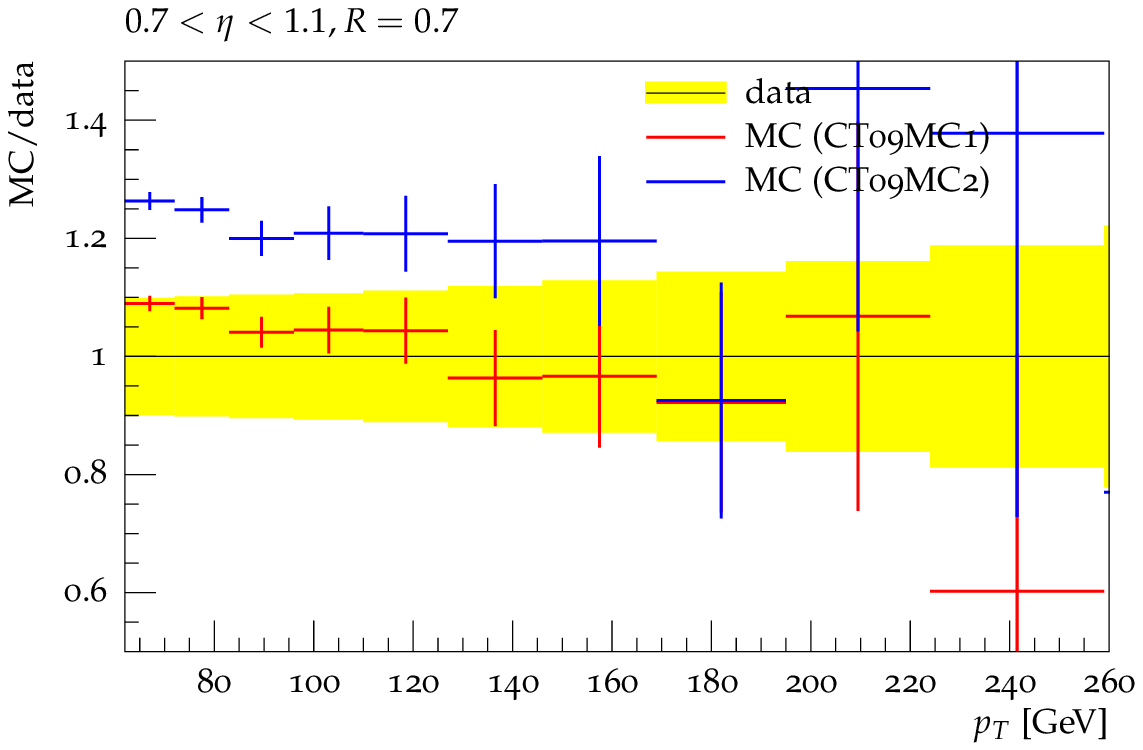}} \\
  \subfloat[]{\label{fig:jetCross5d}\includegraphics[width=0.5\textwidth]
    {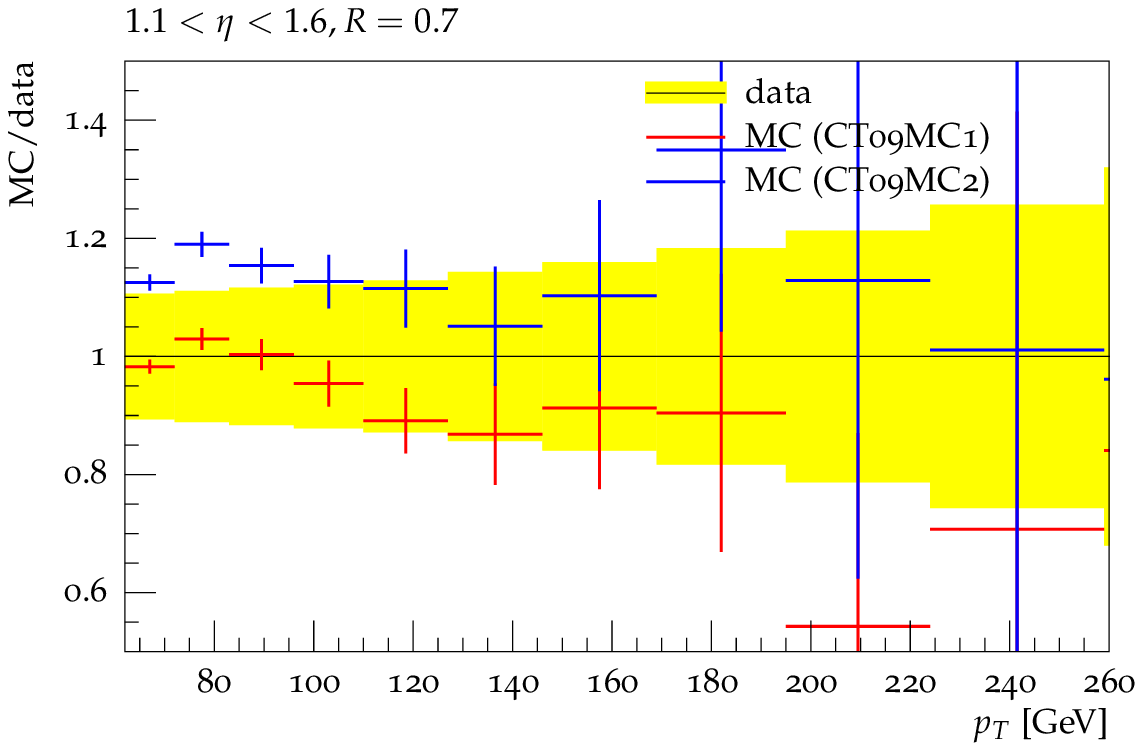}} 
  \subfloat[]{\label{fig:jetCross5e}\includegraphics[width=0.5\textwidth]
    {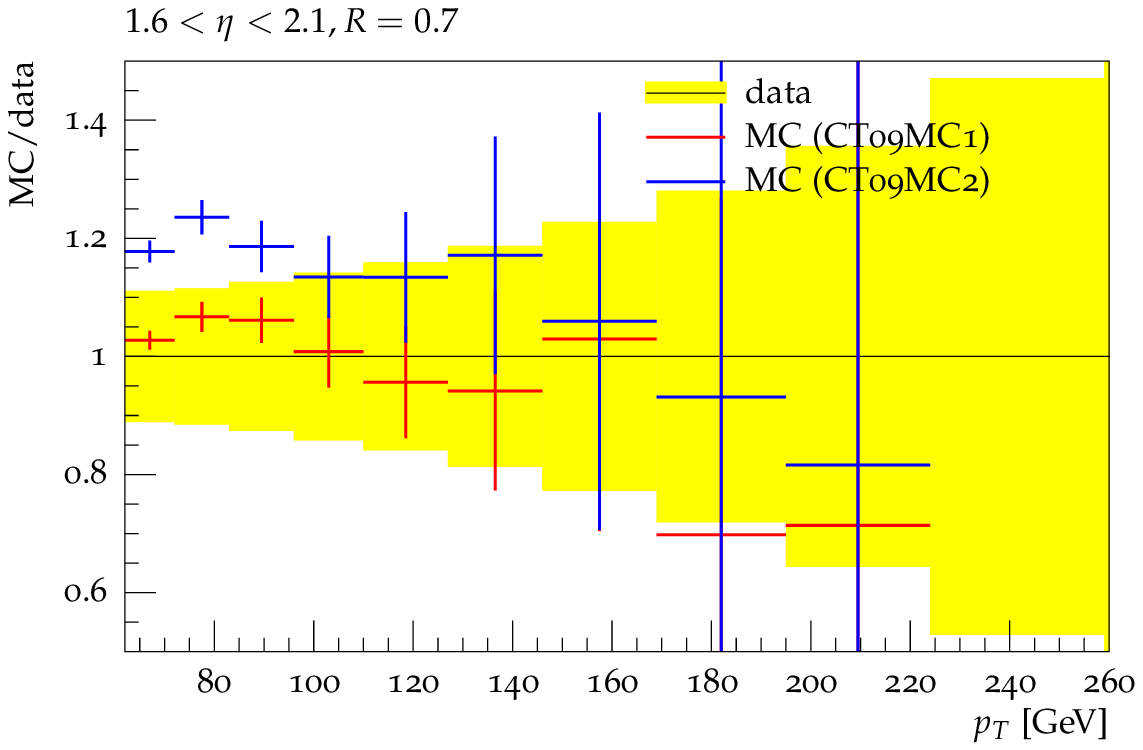}} 
  \caption{Ratio of the inclusive jet cross section, MC/Data.}
  \label{fig:jetCross5}
\end{figure}

  The inclusive jet cross section in Fig.~\ref{fig:jetCrossData}
drops rapidly with increasing $p_{\perp}$ and spans over several orders of
magnitude. The CTEQ5L distribution, shown in Fig.~\ref{fig:jetCrossData},  yields
results which are lower than data, with a ratio between $0.8$ and
$0.9$. The ratio remains fairly constant with $p_{\perp}$ but it is closer
to unity at medium pseudorapidities. Since the differences between experiments and
simulations are hard to see in the main window of Fig.\ref{fig:jetCrossData}
we only show the
$MC/data$ ratio  in the following figures. The results with MRST LO**, MSTW LO,
CTEQ6L1, CTEQ66, CT09MC1 and CT09MC2 are shown in
Fig.~\ref{fig:jetCross2}-\ref{fig:jetCross5}. MRST LO** starts with a much too large cross section and the ratio decreases
when $p_{\perp}$ rises. This behavior is the strongest at low pseudorapidity and as
we move to larger $\eta$ the ratio gets smaller and flatter. All MC-adapted PDFs,
except MCS, show this type of behavior. MC2 and MC1 give results with very
similar shapes but the
MC2 cross section is larger. MRST LO* is related to LO** much in the same fashion as
MC1 to MC2. MSTW LO and CTEQ6L give
cross sections which have similar behavior as with CTEQ5L, i.e. the ratio is less dependent on
$p_{\perp}$ than with the
MC-adapted PDFs. MSTW LO results are less depending on the
pseudorapidity than CTEQ5L, Fig.~\ref{fig:jetCross2}. CTEQ6L1 gives a ratio
which starts to decrease with $p_{\perp}$ at larger rapidities. CT09MCS gives a too low cross section, is once again different from the other MC-adapted PDFs and gives results
which behave in a way more similar to those of the normal leading-order
distributions. Actually, the results with the two NLO PDFs are the closest to
data at central pseudorapidities but the ratios decrease  towards $0.5$ at larger
$\eta$. 

\begin{figure}[tp]
  \centering
  \subfloat[]{\label{fig:Rjet1}\includegraphics[width=0.5\textwidth]
    {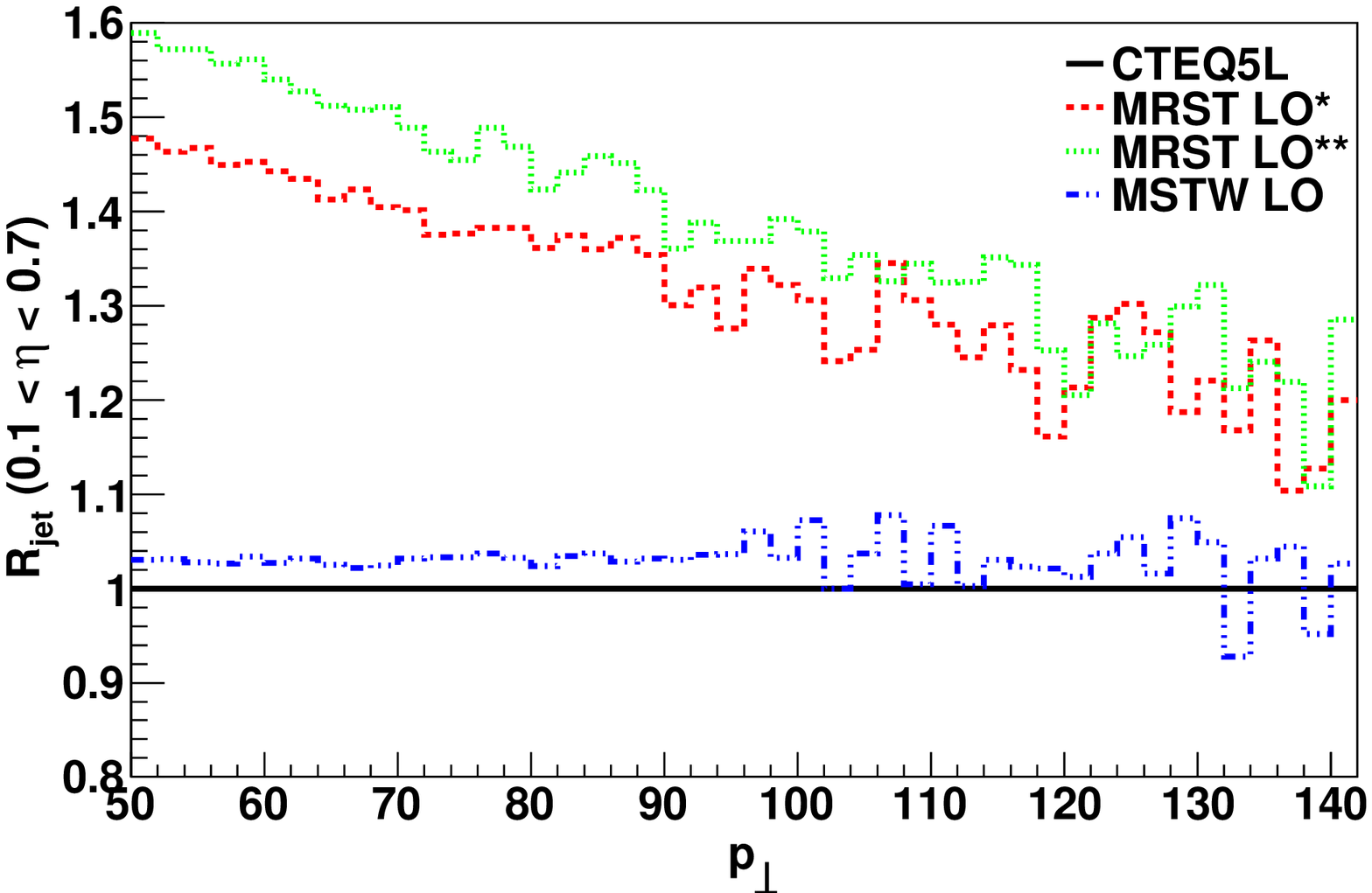}} 
  \subfloat[]{\label{fig:Rjet2}\includegraphics[width=0.5\textwidth]
    {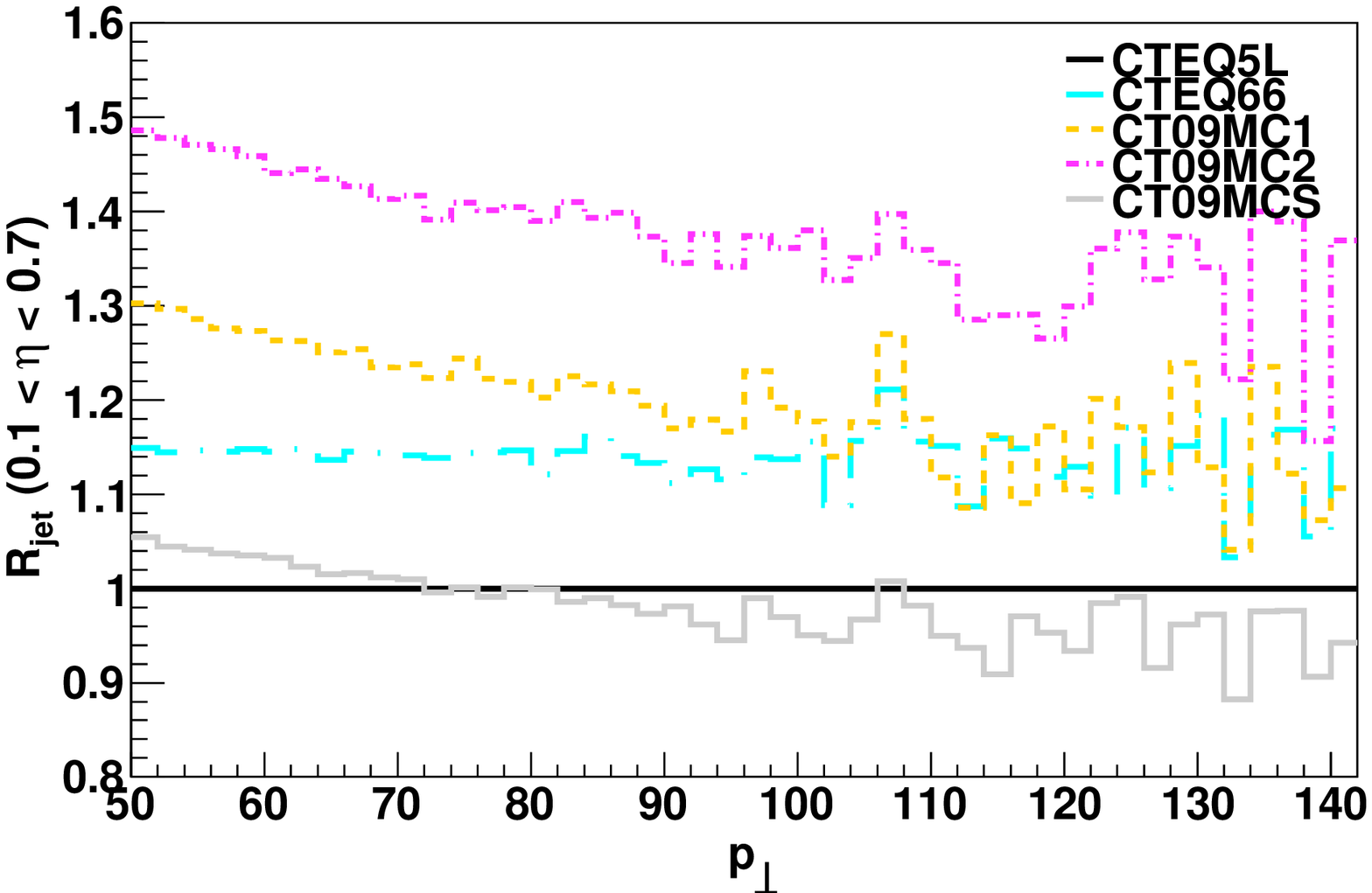}}  
  \caption{Ratio of cross section for the $2 \rightarrow 2$ sub-process with
    the different PDFs over cross section with CTEQ5L.}
  \label{fig:Rjet}
\end{figure}

\begin{figure}[tp]
  \centering
  \subfloat[]{\label{fig:expdfCross-quark}\includegraphics[width=0.5\textwidth]{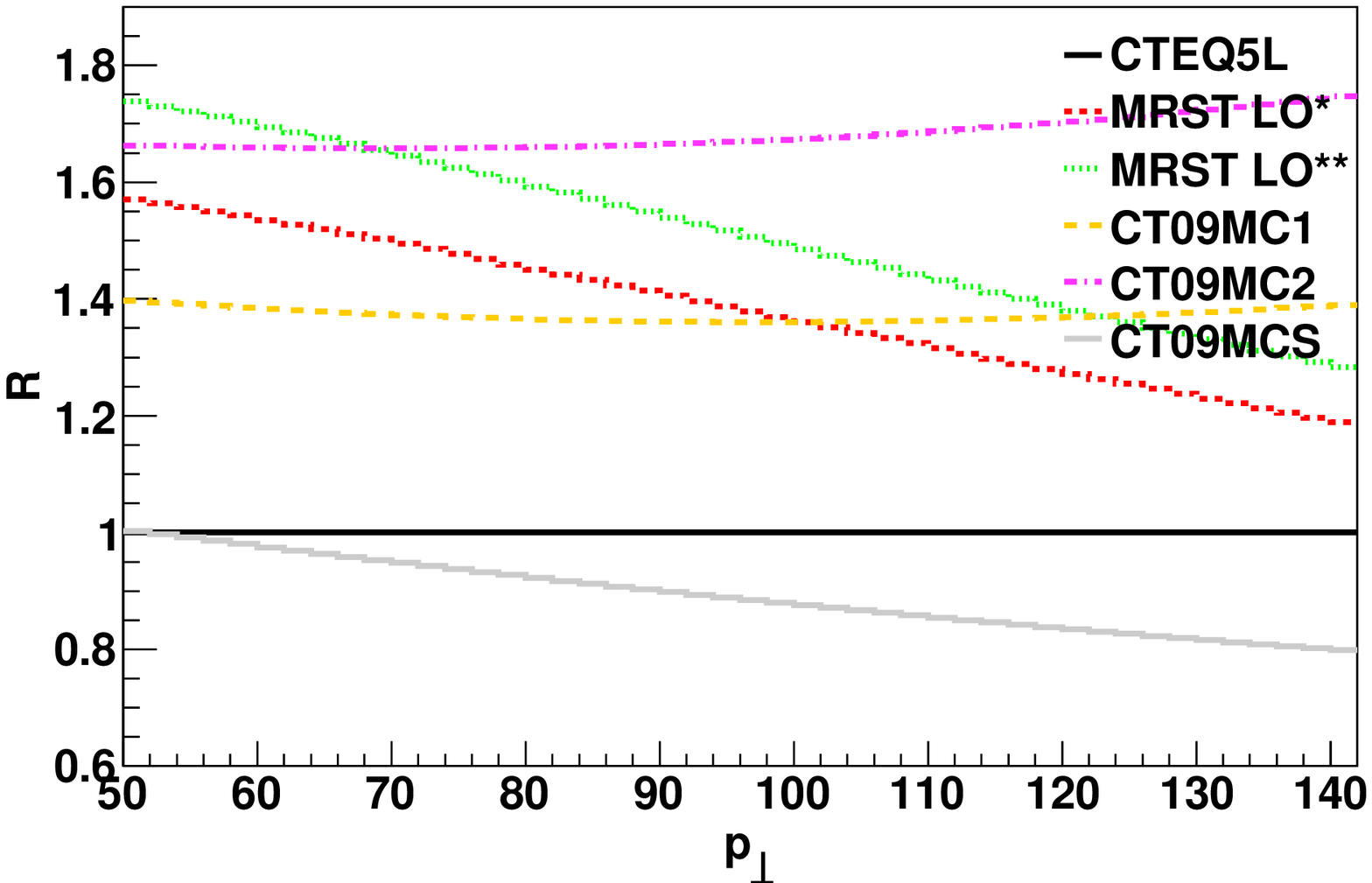}} 
  \subfloat[]{\label{fig:expdfCross-all}\includegraphics[width=0.5\textwidth]
    {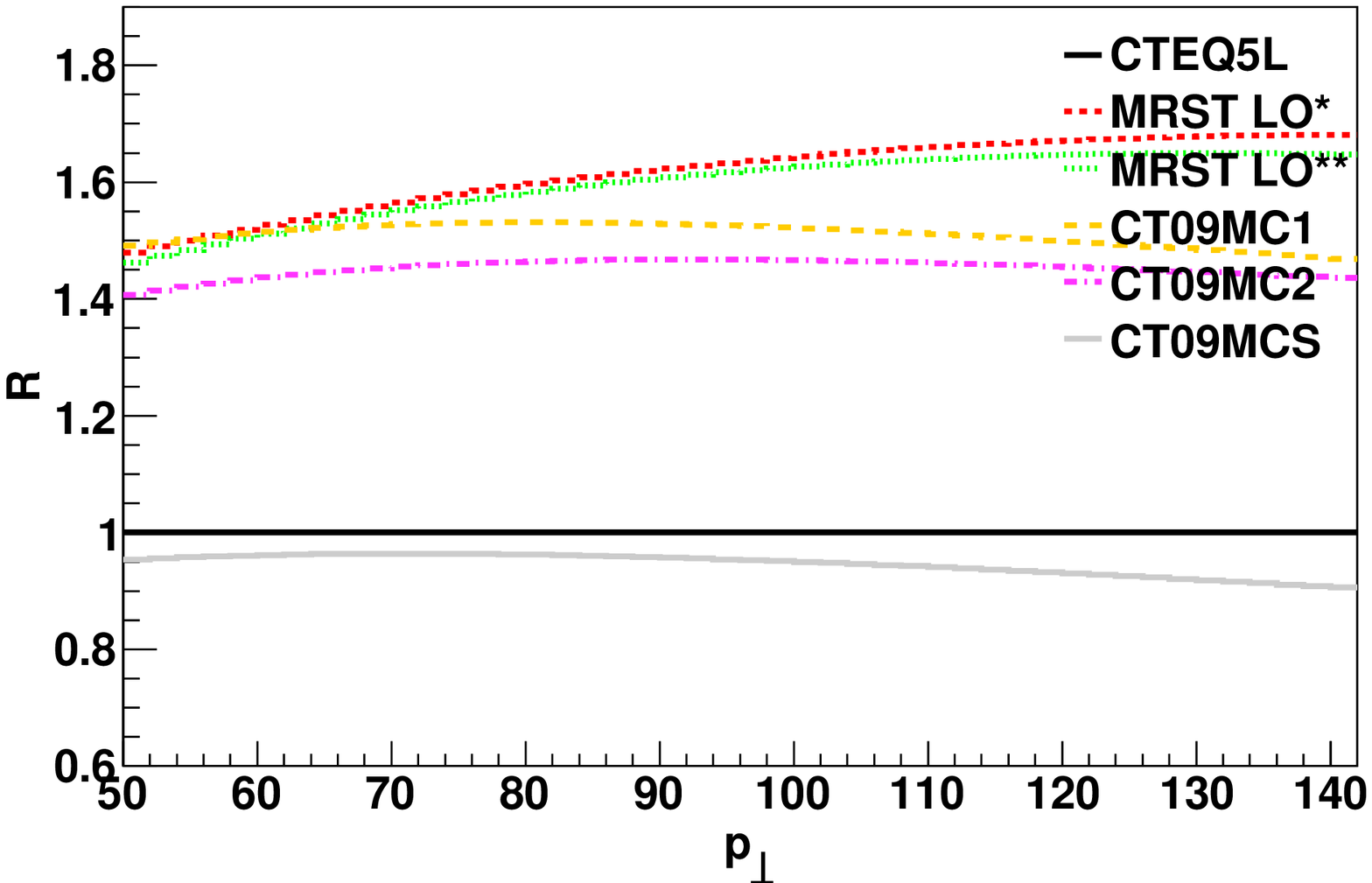}} 
  \caption{Ratio of the approximated cross section with different PDFs to
    CTEQ5L. Left figure show only the $gg$ interactions while the right figure
    also include $qg$ and $qq$ interactions.}
  \label{fig:expdfCross}
\end{figure}

The main feature is the surprising decrease in the cross section ratio from
low to high $p_{\perp}$ with
the MC-adapted PDFs, which we do not see with neither LO nor NLO PDFs. In order to
examine the origin of this difference we look
at the cross section at the parton level, after the hard collision only. Since
we cannot compare to data we choose to compare to CTEQ5L, which gives
a more constant  $p_{\perp}$ evolution of the cross section. The
ratios, $R_{jet}=\sigma_{jet,PDF}/\sigma_{jet,CTEQ5L}$, in Fig.~\ref{fig:Rjet}
show that the relative decrease with LO* and LO** is clearly visible also at parton level. This is also the case for
the three MC1/2/S PDFs in Fig~\ref{fig:Rjet}, but to less extent, while the MSTW LO
results have an almost completely flat ratio. To investigate this further we integrate an approximation of the cross section  
\begin{equation}
\frac{d\sigma}{dp_{\perp}} \sim \frac{d\hat{\sigma}}{dp_{\perp}} \sum_{ij}
\int dx_1dx_2f_i(x_1,Q^2)f_j(x_2,Q^2)
\end{equation}
for only the $gg\rightarrow gg$ interactions
and a clear pattern very similar to the one at parton level become visible, see
Fig.~\ref{fig:expdfCross-quark}. Fig.~\ref{fig:expdfCross-all} shows that this
is no longer the case if
we include the other possible interactions, i.e. $qg\rightarrow qg$ and $qq\rightarrow qq$ but at the low $p_{\perp}$ end the gluon interactions dominate.

\begin{figure}[tp]
  \centering
  \subfloat[]{\label{fig:jetCRap-1}\includegraphics[width=0.5\textwidth]
    {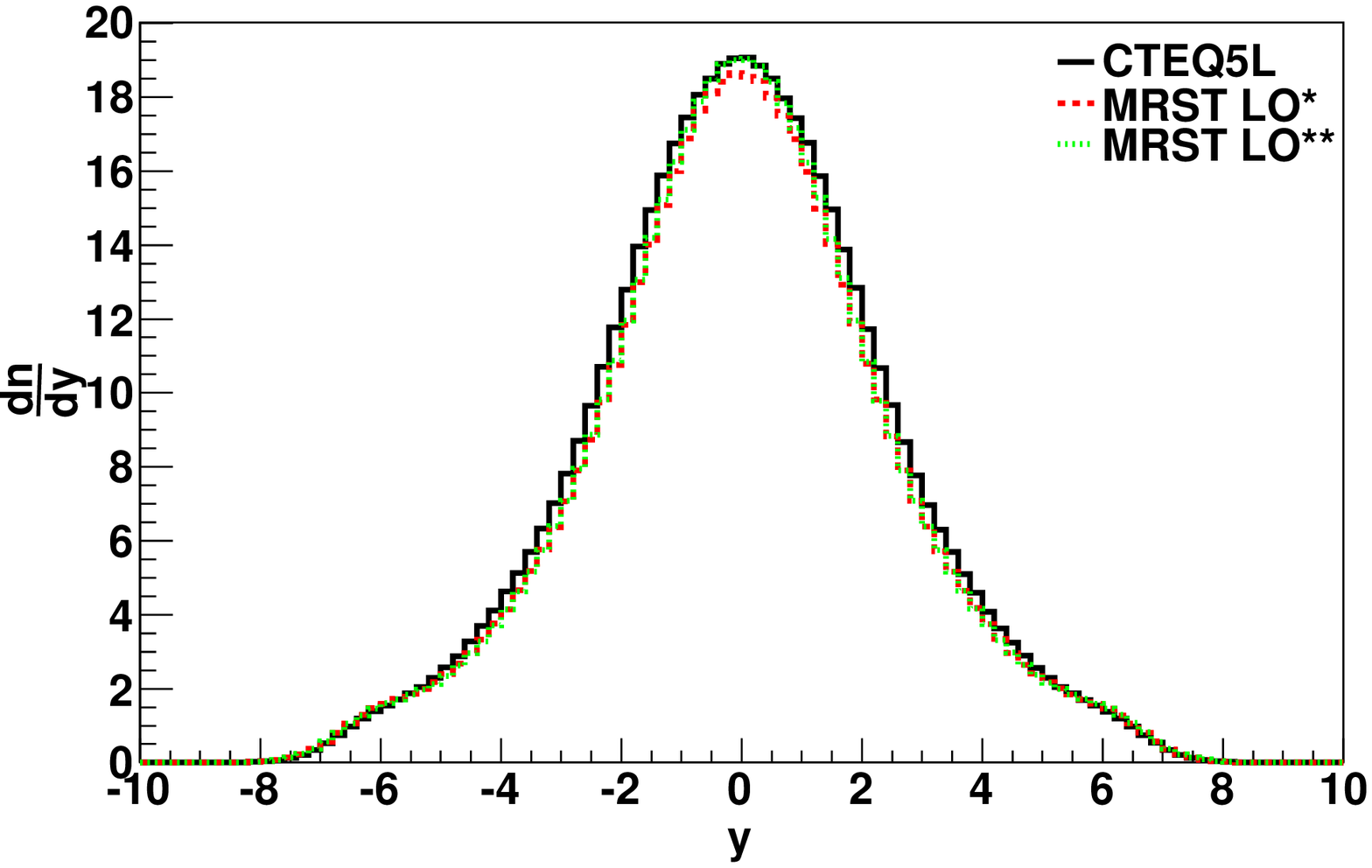}} 
  \subfloat[]{\label{fig:jetCRap-2}\includegraphics[width=0.5\textwidth]
    {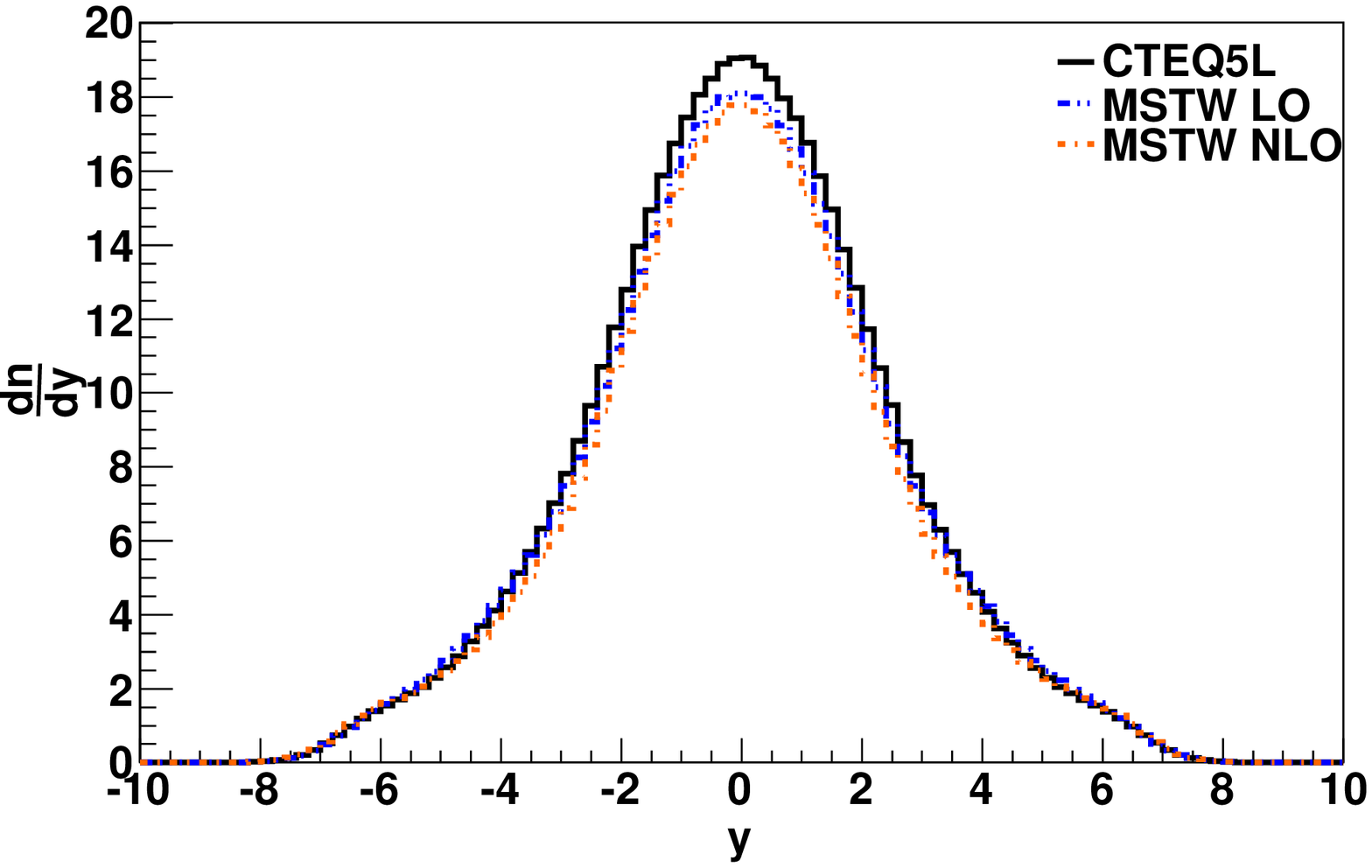}}
  \caption{Hard QCD rapidity distributions.}
  \label{fig:jetCRap}
\end{figure}

\begin{figure}[tp]
  \centering
  \subfloat[]{\label{fig:jetCMult-1}\includegraphics[width=0.5\textwidth]
    {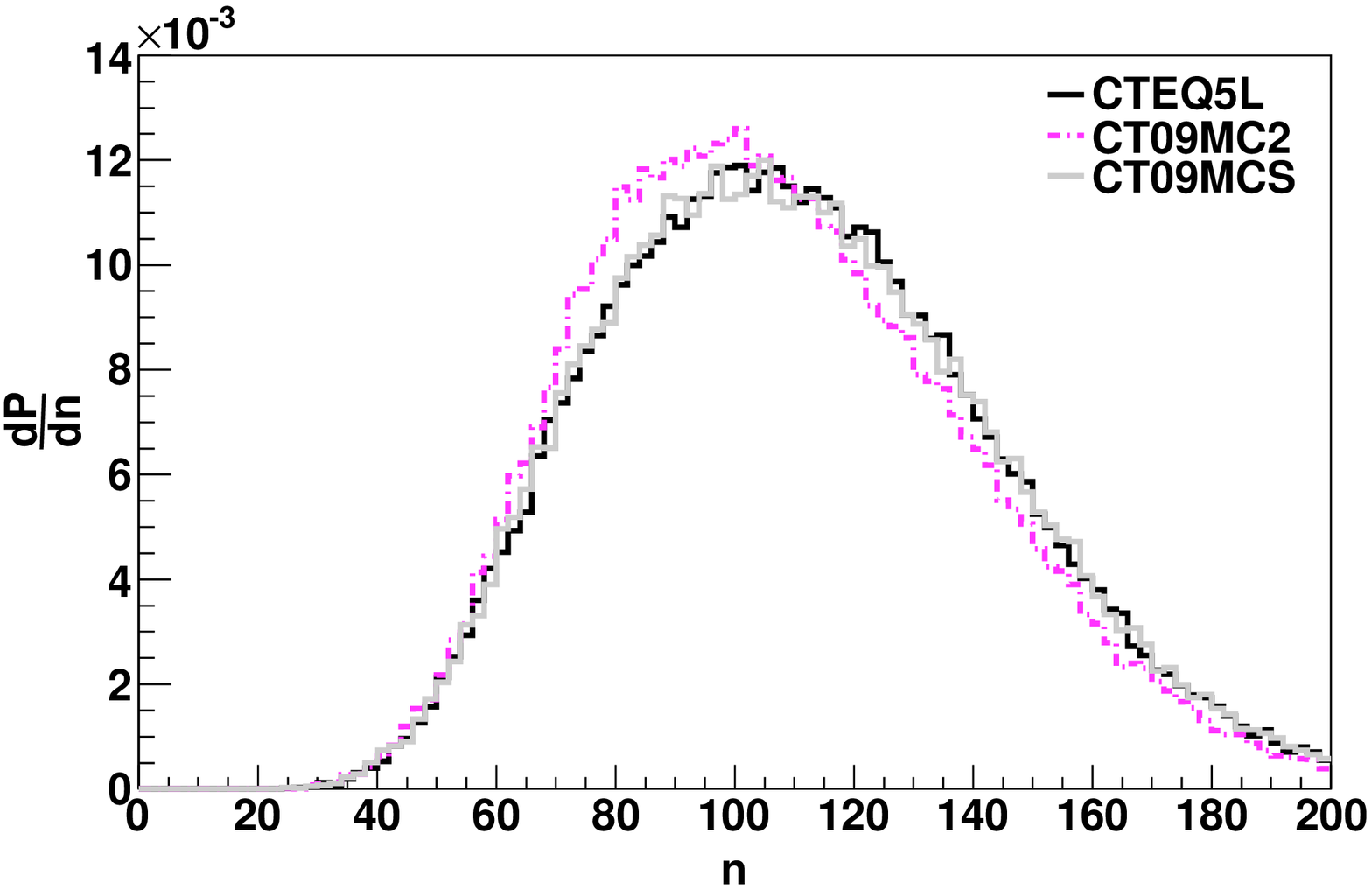}} 
  \subfloat[]{\label{fig:jetCMult-2}\includegraphics[width=0.5\textwidth]
    {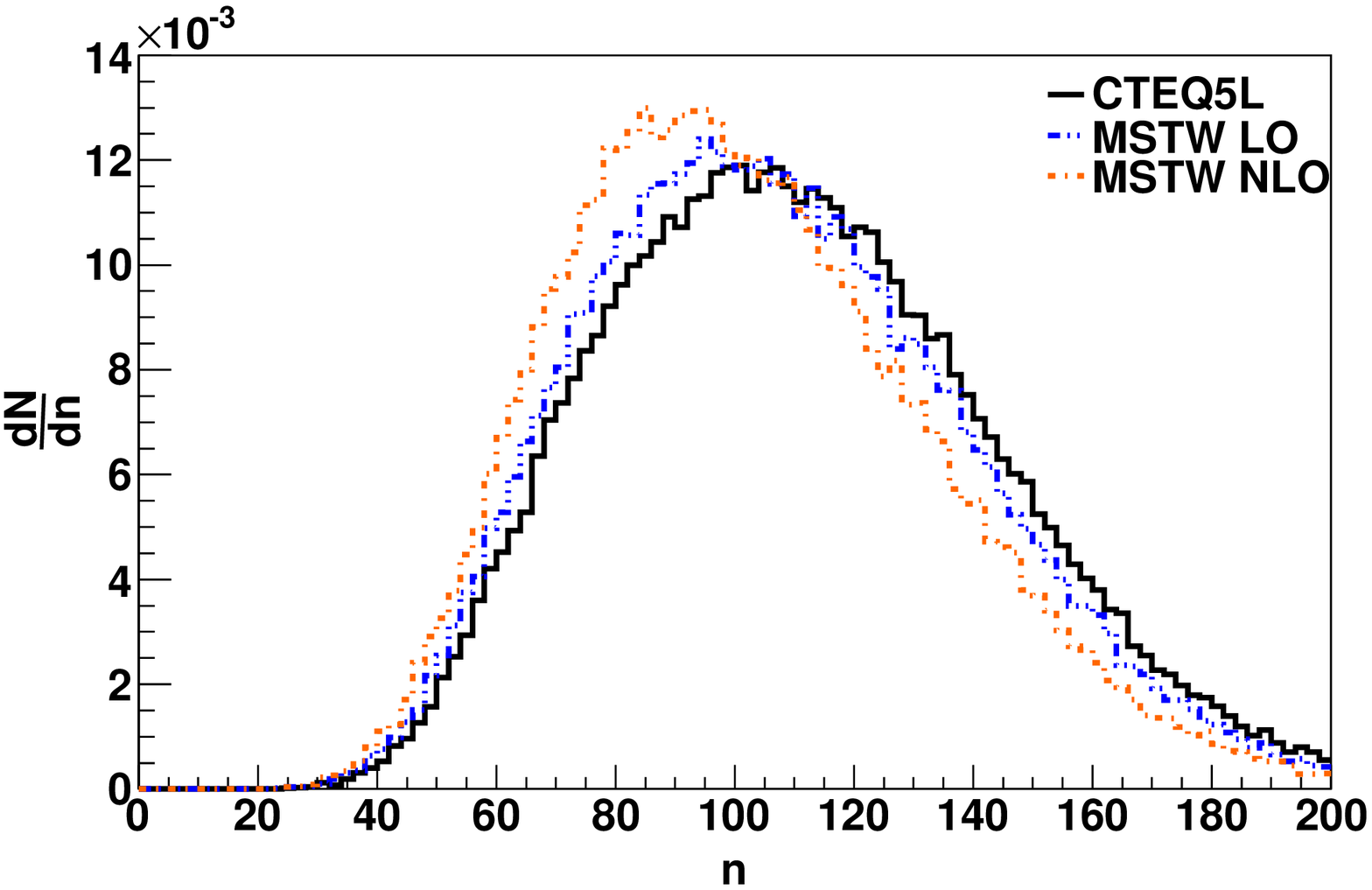}} 
  \caption{Hard QCD multiplicity distributions.}
  \label{fig:jetCMult}
\end{figure}

Turning our attention to the rapidity distribution the results show less
variety, see Fig.\ref{fig:jetCRap}. However, excluding MCS, all the MC-adapted PDFs give a narrower
distribution. One should not be fooled by the steep slope which hides the
differences. MSTW LO and NLO are smaller at central rapidities and actually
have a lower multiplicity while the rest of the CTEQ distributions follow
CTEQ5L. Even though the PDFs have equal charged particle multiplicity for
minbias events this no longer holds true for the hard QCD events. All MC-adapted PDFs except MCS have multiplicity distributions shifted towards lower
multiplicity, which is also the case for MSTW NLO and to some extent for MSTW LO, see
Fig~\ref{fig:jetCMult}.

\section{Summary and Conclusions}
Including the very latest parton distribution functions into \textsc{Pythia8} both caused some troubles and gave some surprises,
especially while venturing outside the grid of the PDFs. At small $x$, where the
gluon distribution dominates, there are large differences between the PDFs,
especially at low $Q^2$. The MRST/MSTW distributions are much larger than the
ones from CTEQ, and MSTW LO goes sky high compared to the rest of the PDFs. At larger $Q^2$
the differences are smaller between the two collaborations, except for the
distributions which freeze their values at the end of the grid, and for MSTW LO
which is
still much larger than the rest. The quarks, and in particular
the up distributions, at large $x$ show smaller differences, but MSTW LO is once
again larger at small $x$. Some of the new MC-adapted PDFs carry a
fraction of the protons momentum larger than unity, in an attempt to compensate
for known shortcomings of leading-order calculations. Therefore these give a
larger activity in the collisions and in order to be able to compare results from the different PDFs
in simulations, we first tuned them to obtain equal charged particle
multiplicity.

The different behavior for MSTW LO was also reflected in rapidity
distributions both at the level of the $2 \rightarrow 2$ sub-process and after
hadronization. In general the differences in the PDFs are blurred down to
parton level and then blurred even more at hadron level, but some large differences
still remain. By comparison to minimum bias data from CDF we saw that the MC
adapted PDFs gave similar results and were in general closer to data than the LO and
NLO PDFs, with the exception of CTEQ6L. Differences between the results with
different PDFs were enhanced when the energy was increased up to the LHC level. We also
discovered that the multiplicity evolved differently with MSTW LO and with the two
next to leading-order PDFs than with the rest.

In the simulations of hard QCD events we saw that the rapidity distributions
with the MC-adapted PDFs were narrower, and that their multiplicity
distributions were shifted to lower multiplicity. The evolution of the
differential jet cross section with $p_{\perp}$ was found to be strongly dependent on
the choice of PDF, both when it came to its size and shape. In the simulations of hard events we saw that the rapidity distribution was
narrower for the MC-adapted PDFs and that their multiplicity
distributions shifted to lower multiplicity. The evolution of the cross
section with $p_{\perp}$ was strongly dependent on the PDF. Comparing the $p_{\perp}$
spectrum ratio (simulation over data) the MC-adapted PDFs showed a rather
surprising change in ratio with $p_{\perp}$ which we examined further. The
same behavior could be seen at the level of the $2 \rightarrow 2$ sub-process
as well and this
was traced to the $gg\rightarrow gg$ interactions, that dominate in this
region. The changes in behavior for the inclusive jet cross section with
different PDFs can also be caused by the change of dominating process in
the $p_{\perp}$, from $gg\rightarrow gg$ to $qg\rightarrow qg$. Simulations of
prompt photon production also showed such decrease in the cross section for
$gg$ but not for $gq$ nor $qq$ interactions.

There is a need for better understanding of parton distribution functions at
small $x$ where the PDFs are 
now very different from each other. In minbias events $x$
values of the order of $10^{-6}-10^{-4}$ are the most common. MC simulations need the PDFs to
range down to $x=10^{-8}$ which so far only the three brand new CT09 PDFs
do, and we would therefore like to encourage MSTW to extend their grid for the
MC-adapted PDFs in their next release. At several occasions we were reminded that it
can be risky to use NLO PDFs in LO MC generators. Our implementation of LO* and LO** with the new grid causes these two distributions to have a
smaller gluon distribution at small $x$, but would otherwise give
results more similar MSTW LO. In addition we found that there is a need for improved numerical stability at large
$x$ in order to keep the leading-order PDFs from going
negative. This could possibly be solved by using less intricate interpolation
routines in this area. Possibly one could choose an $x_{max}$, different
for sea-quarks, gluons and valence quarks,
above which one uses the form $N(1-x)^p$ where $N$ and $p$ are functions of the
virtuality, which would ensure positivity. The large differences in the PDFs get blurred when looking at
simulation results, but nonetheless
do sometimes cause large
variations. 

Changing from the default $\alpha_S$ behavior in \textsc{Pythia8} to
$\alpha_S$ value and running determined by the PDFs did not change the
results, once the multiplicity had been retuned. 

Interesting to note is that the CT09MCS seems to
have some of the features of the other MC-adapted PDFs but in some contexts
gives results more similar to ordinary leading-order PDFs.
The difference in quark rapidity between MC1/2/S and the rest could be due to their
fitting to the NLO pseudo data. The only two other distributions with similar
shape are the two NLO PDFs. 

We could see that a $K$-factor for the
leading-order PDFs could improve the fit to the inclusive jet data, but for the
MC-adapted PDFs the change of the ratio makes it more complicated. Finally, the
differences in the PDFs have a larger impact when the CM-energy of the
collisions increases, and this can cause large uncertainties in
simulations at LHC energies. 

At this point no final answer as to which PDF gives the best results. In order to answer this question one
has to look at a much broader spectrum of observables and also make complete
tunes for the different PDFs.

In the last years there has been a renewed interest in LO tunes with focus on
the applicability in MC generators. The MC-adapted PDFs resulted in some very
interesting differences compared to leading-order PDF but there is still
room for further improvements. With these new PDFs we have gained
a broader spectrum of tools in \textsc{Pythia8} and in examining the origin of
differences and similarities between simulations and experiments.

\section{Acknowledgments}
First and foremost I want to thank my supervisor Torbj\"{o}rn Sj\"{o}strand
for his help and ability to
explain difficult things in an understandable way. I also owe many thanks to
Leif L\"{o}nnblad, Hendrik Hoeth and all the other people at the department of
Theoretical Physics, as well as to Marianne D\"{o}\"{o}s, Sanne Kasemets and Julia
Kryszewska. I would also like to mention Johan Bijnens, Joakim Cederk\"{a}ll, Bo S\"{o}derberg and
Hans-Uno Bengtsson who all played an important role in my previous education.

\end{document}